# Natural Language Processing (NLP) for Requirements Engineering: A Systematic Mapping Study

Liping Zhao†

Department of Computer Science, The University of Manchester, Manchester, United Kingdom, liping.zhao@manchester.ac.uk

Waad Alhoshan

Department of Computer Science, The University of Manchester, Manchester, United Kingdom, waad.alhoshan@postgrad.manchester.ac.uk

Alessio Ferrari

Consiglio Nazionale delle Ricerche, Istituto di Scienza e Tecnologie dell'Informazione "A. Faedo" (CNR-ISTI), Pisa, Italy, alessio.ferrari@isti.cnr.it

Keletso J. Letsholo

Faculty of Computer Information Science, Higher Colleges of Technology, Abu Dhabi, United Arab Emirates, kletsholo@hct.ac.ae

Muideen A. Ajagbe

Department of Computer Science, The University of Manchester, Manchester, United Kingdom, muideen.ajagbe@manchester.ac.uk

Erol-Valeriu Chioasca

Exgence Ltd., CORE, Manchester, United Kingdom, erol@exgence.com

Riza T. Batista-Navarro

Department of Computer Science, The University of Manchester, Manchester, United Kingdom, riza.batista@manchester.ac.uk

† Corresponding author.

***Context:*** NLP4RE – Natural language processing (NLP) supported requirements engineering (RE) – is an area of research and development that seeks to apply NLP techniques, tools and resources to a variety of requirements documents or artifacts to support a range of linguistic analysis tasks performed at various RE phases. Such tasks include detecting language issues, identifying key domain concepts and establishing traceability links between requirements. ***Objective:*** This article surveys the landscape of NLP4RE research to understand the state of the art and identify open problems. ***Method***: The systematic mapping study approach is used to conduct this survey, which identified 404 relevant primary studies and reviewed them according to five research questions, cutting across five aspects of NLP4RE research, concerning the state of the literature, the state of empirical research, the research focus, the state of the practice, and the NLP technologies used. ***Results***: (i) NLP4RE is an active and thriving research area in RE that has amassed a large number of publications and attracted widespread attention from diverse communities; (ii) most NLP4RE studies are solution proposals having only been evaluated using a laboratory experiment or an example application; (iii) most NLP4RE studies have focused on the analysis phase, with detection as their central linguistic analysis task and requirements specification as their commonly processed document type; (iv) 130 new tools have been proposed by the selected studies to support a range of linguistic analysis tasks, but there is little evidence of adoption in the long term, although some industrial applications have been published; (v) 140 NLP techniques (e.g., POS tagging and tokenization), 66 NLP tools (e.g., Stanford CoreNLP and GATE) and 25 NLP resources (WordNet and British National Corpus) are extracted from the selected studies, but most of them – particularly those novel NLP techniques and specialized tools – are not in frequent use; by contrast, frequently used NLP technologies are syntactic analysis techniques, general-purpose tools and generic language lexicons. ***Conclusion***: There is a huge discrepancy between the state of the art and the state of the practice in current NLP4RE research, indicated by insufficient industrial validation of NLP4RE research, little evidence of industrial adoption of the proposed tools, the lack of shared RE-specific language resources, and the lack of NLP expertise in NLP4RE research to advise on the choice of NLP technologies.

# 1. Introduction

The important role of natural language (NL) in requirements engineering (RE) has long been established [1], [2]. In a survey published in 1981, aimed at providing an overview of techniques for expressing requirements and specifications, Abbott and Moorhead stated that "the best language for requirements is natural language [3]." While it is difficult to prove that NL is actually the best option, empirical evidence over the years has shown that it is at least the *most common* notation for expressing requirements in the industrial practice. The online survey of 151 software companies in the early 2000s by Mich et al. [4] concluded that in 95% percent of the cases requirements documents were expressed in some form of NL. This dominance of NL was confirmed by a recent survey of Kassab et al. [5], which involved 250 practitioners. The majority of the participants (61%) in that survey stated that NL was normally used in their companies for describing and specifying software and system requirements. Therefore, based on the past and current empirical evidence, we can safely assume that NL will continue to serve as the *lingua franca* for requirements in the future as well.

The close relationship between NL and requirements has been a source of inspiration for researchers to seek to apply natural language processing (NLP) techniques and tools to processing requirements texts [2]. Among the pioneering researchers are Chen [6] and Abbott [7], who, in the early 1980s, proposed using syntactic features of English sentences for database modeling and program design. These efforts were mostly based on extracting relevant entities from the requirements text based on simple syntactic rules, assuming that NL requirements were expressed in some constrained, predictable format, which, however, is not always the case in practice [8]. After these pioneering works, the beginning of 1990s saw some serious attempts to develop NLP tools for RE, introducing techniques to account for the complexity and variety of NL. Two well-known NLP tools, findphrases by Aguilera and Berry [9] and OICSI by Rolland and Proix [1], were the results of these efforts. While findphrases aimed to identify correlated words and phrases in a requirements text, OICSI intended to find concepts and relationships in the requirements, using lexical affinity and semantic cases [10], respectively.

For the remaining 1990s right up to the beginning of 2000s, a succession of NL tools had been proposed, among which were AbstFinder by Goldin and Berry [11], NL-OOPS by Mich [12], Circe by Ambriola and Gervasi [13], CM-Builder by Harmain and Gaizauskas [14], QuARS by Fabbrini et al. [15], and ARM by Wilson et al. [16]. Most of the works in that period were oriented to identify relevant entities in the requirements, to possibly produce some form of abstract model, and to identify requirements quality defects. The early 2000s appears to be a period of experimentation with new NLP techniques and new ideas. This time witnessed the application of information retrieval (IR) techniques to requirements tracing [17], part-of-speech (POS) tagging to tagging related requirements sentences [18] and statistical NLP techniques to identifying "shallow knowledge" from requirements text [19] as well as to tracing relationships between requirements [20]. Since the late 2000s, NLP supported RE – *NLP4RE* for short – has become a full-fledged research area [21], attracting researchers from the wider RE community. A large number of tools have since been developed, among which are SREE (Tjong and Berry [22]) for ambiguity detection and aToucan (Yue et al. [23]) for model generation. Recent developments include tools for requirements classification [24], detection of defects [25], smells [26] and equivalent requirements [27], glossary extraction [28], and requirements tracing [29].

With the recent widespread availability of NL content relevant to RE, such as feedback from users in app stores and social media, and developers' comments in discussion forums and bug tracking systems, we have observed a rising interest in using NLP techniques, combined with big data analysis, to support data-driven RE [30] and crowd-based RE [31]. These emerging areas aim to leverage information available from stakeholders' implicit and explicit feedback, to improve RE activities such as requirements elicitation and prioritization. Furthermore, given the increasing need to make software systems trustworthy, accountable, legally compliant, as well as security- and privacy-aware, and since most of the legally binding documents are expressed in NL, NLP has been largely applied also to legal documents [32] and privacy policies [33], in the field of RE and Law. Finally, to support Agile software development, requirements expressed in the form of *user stories* have been identified as an interesting area of application for NLP [34].

As a witness of this great interest in NLP applications to RE problems, a dedicated venue has also been set up in the form of workshop, called NLP4RE [35] and co-located with the International Working Conference on Requirements Engineering: Foundations for Software Quality (REFSQ), in both 2018 and 2019. This indicates that not only the research field is growing, but also an active community is emerging in RE.

Some companies have also started to develop NLP tools for RE. For example, Qualicen GmbH[1] developed Requirements Scout, a tool to analyze requirements specifications aiming to uncover requirements *smells*, i.e., defects; thingsThinking[2] proposed Semantha, a tool to perform document comparison on a semantic level; QRA Corp[3] includes QVscribe in its portfolio, a tool for quality and consistency checking; OSSENO Software GmbH[4] developed ReqSuite, a tool to support requirements writing and analysis; and even IBM has recently developed IBM Engineering Requirements Quality Assistant[5], a tool for automated requirements analysis and management that leverages the advanced NLP capabilities of IBM Watson[6]. This suggests that NLP4RE research is being transformed into a practical technology that can serve real world practice of RE.

However, in spite of its long history and this increasing interest, except for a small number of reviews that focus on specific topics of NLP4RE, there has been no effort to provide a comprehensive view of the field as a whole. We believe such an overview is crucial to the further development and success of the field, as it can play an important role in understanding the current state of NLP4RE research and identifying open problems. To this end, this article presents a first ever, large-scale systematic mapping study of NLP4RE research that we conducted in 2019. The mapping study reviewed 404 relevant primary studies reported between 1983 and April 2019, and structured them using a classification scheme to identify research trends and gaps. Five research questions were formulated to shape and steer the mapping study: What is the state of the literature on NLP4RE? What is the state of empirical research in NLP4RE? What is the focus of NLP4RE research? What is the state of the practice in NLP4RE? What are the enabling NLP technologies for NLP4RE research? The answers to these questions – that is, the results of the mapping study – are reported in this article.

To continue, Section 2 sets the scene by introducing the concepts of NLP and NLP4RE. Section 3 presents the related reviews to show the need for this mapping study. Section 4 describes the method for our mapping study while the mapping results are analyzed in Section 5. Section 6 reflects on the key findings and their implications for future research and practice. Section 7 discusses the validity threats to this mapping study and our countermeasures. Finally, Section 8 draws some conclusions based on the mapping study.

# 2. Concepts and Definitions

## 2.1 Natural Language Processing (NLP)

NLP is a field that employs computational techniques for the purpose of learning, understanding and producing human language content [36]. Liddy [37] provided this definition:

**Definition 1:** *Natural Language Processing is a theoretically motivated range of computational techniques for analyzing and representing naturally occurring texts at one or more levels of linguistic analysis for the purpose of achieving human-like language processing for a range of tasks or applications [37].*

[1] https://www.qualicen.de/en/
[2] https://www.thingsthinking.net
[3] https://qracorp.com
[4] https://www.osseno.com/en/
[5] https://www.ibm.com/us-en/marketplace/requirements-quality-assistant
[6] https://www.ibm.com/watson

In this definition, the notion of "*levels of linguistic analysis"* refers to the *phonetic, morphological*, *lexical, syntactic, semantic, discourse,* and *pragmatic analysis* of language [37], the assumption of which is that humans normally utilize all of these levels to produce or comprehend language [38]. NLP systems may support different levels, or combinations of levels of linguistic analysis. The more levels of analysis NLP systems support, the stronger or more capable these systems are supposed to be. Today, apart from a few pioneering efforts on discourse and pragmatic processing, start-of-the-art NLP technologies have only reached the lexical and syntactic processing levels for full-fledged English, with limited semantic capabilities [39].

The approaches to NLP can be broadly classified into *symbolic NLP* and *statistical NLP* [37]. Although both types of NLP have been investigated at the same time since the early days of the NLP field (circa 1950s), until the 1980s, it was the symbolic NLP that dominated the field.

Symbolic NLP emerged from artificial intelligence (AI). It is based on explicit representation of facts about language and associated algorithms and uses this knowledge to perform deep analysis of linguistic phenomena [37]. Symbolic NLP approaches include logic or rule-based systems, and semantic networks. In rule-based systems, linguistic knowledge is represented as facts or production rules, whereas in semantic networks, this knowledge is represented as a network of interconnected concepts.

However, symbolic NLP approaches lack the flexibility to adapt dynamically to new language phenomena, because they use the handwritten rules or the explicit representations built by human analysis of well-formulated examples to analyze input text [37]. Such rules may become too numerous to manage [40]. In addition, symbolic approaches may be frail when presented with unfamiliar, or ungrammatical input [37]. Beginning in the 1980s, but more widely in the 1990s, statistical NLP had regained popularity, as a result of the availability of critical computational resources and ML methods [36]. Since then, statistical NLP has been the mainstream NLP research and development [39]. For example, many of today's NLP tools such as POS taggers and syntactical parsers are based on statistical NLP [41].

In contrast to symbolic NLP, which uses detailed handwritten rules, statistical NLP employs various machine learning (ML) methods and large quantities of linguistic data (text corpora) to develop approximate, probabilistic models of language. These statistical models are simple and yet robust, because they are based on actual examples of linguistic phenomena provided by the text corpora, rather than deep analysis of linguistic phenomena as in symbolic NLP. When trained with large quantities of annotated language data, statistical NLP can produce good results, because it can learn most common cases in the copious data. Furthermore, the more abundant and representative the data, the better statistical NLP becomes.

However, statistical NLP can also degrade with unfamiliar or erroneous input [40], the problem similar to that of symbolic NLP. Furthermore, statistical NLP has been mainly useful for low-level NLP tasks such as lexical acquisition, parsing, POS tagging, collocations, and grammar learning [37]. Today, many text and sentiment classifiers of statistical NLP are still solely based on the use of the words of a text to ascertain the meaning of the text, rather than using structure and semantics of the sentences or discourses of the text. Also, most statistical models are trained with text corpora of everyday usage language, such as WSJ (Wall Street Journal) articles. Consequently statistical NLP can be unreliable for domain-specific text such as software requirements.

Most recently, circa 2012, deep learning (DL) methods began to emerge in the NLP scene [42]. The central idea of DL is that it allows a machine to be fed with a large amount of raw data and to automatically discover the representations or features needed for detection or classification [43]. Thus DL requires very little feature engineering by hand. Furthermore, the features learned by DL models are high-level, allowing for better generalization even over new, unseen data. By contrast, conventional ML techniques used in NLP are limited in their ability to process natural data in their raw form. This means that constructing a statistical NLP system requires careful engineering and considerable domain expertise to design a feature extractor that transforms the raw text into a suitable internal representation (i.e., feature vector), from which the ML subsystem, often a text classifier, can detect or classify patterns in the input. Both NLP and deep learning experts predict that NLP is an area in which deep learning could make a large impact over the next few years [36], [43]. Nonetheless, recent trends in deep learning based NLP show that coupling symbolic AI will be key for stepping forward in the path from NLP to natural language understanding [42]. This reaffirms the view that symbolic approaches and statistical approaches are complementary.

Our mapping study will focus on the application of NLP technologies to NLP4RE, regardless whether they are based on symbolic or statistical NLP.

## 2.2 Natural Language Processing for Requirements Engineering (NLP4RE)

Based on the definition of NLP (Definition 1), we define NLP4RE as follows:

**Definition 2:** *Natural language processing supported requirements engineering (NLP4RE) is an area of research and development that seeks to apply NLP technologies (techniques, tools and resources) to a variety of requirements documents or artifacts to support a range of linguistic analysis tasks performed at various RE phases.*

This definition has a number of key elements. First, we establish NLP4RE as *an area of research and development that seeks to apply NLP technologies*. This has to be the precondition for NLP4RE, because NLP4RE is motivated and enabled by NLP. We differentiate between three types of NLP technology: NLP technique, NLP tool and NLP resource. A *NLP technique* is a practical method, approach, process, or procedure for performing a particular NLP task, such as POS tagging, parsing or tokenizing. A *NLP tool* is a software system or a software library that supports one or more NLP techniques, such as Stanford CoreNLP[7], NLTK[8] or OpenNLP[9]. A *NLP resource* is a linguistic data resource for supporting NLP techniques or tools, which can be a language *lexicon* (i.e., dictionary) or a *corpus* (i.e., a collection of texts). Existing lexicons include WordNet[10] and FrameNet[11], whereas examples of corpus include British National Corpus[12] and Brown Corpus[13].

Second, NLP4RE deals with *a variety of requirements documents or artifacts*. Most of requirements documents are expected to be in NL. This is particularly so in early phase RE, in which requirements analysts may have to consult a wide variety of documents in order to develop an understanding of the problem domain. Such documents include interview scripts, legal documents, standards, and operational procedures [19]. More recently, online product reviews [44] have been found useful for understanding the needs and wants of end users. Consequently, the types of input to NLP4RE are broad and diverse.

Third, while NLP strives for *human-like language processing,* to achieve human-like performance [37], NLP4RE has a less ambitious goal, as its main objective is to *assist* requirements analysts in performing various linguistic analysis tasks [45], [46] for different RE areas or phases. Such tasks include detecting language issues, identifying key domain concepts and establishing traceability links between requirements, etc. As highlighted by Berry et al. [45], the goal of NLP4RE tools is not to replace the human analyst but to complement their work with those clerical or data intensive activities in which a computerized system can be more effective than a human.

We use this broad definition of NLP4RE to delineate the scope of our mapping study and to help identify relevant studies to NLP4RE.

# 3. Related Reviews

The RE literature counts several surveys with various degrees of relevance and quality, and covering some specific areas of interest. A recent survey of empirical RE research by Daneva et al. [47] identified 7 mapping studies and 49 systematic reviews, but none of them addresses the whole research field of NLP4RE targeted by our work. In conducting the literature search for this mapping study, we identified 18 reviews relevant to NLP4RE. Here, we provide a brief overview of these reviews, to show why we need to conduct the mapping study and why it is timely to do so.

Among those 18 reviews, four of them focus on *modeling* activities in software engineering. In particular, Loniewski et al. [48] present a survey of RE techniques in the context of model-driven development, including the cases where model transformation involved the requirements expressed in NL. Yue et al. [49] provide a review of different techniques for transforming textual requirements into analysis models. In their review, NLP support for model transformation is also considered. On the other hand, the review presented by Nicolás and Toval [50] focuses on the techniques used to generate NL or formal requirements from models. The review by Dermeval et al. [51] covers the studies that use ontologies for requirements modeling and shows that most of the reviewed studies dealt with textual requirements by means of artificial intelligence techniques, including NLP, to support different language analysis tasks.

[7] https://nlp.stanford.edu/software/
[8] https://www.nltk.org
[9] https://opennlp.apache.org
[10] https://wordnet.princeton.edu/
[11] https://framenet.icsi.berkeley.edu/fndrupal/
[12] http://www.natcorp.ox.ac.uk/
[13] http://clu.uni.no/icame/manuals/

Another group of reviews is concerned with topics related to *requirements management*, including retrieval, tracing and classification of requirements. Specifically, Irshad et al. [52] reviewed the papers on requirements reuse, including the work that applied NLP and IR techniques to text similarity evaluation to match input queries with existing requirements in a repository. Torkar et al. [53] reviewed the different methodologies to support traceability, considering also approaches that use some textual analysis support. On a different note, but still related to requirements management, Binkhonain and Zhao [54] reviewed research on applying ML and NLP techniques to the classification of non-functional requirements. More oriented towards software product line engineering, Bakar et al. [55] surveyed the usage of NLP techniques in the identification of common and variant features in NL requirements documents as well as other NL sources, such as NL product descriptions. Building on this survey, Li et al. [56] focused on identifying features and analyzing their relationships in textual requirements.

A rather recent, yet lively, group of reviews is concerned with analysis of publicly available *feedback* produced by users and developers. Among these reviews, Martin et al. [57] reported a mapping study of the research on app store analysis in software engineering and identified NLP as relevant tools for feature analysis and app review analysis. Within the field of app review analysis, Tavakoli et al. [58] reviewed the application of ML and NLP techniques to extracting and classifying useful information from users' feedback, so as to distinguish between requirements-relevant information and other types of users' comments. Two recent reviews by Santos et al. [59], [60] were concerned with classification of app reviews and users' feedback. Finally, with a broader focus on developers' feedback, Nazar et al. [61] reviewed works on summarization of the various data sources, including bug reports, mailing lists and developer forums, to extract requirements related information.

Other reviews related to ours, but more limited in scope and extension, are those by: Ahsan et al. [62], on test generation from requirements; Shah et al. [63] on ambiguity detection in requirement also by means of NLP; Casamayor et al. [64], presenting a non-systematic survey on text mining and NLP in the field of model-driven design; and Nazir et al. [65], pursuing a mapping study on NLP for RE analogous to ours, but considering only 27 primary studies. Recently, an MSc Thesis reported on a systematic literature review on using NLP for requirements elicitation and analysis [66]. Although very rigorous, the review is limited to both its scope, as it only focused on requirements elicitation and analysis, and the number of studies included, as it surveyed 144 studies.

On the one hand, this vast landscape of recent reviews touching the field of NLP4RE shows that specific sub-fields have attracted increasing interest in the last years, raising the need for secondary studies to scope and guide the research in these areas. On the other hand, while specific NLP4RE topics are addressed, none of the existing reviews provides a comprehensive mapping study on NLP4RE, therefore making the current work particularly timely and useful to better conceptualize and identify potential synergies among the different areas of investigation, and establish novel research direction in this growing research field.

# 4. Review Method

To achieve our goal, we have carried out a systematic mapping study [67] using the basic method presented in [68]. This section presents our research questions and describes the main activities involved in the mapping study; we report our mapping results in Section 5.

## 4.1 Research Questions

The research questions (RQs) for our mapping study are stated as follows:

*RQ1: What is the state of the literature on NLP4RE?* Specifically, what is the population of the published literature on NLP4RE? What is the publication timeline? What are the leading publication venues?

*RQ2: What is the state of empirical research in NLP4RE?* Specifically, what types of research have been carried out in the area of NLP4RE? What types of evaluation method have been used in the research? What relationships can be observed between these research types and evaluation methods?

*RQ3: What is the focus of the NLP4RE research?* Specifically, what RE phases have been addressed? What linguistic analysis tasks have been investigated for these phases? What is the relationship between these RE phases and tasks? What types of input document have been considered?

*RQ4: What is the state of the practice in NLP4RE?* Specifically, what new tools have been developed? Which RE phases and tasks do these tools support? Which of these tools are available to the public?

*RQ5: What are the enabling NLP technologies for the NLP4RE research?* Specifically, what NLP techniques, tools and resources have been employed? Which ones are most popular? What relationships can be observed between NLP technologies and NLP4RE tasks?

These five main RQs are interrelated, designed to interrogate the NLP4RE literature progressively, from the state of the literature (RQ1), to the state of empirical research reported in this literature (RQ2), to the focus of the NLP4RE research (RQ3), to the state of the practice in the NLP4RE area (RQ4), and finally, to the NLP technologies that enable NLP4RE research and practice (RQ5). Collectively, these RQs define the scope of our mapping study and articulate what we want to accomplish by doing this study. Each RQ is further elaborated into a set of specific questions, allowing us to survey the NLP4RE literature in great detail. Consequently, the answers to these specific questions collectively provide the answers to the five main RQs.

## 4.2 Study Selection Process

### 4.2.1 Determining the Digital Libraries

We have identified the following digital libraries as the data sources for our mapping study: ACM Digital Library (ACM); IEEE Xplore Digital Library (Xplore); ScienceDirect (SD); SpringerLink (SL). These libraries were chosen because they host the major journals and conference proceedings related to software engineering (SE) and RE. To complement these libraries, we have also selected Association for Computational Linguistics Library (ACL), where the major contributions to NLP are likely to be published. These five libraries serve as our data sources for identifying the relevant literature.

### 4.2.2 Formulating the Search Strategy

Our search strategy was based on the direct search of the electronic databases of the aforementioned five digital libraries. The search terms for querying these libraries were constructed using the steps presented in [69]. Specifically, we used the major terms "requirements engineering" (representing the context of the research) and "natural language processing" (representing the intervention in this context) as the base terms; elaborate each base term with alternative spellings and synonyms; use the Boolean OR to incorporate synonyms, alternative spellings, alternative terms, and sub-field terms into each base term set, and Boolean AND to link the two sets of terms. Several iterations were performed to identify and refine the keywords. The complete set of the search terms is presented in Table 1.

**Table 1: Keywords for Identifying the NLP4RE Literature**

| Main Keywords | Derived Keywords |
|---|---|
| Requirements engineering | Requirements elicitation, requirements analysis, requirements specification, requirements modeling, requirements modeling, requirements validation, requirements verification, requirements management, requirements traceability, requirements classification, requirements document, requirements specification |
| Natural language processing | NLP, statistical NLP, machine learning, deep learning, information extraction, information retrieval, text mining, text analysis, linguistic instruments, linguistic approaches |

### 4.2.3 Performing the Literature Search

The library search was completed in April 2019 and the results were downloaded and imported into an Endnote library where all duplicates were automatically removed by Endnote. Multiple pilot studies were performed to adjust the string and the categories before the final search. For the first four libraries, no publication timeline was specified in the queries; however, the search of ACL was restricted to the last 10 years, as NLP4RE research has only established itself as a subfield in RE in the recent years and we wanted to check whether it had some recent impact also on the NLP community. By default, only the publications written in English were retrieved from all libraries except SL and ACL, for which we explicitly set the language to English. The final results from the main search were 11,406. These results were then combined with the initial search results. Endnote automatically removed the duplicates and the remaining combined results are 11,489. Finally, we conducted a complementary search on Google Scholar, to check for additional studies not included in the main search results. This search retrieved 51 studies. Adding them to the search results in our Endnote library gave us a total of 11,540 items for study selection.

#### 4.2.4 Selecting the Relevant Studies

We defined a set of inclusion and exclusion criteria to guide us in study selection. These criteria are shown in Table 2. Our criteria explicitly exclude the short papers, because in general such papers lack detailed description of their contributions and including them can potentially skew the results of the mapping study.

**Table 2: Inclusion (I) and Exclusion (E) Criteria for Selecting Relevant Studies**

| I/E | No. | Criteria |
|---|---|---|
| I | 1 | Include peer-reviewed primary studies that are relevant to NLP4RE (cross-checking and validation needed for such studies). |
| I | 2 | If there are multiple relevant studies that report the same research, include the longest study only and exclude the rest of them. |
| E | 1 | Exclude tables of contents, editorials, white papers, commentaries, extended abstracts, communications, books, tutorials, non-peer reviewed papers, and duplicate papers. |
| E | 2 | Exclude short papers that have fewer than 6 pages if they are in a single column format. |
| E | 3 | Exclude reviews or secondary studies. |
| E | 4 | Exclude papers that are clearly not relevant to NLP4RE (cross-checking and validation needed for such papers). |

Based on these criteria, a team of four *data inspectors* (Alhoshan, Letsholo, Ajagbe, Chioasca) and three *supervisors* (Zhao, Ferrari, Batista-Navarro) carried out study selection. The study selection process, depicted in Figure 1, is as follows.

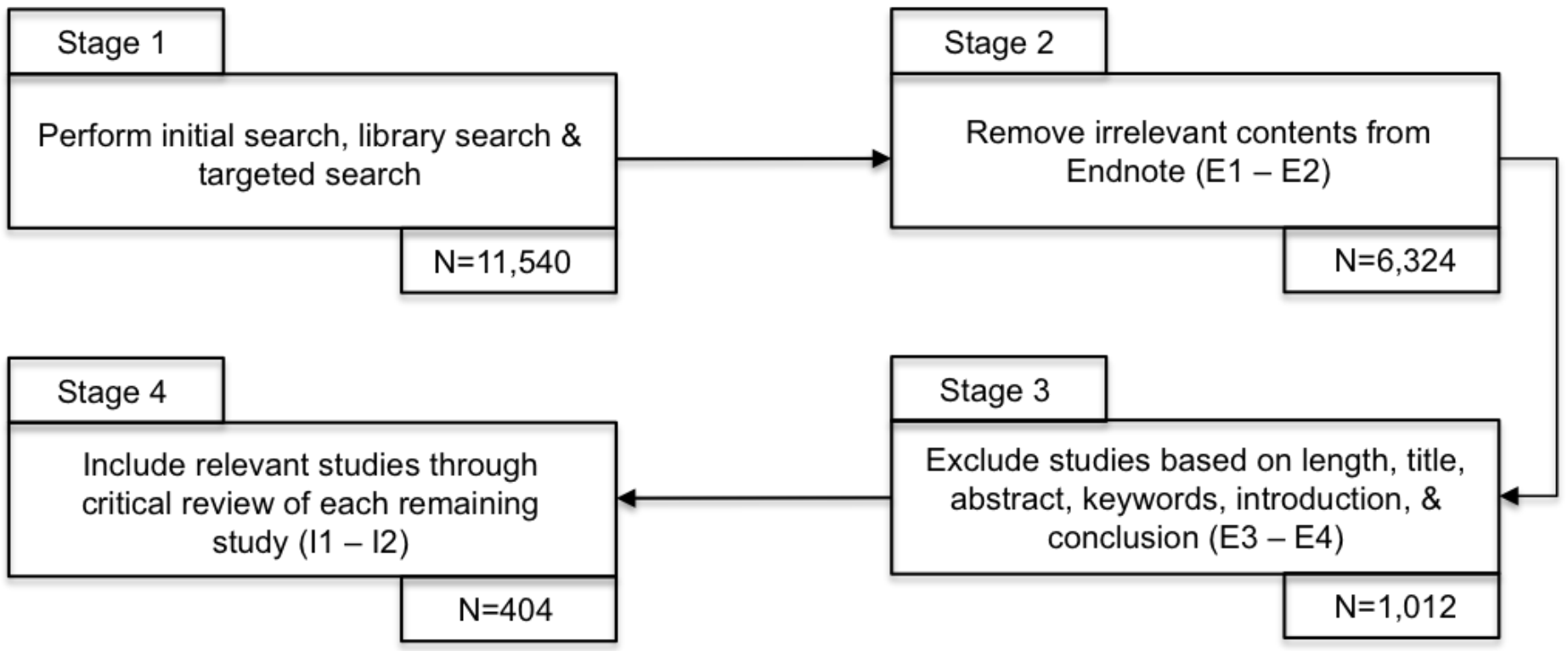

**Figure 1: Study selection process.**

The inspectors initially applied E1 and E2 (Table 2) to the Endnote library, to remove irrelevant contents such as editorials, commentaries and so on. They also performed a series of checks to ensure no identical papers remained in the library. After this filtering, there were 6,324 papers left in the process. In the next step, the inspectors checked the length, title, abstract, introduction, and conclusion of each study, and removed short papers, irrelevant papers and secondary studies in accordance with E3 and E4. This step was autonomously performed by the inspectors, with the support of the lead supervisor (Zhao) for undecided cases.

The remaining 1,012 primary studies were divided between the four data inspectors, who independently reviewed their allocated studies according to I1 and I2 to determine if each of these studies should be included or excluded. This involved carefully reading the full text of each study to establish its relevance and to identify its key components with respect to our predefined categories. During the selection, the supervisors performed regular and random checking on the selected and deselected studies to ensure they were correctly included or excluded. Any discrepancies and inconsistencies were identified and immediately notified to the responsible inspectors. After individual selection, the inspectors crosschecked each other's results. Undecided cases were resolved with the involvement of all three supervisors. The selected studies by individual inspectors were then combined, resulting in a total of 416 studies. The lead supervisor carried out the final check on these 416 studies, by examining the title and abstract of each study, and in undecided cases, by reading the text of the study. This identified 4 duplicate studies and 8 irrelevant studies. The remaining 404 studies were included in our mapping study. The references of these studies are given in Appendix 1.

## 4.3 Data Extraction and Classification

### 4.3.1 Defining the Classification Scheme

Building a classification scheme is the central task of any mapping study, as the main purpose of a mapping study is to classify the literature [68], [70]. Our classification scheme reflects the five research questions. It is made up of four facets, each containing a set of categories. Figure 2 depicts this classification scheme, where the number associated with each category is the number of occurrences in that category, to be discussed in Section 5. The four facets and their categories are described as follows.

1. *Publication Facet.* This facet is for classifying the publication information. The resulting classification will be used to answer RQ1. This facet contains the following three categories: *Publication Types, Publication Venues* and *Publication Years*.
2. *Research Facet.* This facet is for classifying the selected studies according to their research types and evaluation methods. The resulting classification will be used to answer RQ2. This facet contains the following two categories:
   - *Research Types.* Five types of research are considered, using the taxonomy of Wieringa et al. [71] (see Table 3).
   - *Evaluation Methods.* Nine types of evaluation methods are considered, according to Chen and Babar [72] (see Table 4).

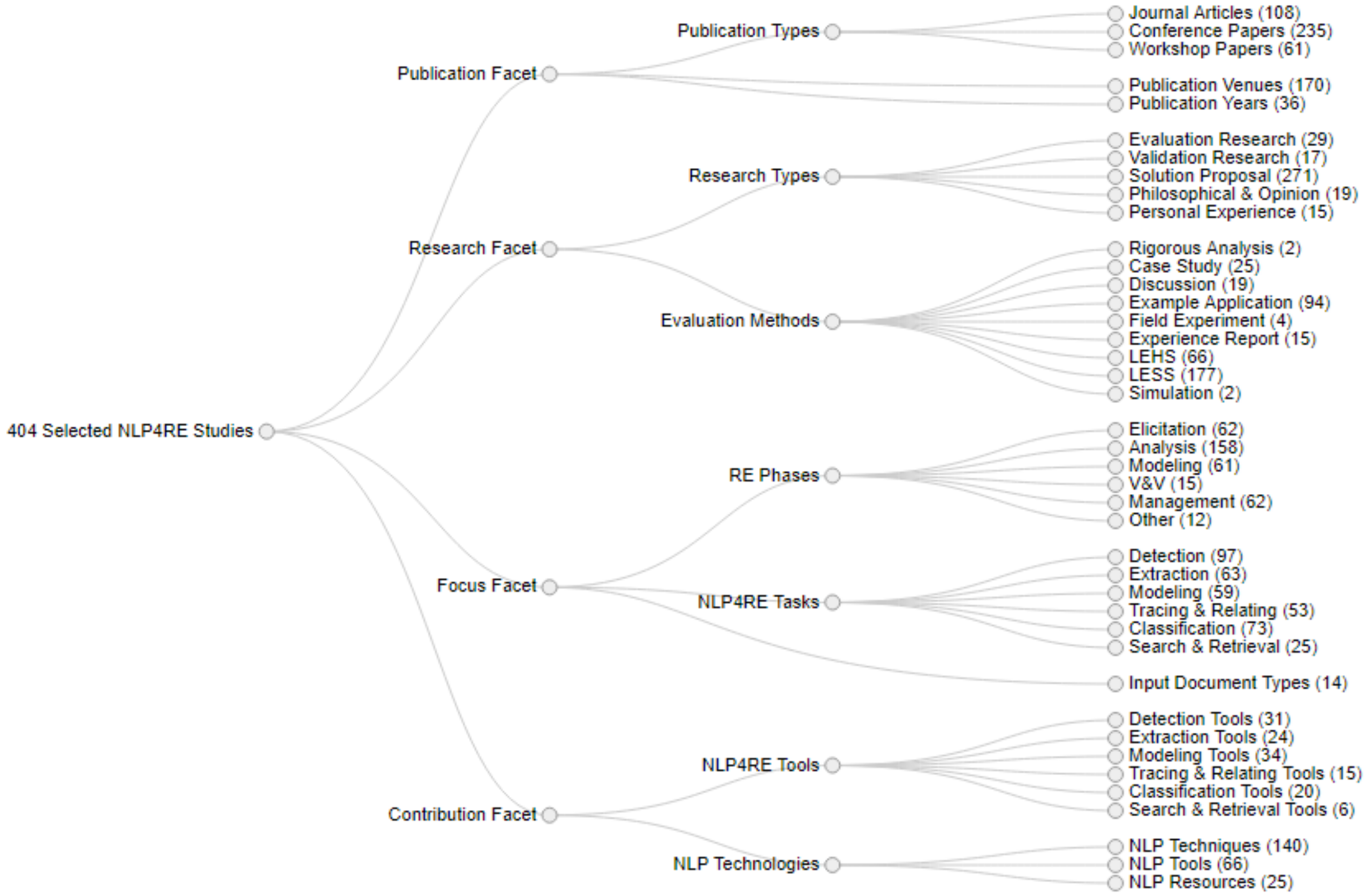

**Figure 2: Faceted classification scheme for mapping the NLP4RE literature.**

3. *Focus Facet.* This facet is for classifying the selected studies according to their RE phases, NLP4RE tasks and input documents. The resulting classification will be used to answer RQ3. This facet contains the following three categories:
   - *RE Phases.* The RE phases are defined in Table 5, where "Other" is a placeholder, which can be replaced by Testing or Design, or other phases of the software process. The first five RE phases are based on the work of Cheng and Atlee [73].
   - *NLP4RE Tasks.* NLP4RE tasks are linguistic analysis tasks performed during RE process. We use six general types NLP4RE task for our mapping study, as shown in Table 6. The first four tasks are based roughly on the work of Berry et al. [45], whereas the last two are defined by us.
   - *Input Document Types.* The selected studies are classified into categories according to the input document type they use. The document types used in this classification have not been predefined; they will be synthesized from the extracted data.
4. *Contribution Facet.* This facet is for classifying the contribution and the underlying technologies of the selected studies. The resulting classification will be used to answer RQ4 and RQ5. This facet contains the following two categories:
   - *NLP4RE Tools.* These tools are the results reported by the NLP4RE literature.
   - *NLP Technologies.* NLP technologies used by the selected studies are classified into three types: NLP Technique, NLP Tool and NLP Resource.

**Table 3: Types of Research for Classifying the NLP4RE Literature (Adapted from Wieringa et al. [71])**

| Research Type | Explanation |
|---|---|
| Evaluation Research | This type of research involves empirical evaluation from industries, often in the form of case study or field study. |
| Validation Research | This type of research does not involve new technologies; instead, it is concerned with validating existing technologies (methods, tools, algorithms etc.). Comparative studies are typically validation research. |
| Solution Proposal | This type of research typically involves the development of a novel solution (e.g., a new method, technique or tool) to a problem. This type of research may contain validation research. |
| Philosophical/ Opinion | This type of research covers a broad range of work, under the guise of vision statements, position papers, opinions, viewpoints etc. This type of research is conceptual or theoretical, so its outcome is often a new understanding or a new perspective of some research area. |
| Personal Experience | This type of research is reflectional, based on personal experience of applying an existing technology to a real-world problem. |

**Table 4: Types of Evaluation Method for Classifying the NLP4RE Literature (based on Chen and Babar [72])**

| Evaluation Method | Explanation |
|---|---|
| Rigorous Analysis | Rigorous derivation and proof, suited for formal model. |
| Case Study | An empirical inquiry that investigates a contemporary phenomenon within its real-life context; when the boundaries between phenomenon and context are not clearly evident; and in which multiple sources of evidence are used. |
| Discussion | Provided some qualitative, textual, opinion. |
| Example | Authors describing an application and provide an example to assist in the description, but the example is "used to validate" or "evaluate" as far as the authors suggest. |
| Experience Report | The result has been used on real examples, but not in the form of case studies or controlled experiments, the evidence of its use is collected informally or formally. |
| Field Experiment | Controlled experiment performed in industry settings. |
| Laboratory Experiment with Human Subjects (LEHS) | Identification of precise relationships between variables in a designed controlled environment using human subjects and quantitative techniques. |
| Laboratory Experiment with Software Subjects (LESS) | A laboratory experiment to compare the performance of newly proposed system with other existing systems. |
| Simulation | Execution of a system with artificial data, using a model of the real word. |

**Table 5: RE Phases for Classifying the NLP4RE Literature (Based on Cheng and Atlee [73])**

| RE Phase | Explanation |
|---|---|
| Elicitation | This phase comprises activities that enable the understanding of the goals, objectives, and motives for building a proposed software system. |
| Analysis | This phase involves evaluating the quality of recorded requirements and identifying anomalies in requirements such as ambiguity, inconsistency and incompleteness. |
| Modeling | This phase involves building conceptual models of requirements that are amenable to interpretation. |
| Validation & Verification (V&V) | Requirements validation ensures that models and documentation accurately express the stakeholders' needs. Validation usually requires stakeholders to be directly involved in reviewing the requirements artifacts. Verification entails proving that the software specification meets these requirements. Such proofs often take the form of checking that a specification model satisfies some constraint (model checking). |
| Management | This is an umbrella activity that comprises a number of tasks related to the management of requirements, including the evolution of requirements over time and across product families, and the task of identifying and documenting traceability links among requirements artifacts and between requirements and downstream artifacts. |
| Other | This is an open-end category that allows us to record other NLP4RE related software development activities. For example, during software testing, NLP may be used to analyze requirements to generate test cases. In this case "Other" will be replaced by "Testing". During software design, NLP may be used to transform requirements into design artifacts. "Other" will be replaced by "Design". |

**Table 6: NLP4RE Tasks for Classifying the NLP4RE Literature**

| NLP4RE Task | Meaning | Explanation |
|---|---|---|
| Detection | Detect linguistic issues in requirements documents | This task is typically to support manual review activities to make the requirements, or requirements-related artifacts, clear and unequivocal. The linguistic issues to be detected may range from the controversial usage of passive voice, to the occurrence of typically vague phrases (e.g., *as soon as possible*, *after some time*) or weak verbs (e.g., *may*, *could*), to the presence of syntactic and pragmatic ambiguities. Also checking the adherence to pre-defined requirements templates, and identifying equivalent requirements, can be included in this task, as the goal is still to enforce rigor in requirements texts. |
| Extraction | Identify key domain abstractions and concepts | This task normally aims to extract single or multi-word terms from requirements texts to establish domain-specific and project-specific *glossaries*, as requirements often include domain-specific, compound terms that are not commonly used. The extracted glossaries can be exploited for other objectives, including completeness or consistency checking, product comparison, classification, and modeling. |
| Classification | Classify requirements into different categories | This task aims to classify requirements into different types, base on the purpose for which the task is applied. For example, requirements can be categorized based on their *functional* category, to ease requirements apportionment and reuse or based on its *quality* category, to identify non-functional requirements that may be hidden within functional ones. Also, when applied to users' feedback and online discussions, classification can help identifying feedback that is specifically concerned with new requirements, or feedback referring to specific features of interest, possibly with the sentiment expressed by the product's users. |
| Modeling | Identify modeling concepts and constructing conceptual models | The task typically makes use of the extraction task, and can take different flavors, from the generation of UML models to support analysis and design, to the synthesis of feature models in a product-line engineering context, to the generation of high-level models of early requirements or user stories to support project scoping. |
| Tracing & Relating | Establish traceability links or relationships between requirements, or between requirements and other software artifacts such as models, code, test cases, and regulations | This task mainly aims to support manual tracing activities oriented to enforce and demonstrate process consistency, especially in a regulated context or in large-scale enterprise software. We include in this class also those works dealing with change impact analysis, as they also address the problem of identifying relationships between requirements or other artifacts. |
| Search & Retrieval | Search and retrieve requirements or requirements sets from existing repositories | The goal of this task can be to reuse existing requirements assets to match with the needs of novel customers, or to support domain scoping towards the development of new product, by recommending specific features based on existing software descriptions available online |

#### 4.3.2 Extracting and Aggregating the Data

Our data extraction process was organized into four separate phases, described as follows. In Phase 1, we extracted the data for the categories and subcategories of the publication facet. These data were automatically obtained from our Endnote library where the selected studies were stored.

Phase 2 involved extracting the data for the categories and subcategories of the remaining three facets. To do so we first created a data extraction form on Google Sheets. The categories of the three facets (research, focus and contribution) were mapped onto the columns in the form. Some categories, i.e., RE phases and NLP4RE tasks, were given extra columns, to allow additional RE phases or tasks to be recorded. The rows of the form were used to record the data related to the selected studies, one row per study, and each row was identified by Study ID of a study. Next, we divided the selected studies evenly among the four data inspectors, who reviewed each study and performed data extraction for each study.

In parallel with the inspectors, the supervisors performed data extraction on the randomly selected studies and then compared the results with those obtained by the inspectors. Any discrepancy and inconsistency were discussed with the inspectors and actions taken to ensure the accuracy and consistency of the extracted data across the board.

After the completion of data extraction, in Phase 3, we carried out thematic synthesis on the descriptions related to input document types to identify specific input document types. We also performed thematic synthesis on the descriptions related to NLP techniques, NLP tools and NLP resources to create a coherent set of names for these technologies. The steps recommended by Cruzes and Dybå [74] were adopted for our thematic synthesis: code each piece of text with a label or term; translate related codes into the themes; organize the themes into high-level themes.

Finally, in Phase 4, we carried out data cleaning for each category to harmonize and standardize the terms in it; we inspected the members in each category to ensure our classification was accurate and consistent; we conducted statistical analysis of the categories; we employed a variety of visual tools [75], including tables, bar charts and dendrograms, to produce the visual representations of the analysis results [75], [76]. In the next section, we answer our five RQs based on the mapping results.

# 5. Mapping Results

## 5.1 RQ1: State of NLP4RE Literature

*Population of the NLP4RE literature.* A total of 404 primary studies relevant to NLP4RE have been identified by our mapping study, comprising 26.75% (108) journal articles, 58.17% (235) conference papers and 15.10% (61) workshop papers. Such a trend, as we observed, is consistent with other areas in RE [77] and SE [78].

*Publication timeline.* These studies have been published over the past 36 years, from 1983 to 2019 as Figure 3 shows (NB. The number of studies published in 2019 is incomplete, as our library search ended in April 2019); however, the majority of these studies (88.61% or 358 out of 404) have actually been published since 2004. This means that before 2004, an average publication rate was roughly two papers per year, whereas after 2004 it was about 24 papers per year. This rapid growth can be attributed to technological advances in NLP over the past 15 years or so [36].

*Publication venues.* These studies were found from 170 different publication venues (see Appendix 2). Of these, 12 venues are identified as the leading venues for these studies (see Figure 4), as they have collectively published 45.05% (or 182) of the selected studies. Thus, on average, each of these 12 venues has published 15.17 papers, whereas each of the remaining 158 venues has only published 1.41 papers. Although this extensive number of diverse venues indicates that NLP4RE is a topic of general interest – though not necessarily central – to diverse communities, such a wide spread of the venues also has its downside, as it makes locating relevant NLP4RE work and building a consolidated NLP4RE knowledge base a difficult task.

These 12 leading venues count some top RE and SE conferences and journals, as Figure 4 shows. It can be observed that the three top RE venues (RE, REFSQ and REJ) have published a quarter of all selected studies, indicating that these are the core publication channels for NLP4RE research. As it can also be observed from Figure 4, the five top SE conferences and journals (JSS, ASE, ICSE, IST, and TSE) have published more than a 10 percent of the studies, indicating that NLP4RE research has a broad audience in the SE community. To assess the impact of NLP4RE research in the SE community, we compared the proportion of the NLP4RE papers published at ICSE with the proportion of the GORE (Goal-Oriented Requirements Engineering) papers published at ICSE [77]. We found that NLP4RE has produced 9/404 (2.25%) ICSE papers, whereas GORE has produced 5/246 (2.03%), indicating that NLP4RE has made a similar impact on SE as CORE. Given that GORE is a well-established subfield in RE with a long research tradition, these comparable results show that NLP4RE has now become an important research area in RE, with relevant impact outside the RE field.

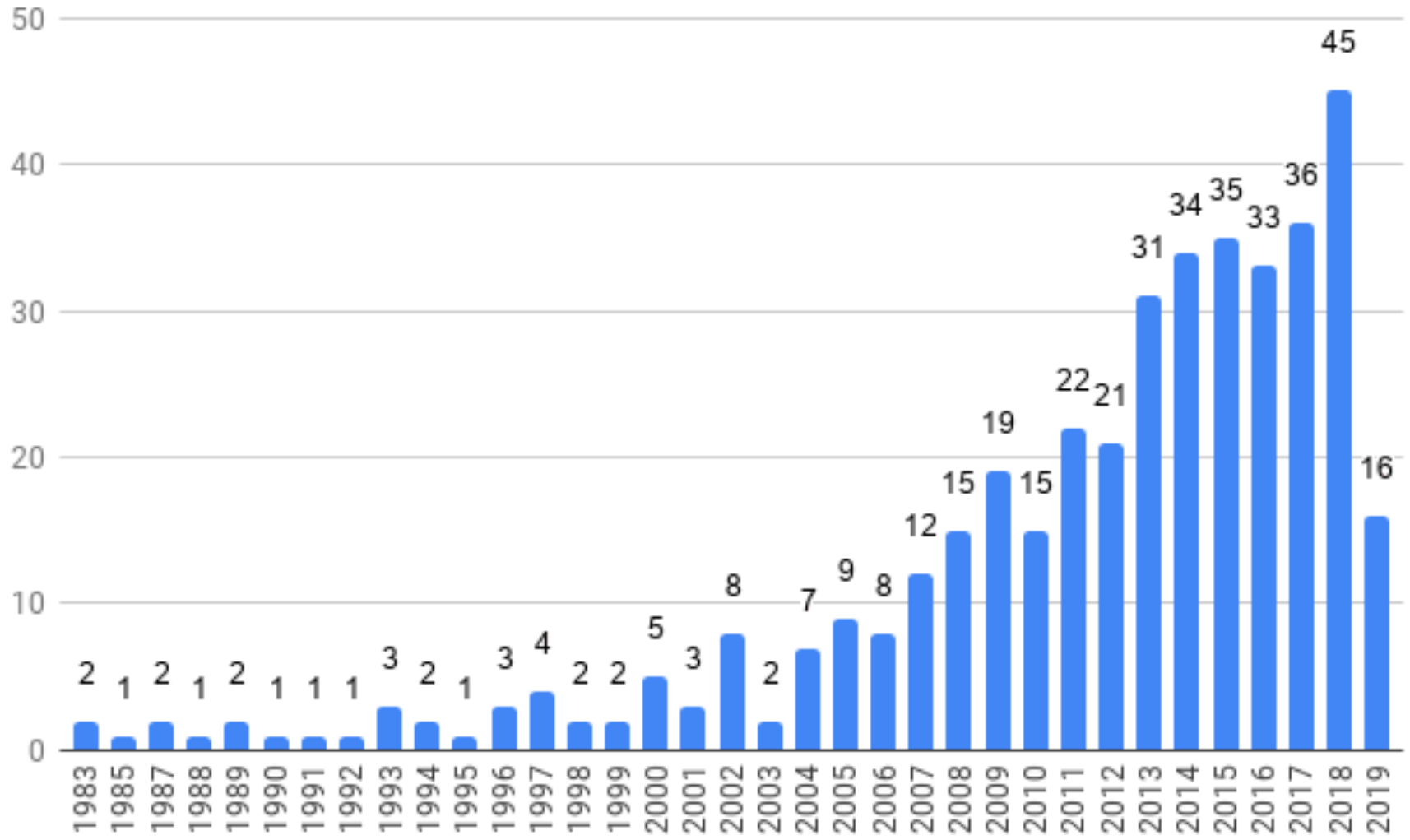

**Figure 3: Publication timeline of the 404 selected studies.**

Although the ACL library was included in our library search, we only found two relevant studies from it. The reason might be that the main publication venues for NLP4RE still focus on RE and SE, and that the interest of the NLP community in RE-related studies is still limited. As NLP represents one crucial facet of NLP4RE, engagement with the NLP research community is important for NLP4RE researchers. One way of doing this is to take NLP4RE research results to NLP conferences or journals for validation and feedback.

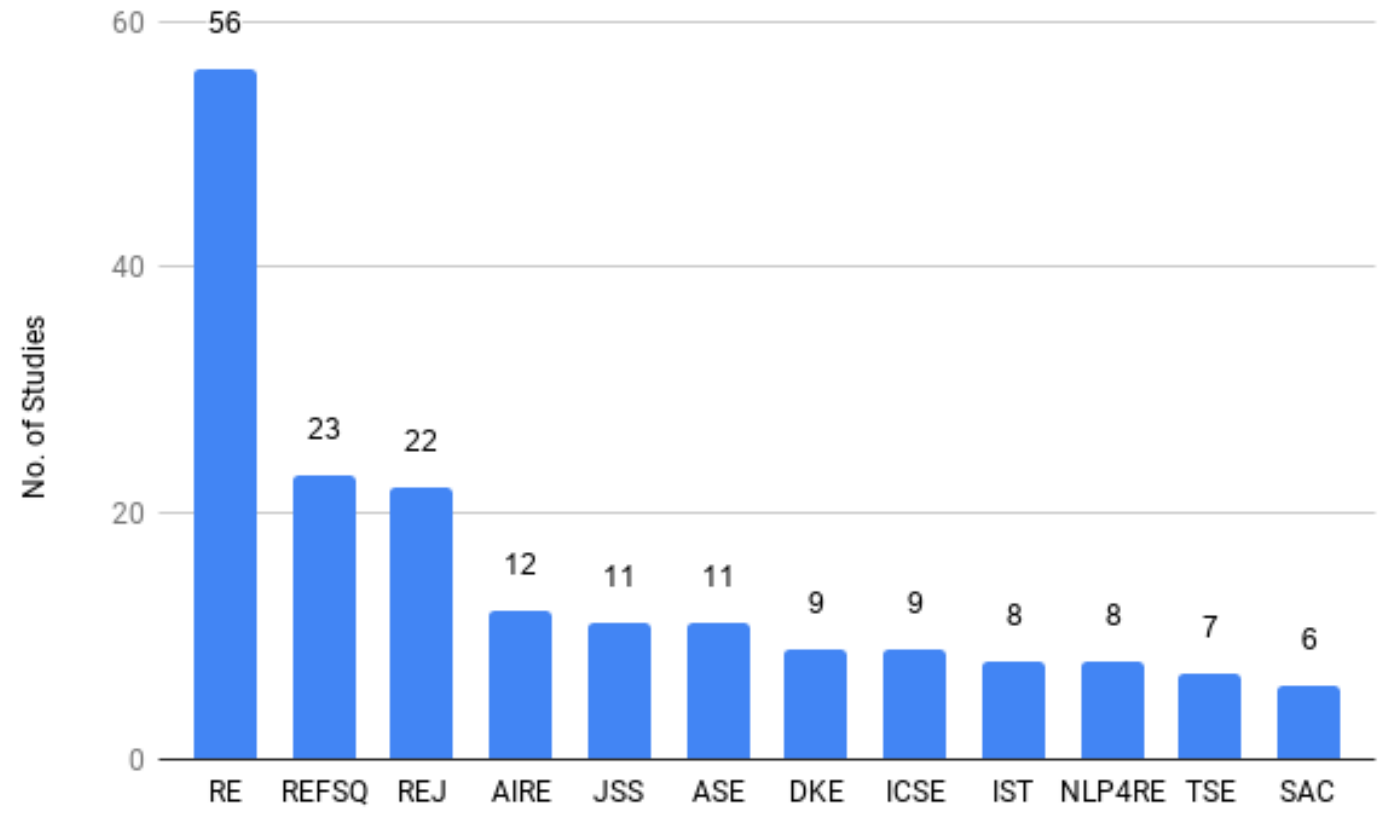

**Figure 4: Leading publication venues for NLP4RE research.**

## 5.2 RQ2: State of Empirical Research in NLP4RE

To understand the state of the empirical research in NLP4RE, we analyze the types of research reported in the 404 studies, the evaluation methods used to validate these studies and the relationships between them. Figure 5 shows the distribution of the selected studies among different research types, while Figure 6 depicts the distribution of the selected studies among different evaluation methods.

*Types of NLP4RE research.* Figure 5 shows that Solution Proposal is the most frequently used research type among the NLP4RE studies (67.08%). Validation Research is in a distant second, only on 17.33% studies. The remaining research types are less frequently used, ranked as follows: Evaluation Research (7.18%), Philosophical & Opinion (4.70%) and Personal Experience (3.71%). Such a trend was also reported in systematic reviews on other RE and SE areas [77], [78].

*Types of evaluation method for NLP4RE research.* From Figure 6, it is evident that Laboratory Experiment with Software Subjects (LESS) is the most frequently used evaluation method, followed by Example Application and Laboratory Experiment with Human Subjects (LEHS). Other evaluation methods are used less. In particular, Case Study, Discussion and Experience Report are only used by a small number of studies, while Field Experiment, Rigorous Analysis and Simulation are rarely used.

*State of empirical research in NLP4RE.* Empirical Research is the research that uses empirical evidence. Based on the types of empirical method used in SE [79], for the identified evaluation methods used for NLP4RE research, only LESS, LEHS, Case Study, and Field Experiment are counted as empirical methods, whereas the remaining methods are not. These four empirical methods form an experimental path that starts with laboratory experiments (LESS or LEHS) and ends with real-world validation using Case Study or Field Experiment [80]. While LESS and LEHS are conducted in a very controlled environment with simplified reality, Case Study and Field Experiment are performed in a less controlled and real environment. As more than 60% (243) studies are evaluated by a laboratory experiment (LESS or LEHS) as opposed to 7% by a Case Study or Field Experiment, Empirical Research in NLP4RE is still at the early stage of the experimental path, waiting to be evaluated rigorously by industrial case studies or field experiments.

*Relationships between NLP4RE research types and evaluation methods.* We now examine the relationships between the research types and the evaluation methods found in the selected studies. Using the pivot table (Table 7), we can observe an alignment between the research types and the evaluation methods applied. This alignment suggests that the selected studies have used appropriate evaluation methods for research validation. A further insight from Table 7 is that *the typical NLP4RE study is a solution proposal, possibly evaluated through an experiment or example application, but without an evaluation in an actual industrial context.*

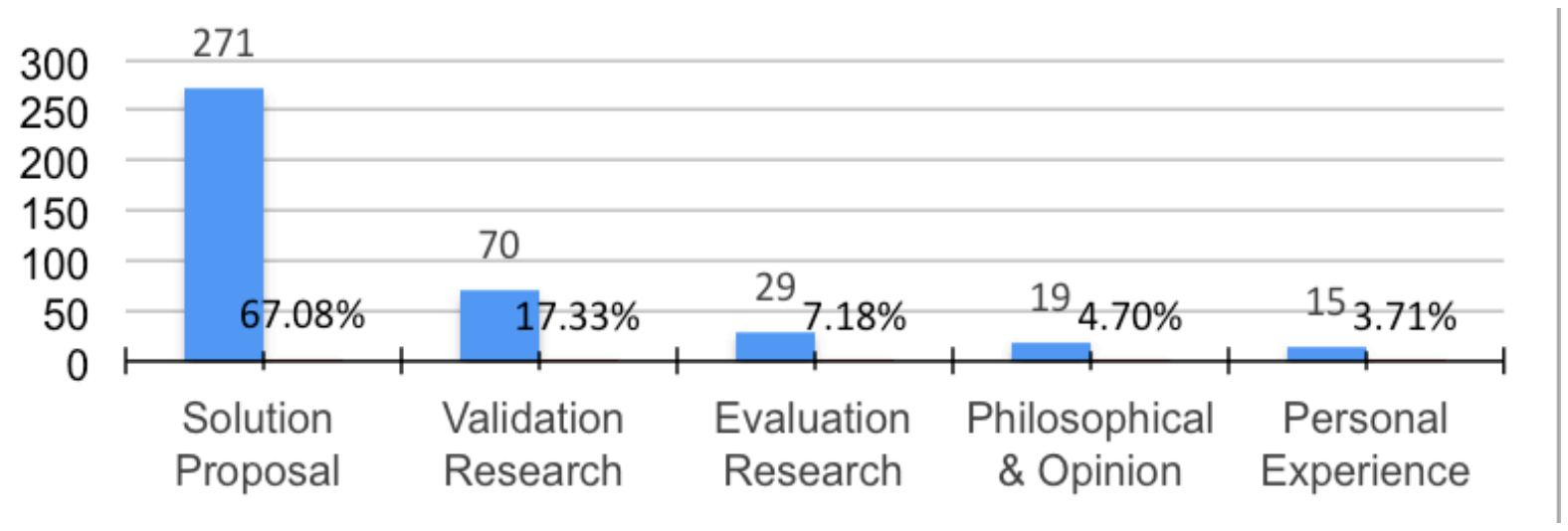

**Figure 5: Distribution of the selected studies to different research types.**

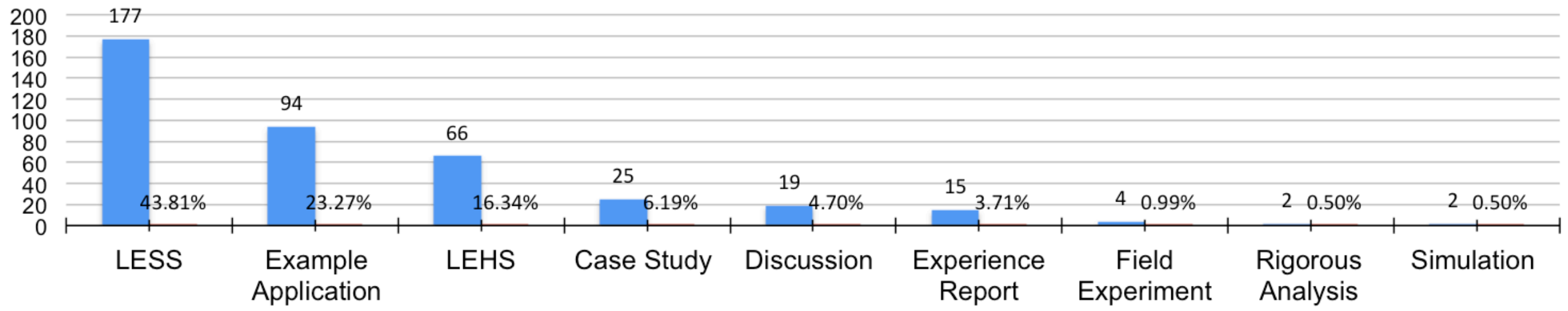

**Figure 6: Distribution of the selected studies to different evaluation methods.**

## 5.3 RQ3: Focus of NLP4RE Research

To address this research question, we exclude personal experience and philosophical & opinion papers, because they do not provide data pertinent to this question. We are therefore left with 370 studies for RQ3. These studies consist of 271 solution proposals, 70 validation research studies and 29 evaluation research studies.

*Most targeted and least targeted RE phases.* Figure 7 shows the distribution of the 370 studies by the RE phases. Clearly, Analysis is by far the most targeted RE phase, followed by Management, Elicitation and Modeling. V&V (validation & verification), Testing and Design are the least targeted phases. Analysis as the most targeted phase by NLP4RE research may be due to its broad role in RE. The lack of attention to V&V may be because V&V is a phase at the boundary between requirements and the rest of the process. The same can be said to Testing and Design phases. This shows that the main focus of NLP4RE is the phases within the RE process, with limited attention to the relationship with the whole software process. This indicates an additional space for further research.

**Table 7: Relationships between Research Types and Evaluation Methods**

| | Solution Proposal | Validation Research | Evaluation Research | Philosophical & Opinion | Personal Experience | **Row Total** |
|---|---|---|---|---|---|---|
| LESS | 140 | 37 | 0 | 0 | 0 | **177** |
| Example Application | 94 | 0 | 0 | 0 | 0 | **94** |
| LEHS | 35 | 31 | 0 | 0 | 0 | **66** |
| Discussion | 0 | 0 | 0 | 19 | 0 | **19** |
| Case Study | 0 | 0 | 25 | 0 | 0 | **25** |
| Experience Report | 0 | 0 | 0 | 0 | 15 | **15** |
| Field Experiment | 0 | 0 | 4 | 0 | 0 | **4** |
| Rigorous Analysis | 0 | 2 | 0 | 0 | 0 | **2** |
| Simulation | 2 | 0 | 0 | 0 | 0 | **2** |
| **Column Total** | **271** | **70** | **29** | **19** | **15** | **404** |

Note that about a third of the studies have considered two separate RE phases in their research, but the investigation of the second phase was often brief and secondary. In order to not skew the statistics, we decided to focus only on the main RE phase targeted by each study.

*Most studied and least studied NLP4RE tasks.* Figure 8 shows the distribution of the 370 studies by the NLP4RE tasks. Evidently, Detection, Classification, Extraction, Modeling, and Tracing & Relating are the most studied tasks, whereas Search & Retrieval is the least studied task. In what follows, we identify the focus of NLP4RE research through the relationships between these tasks and the RE phases where these tasks are performed.

*Relationships between NLP4RE tasks and RE phases.* For each RE phase, we count the number of studies that investigate each NLP4RE task. The results are presented in Table 8 and briefly explained as follows:

- Analysis phase: The NLP4RE tasks investigated for this phase are broad, including Detection, Classification, Extraction, Tracing & Relating, and Search & Retrieval. This suggests that there are a variety of activities on this phase that can be supported by NLP techniques. This also shows why Analysis is the most targeted phase. Table 8 shows that Detection is the central task for this phase, suggesting that the main role of NLP in the Analysis phase is to help detect language issues in requirements documents.
- Management phase: The NLP4RE tasks investigated for this phase are also Detection, Classification, Extraction, Tracing & Relating, and Search & Retrieval, suggesting that there are a variety of activities on this phase that can be supported by NLP techniques. The central task on this phase is Tracing & Relating, indicating that the main role of NLP in this phase is to help identify traceability relationships between requirements.
- Elicitation phase: The NLP4RE tasks investigated for this phase are also Detection, Classification, Extraction, Tracing & Relating, and Search & Retrieval, suggesting that there are a variety of activities on this phase that can be supported by NLP techniques. The central task on this phase is extraction, indicating that the main role of NLP in this phase is to help extract requirements concepts.

– Modeling phase: In contrast to the aforementioned RE phases, the NLP4RE tasks investigated for this phase are much narrower, including only Extraction and Modeling. Clearly, the studies investigating the Modeling task have outnumbered the studies investigating the Extraction task, making Modeling the central task for this phase.
– V&V and Testing phases: The NLP4RE tasks investigated for these phases are the same, comprising Detection, Classification, Extraction, and Tracing & Relating. While this suggests that there are a variety of activities on these phases that can be supported by NLP techniques, from Table 8, it is difficult to deduce which of these tasks is central on these phases.
– Design phase: The NLP4RE tasks investigated for this phase are exactly the same as those for the Modeling phase. This suggests that the main role of NLP in this phase is also to help identify requirements concepts and relationships.

As it can be observed from Table 8, Analysis, Management, Elicitation, and Modeling are the focus of NLP4RE research, and each of these phases has a central or *typical* NLP4RE task that represents the main role of NLP in that phase. Though not a central task for any RE phase, Classification has received more attention than Tracing & Relating, which is a central task for Management, perhaps due to its importance and general applicability in RE.

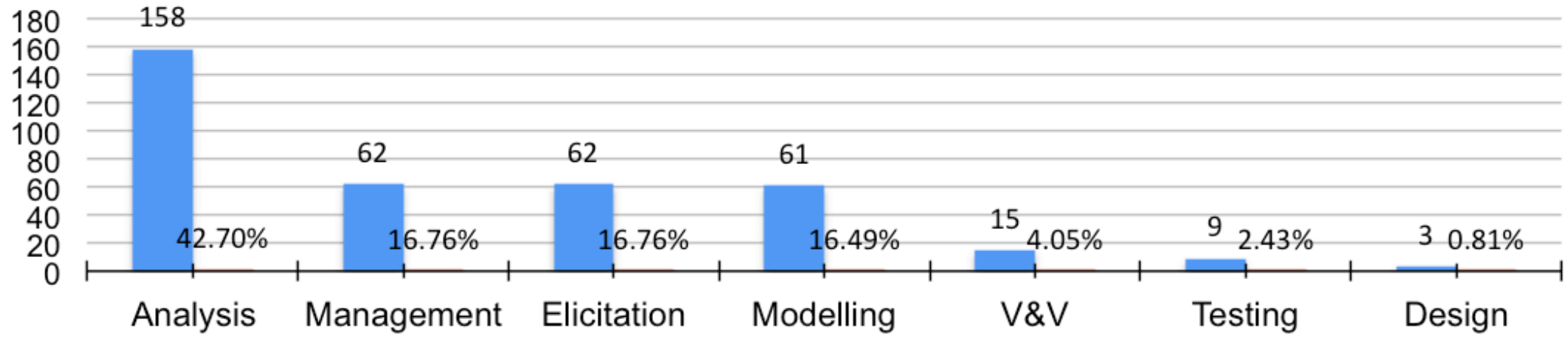

**Figure 7: Distribution of the selected studies to different RE phases.**

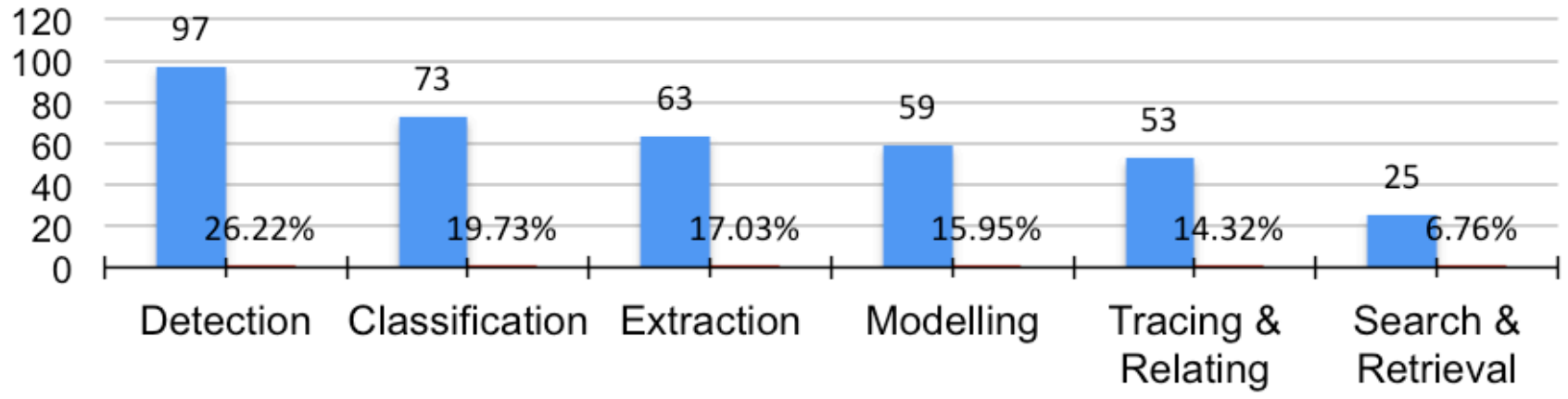

**Figure 8: Distribution of the selected studies to different NLP4RE tasks.**

**Table 8: RE Phases and Corresponding NLP4RE Tasks**

| | Detection | Classification | Extraction | Modeling | Tracing & Relating | Search & Retrieval | **Row Total** |
|---|---|---|---|---|---|---|---|
| Analysis | **75** | 37 | 16 | 0 | 19 | 11 | **158** |
| Management | 7 | 16 | 4 | 0 | **26** | 9 | **62** |
| Elicitation | 8 | 14 | **32** | 0 | 3 | 5 | **62** |
| Modeling | 0 | 0 | 3 | **58** | 0 | 0 | **61** |
| V&V | 5 | 4 | 2 | 0 | 4 | 0 | **15** |
| Other (Testing) | 2 | 2 | 4 | 0 | 1 | 0 | **9** |
| Other (Design) | 0 | 0 | 2 | 1 | 0 | 0 | **3** |
| **Column Total** | **97** | **73** | **63** | **59** | **53** | **25** | **370** |

*Types of input document processed by NLP4RE studies.* From the identified input documents, we synthesized them into 14 different types, as Figure 9 shows. As the description of input documents in our reviewed studies is often brief and vague, we established the nature of some input documents through deduction and interpretation. Evidently, Requirements Specification is the most dominant input document type for NLP4RE research. We noticed that some document types are more recent, such as User Feedback, User-Generated Content, Legal & Policy, User Story, and Domain Document. We observed that mining requirements related information from user-generated content such as app reviews [44] and tweets [81] has become a growing trend in NLP4RE research, as the widely availability of this type of text can make it easy to validate and replicate NLP4RE results, a possible solution to the shortage of real requirements data [21].

In addition to different types of textual document, Figure 9 also depicts non-textual documents such as Other Model, Code and UML Diagram, but these types of document are not commonly used. Typically, these documents would need to be pre-processed and paraphrased into a structured textual form before they can be treated by NLP techniques.

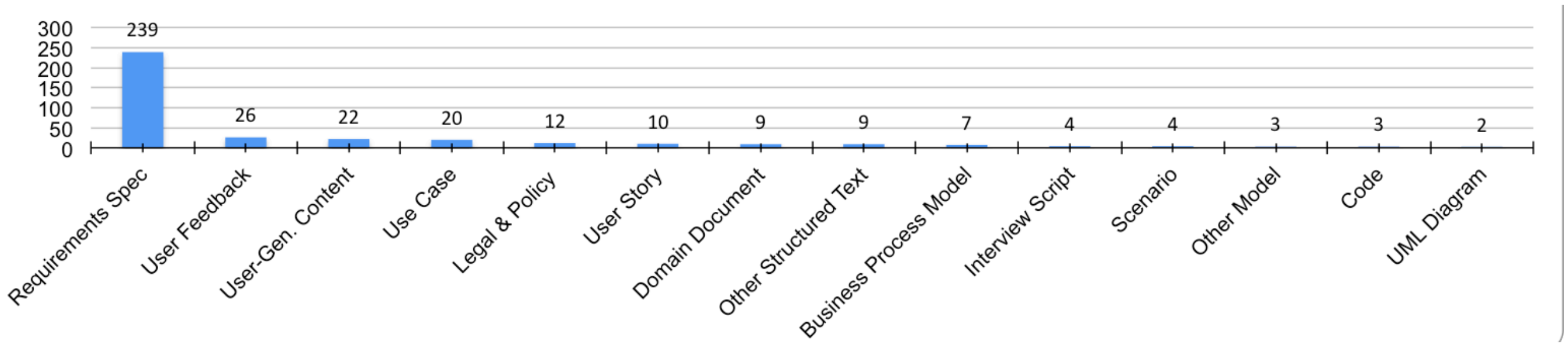

**Figure 9: Input document types and the number of studies using each type.**

## 5.4 RQ4: State of Practice in NLP4RE Research

*Types of new tool developed*. From the reviewed studies that explicitly report new tools, we found 130 named tools. These tools, hence called NLP4RE tools, are presented in Table 9 where they are categorized according to their main NLP4RE tasks. In Table 9, each tool is represented by its name (e.g., OICSI) and indexed by a study identifier (e.g., S678). Through the study identifier of each tool, the reader can locate the reference of the study that reports the tool in Appendix 1. Evidently, the top-ranked tools are modeling (34) and detection (31), followed by extraction (24), classification (20) and tracing & relating (15). At the bottom of the rank is search & retrieval (6).

The greater attention given to the modeling tools reflects the uniqueness and importance of modeling task, as modeling is a fundamental activity in RE and so much so there is a RE phase dedicated entirely to modeling. In addition, modeling entails the combination of both NLP and knowledge representation techniques, thus new tools are needed to support this task. By contrast, other types of tool, especially classification, detection and extraction, may be more readily composed from available NLP technologies and tools. For example, WEKA can be used to support classification, and GATE can be used to support extraction and detection. Consequently research on these tasks may focus on exploring different NLP techniques and tools, rather than on developing new tools. Finally, the limited number of search & retrieval tools can be attributed to the limited attention given to the search & retrieval task.

During our categorization of different tools, we noticed that some extraction tools also support other NLP4RE tasks such as paraphrasing and summarization. For example, extraction tools NAT2TEST and GuideGen perform both extraction and summarization tasks. The latter task is used to compose test cases or test guidelines.

*Relationships between NLP4RE tools, NLP4RE tasks and RE phases.* We use a circular dendrogram (Figure 10) to show the relationship between these tools and their corresponding RE phases. In this diagram, the NLP4RE tools are first grouped into the clusters by their NLP4RE tasks (the middle layer of the diagram) and then by their targeted RE phases (the inner layer of the diagram). This diagram can be used as a roadmap for us to navigate in both directions: from a given NLP4RE tool to its NLP4RE task and phase, and from a given RE phase to its tasks and available NLP4RE tools. Using this map, we can ask, for example, which NLP4RE tool is developed for which NLP4RE task and in which RE phase. Based on our results no tools have been found for design phase. Further, this map clearly illustrates the aforementioned *typicality* of NLP4RE tasks versus RE phases.

**Table 9: Categories of NLP4RE Tools**

| Tool Type | Tool Name (Study ID) | No. Tools | Percent |
|---|---|---|---|
| **Modeling** | OICSI (S678), NL-OOPS (S553), EA-Miner (S499), CM-Builder (S343), Circe (S34), LIDA (S623), NIBA Toolset (S272), RETNA (S108), aToucan (S909), DBDT (S31), Cico (S34), NL2UMLviaSBVR (S70), RADD-NLI (S121), SUGAR (S190), GRACE (S208), AREMCD (S219), RUCM (S227), RSLingo (S266), Zen-ReqConfig (S482), TREx (S496), NAPLES (S499), GeNLangUML (S551), ConstraintSoup (S600), C&L (S707), AnModeler (S799), SBEAVER (S813), KCMP Dynamisch (S272), Xtext (S20), Kheops (S35), Visual Narrator (S683), ProcGap (S800), FeatureX (S772), CMT & FDE (S261), VoiceToModel (S765) | 34 | 26.15% |
| **Detection** | ARM (S861), SREE (S812), RQA (S903), AnaCon (S41), REGICE (S55), NARCIA (S56), LELIE (S75), SRRDirector (S86), MIA (S114), KROSA (S178), NAI (S226), QuARS (S232), CAR (S252), CARL (S298), RAVEN (S303), ReqSAC (S370), RAT (S376), MaramaAIC (S395), RESI (S432), RECAA (S447), DeNom (S448), RETA (S450), AQUSA (S501), Dowser (S644), QAMiner (S661), LeCA (S701), S-HTC (S258), CNLP(S464), Pragmatic Ambiguity Detector (S256), ReqAligner (S663), REAssistant (S662) | 31 | 23.85% |
| **Extraction** | findphrases (S13), AbstFinder (S307), FENL (S71), NAT2TESTSCR (S131), NLP-KAOS (S132), SAFE (S385), AUTOANNOTATOR (S433), UCTD (S453), GUEST (S598), Guidance Tool (S688), SpecQua (S743), NAT2TEST (S744), semMet (S777), Test2UseCase (S810), OCLgen (S845), Text2Policy (S872), GaiusT (S888), SNACC (S891), Doc2Spec (S897), ARSENAL (S915), MaTREx tool (S284), ELICA (S2), CHOReOS (S520), GuideGen (S907) | 24 | 18.46% |
| **Classification** | ASUM (S129), RUBRIC (S223), WCC (S257), NFR2AC tool (S306), ALERTme (S332), PUMConf (337), FFRE (S341), AUR-BoW (S500), SEMIOS (S550), CRISTAL (S629), CoReq (S672), SD (S674), ACRE (S757), SOVA R-TC (S778), SMAA (S788), CSLabel (S892), HeRA (S718), NFR Locator (S758), SURF (S910), NFRFinder (S647) | 20 | 15.38% |
| **Tracing & Relating** | Coparvo (S24), Trustrace (S25), Histrace (S25), CoChaIR (S26), HYPERDOCSY (S38), ReqSimile (S171), LGRTL (S198), CQV-UML (S400), TiQi (S651), REVERE (S717), LiMonE (S723), ESPRET (S792), COCAR (S805), RETRO (S934), WATson (S302) | 15 | 11.54% |
| **Search & Retrieval** | RE-SWOT (S174), IntelliReq (S602), ReqWiki (S711), iMapper (S784), PriF (S802), WIKINA (S686) | 6 | 4.62% |
| **Total** | | **130** | **100%** |

*Tool development timeline.* These NLP4RE tools were proposed between 1990 and 2019 and their development timeline is shown in Figure 11. Clearly, before 2004, the development had been patchy, with just 18 tools produced; from 2004 onwards, however, there has been a year-on-year growth of NLP4RE tools, with only a brief dip in 2007. We observe that this growth period corresponds to the strong growth period of NLP4RE research (Figure 3). The number of tools in 2019 is incomplete as our search was completed in April 2019.

*Tool availability.* Only 15 (11.54%) of these tools can be found online (See Table 10). However, of these 15 tools, aToucan requires access permission while IntelliReq is not accessible. For the remaining 13 tools, seven are open source tools hosted respectively on GitHub, SourceForge and a Semantic Wiki; three provide web-interface for users to try and another three can be downloaded from the Internet and installed on user computers. Among these accessible tools, only ReqSimile has stood the test of time, while the others were developed recently. But even ReqSimile is still in Beta version, meaning that it is available for testing before its general release. Clearly, the state of the availability of NLP4RE tools is rather poor.

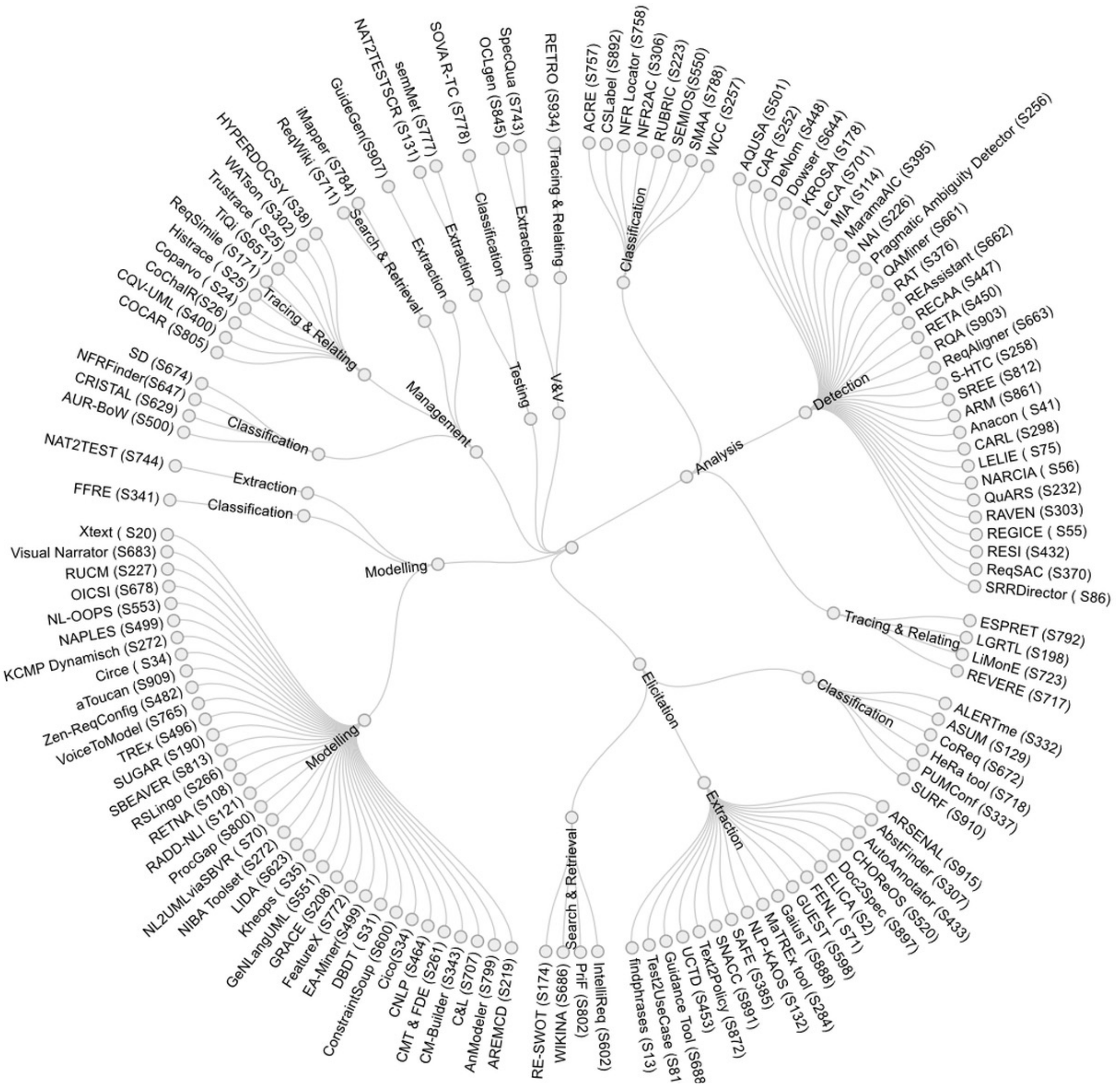

**Figure 10: The 130 NLP4RE tools clustered by NLP4RE tasks and then by RE phases.**

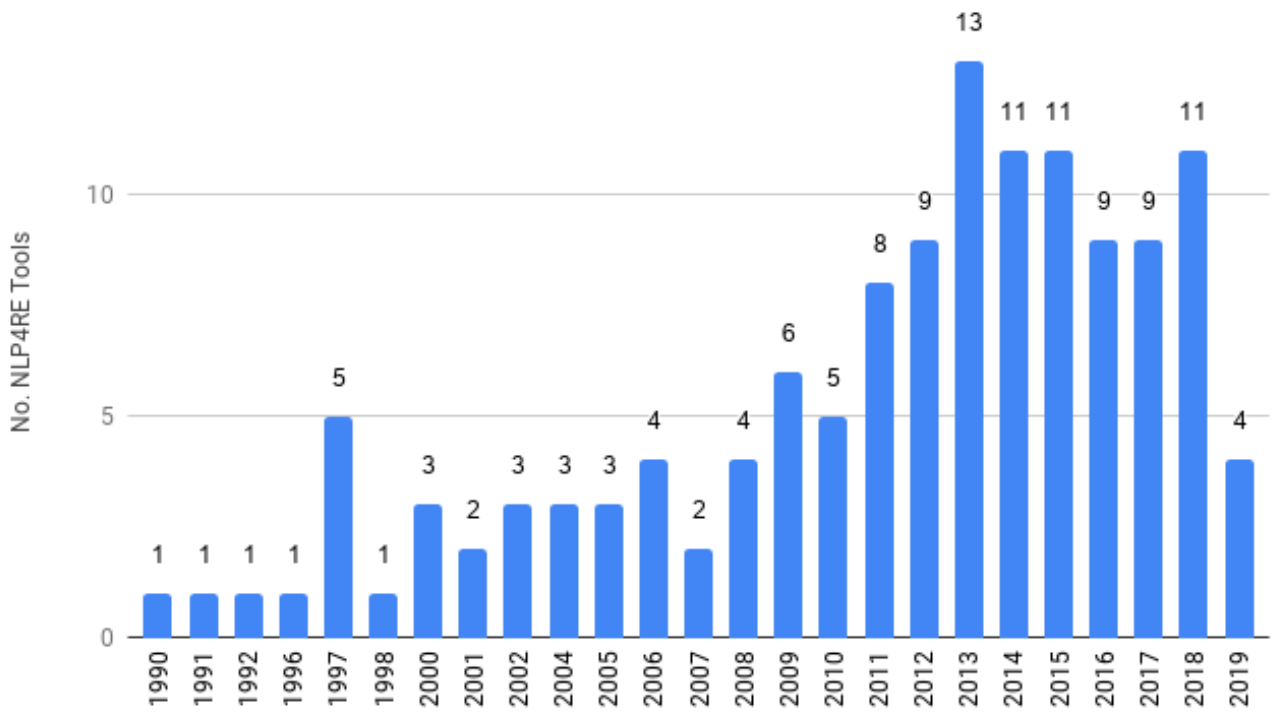

**Figure 11: Development timeline of NLP4RE tools.**

Table 10: NLP4RE Tools Available Online

| Tool Name | Tool Type | Year | Web Address | Status |
|---|---|---|---|---|
| aToucan (S909) | Modeling | 2015 | https://sites.google.com/site/taoyue/atoucan-models | Need access permission |
| CMT & FDE (S261) | Modeling | 2015 | https://github.com/isti-fmt-nlp/tool-NLPtoFP | Free open source |
| Visual Narrator (S683) | Modeling | 2016 | http://www.staff.science.uu.nl/~dalpi001/revv/ | Free to try only, with a simple UI |
| AnModeler (S799) | Modeling | 2016 | https://sites.google.com/site/anmodeler/ | Software can be downloaded |
| FeatureX (S772) | Modeling | 2018 | https://github.com/5Quintessential/FeatureX | Open source software |
| SpecQua(S743) | Extraction | 2014 | http://specqua.apphb.com | Free to try only, with a simple UI |
| Text2UseCase (S810) | Extraction | 2019 | https://sites.google.com/view/text2usecase/home | Web-based application, free to try, professional look and feel |
| GuideGen (S907) | Extraction | 2019 | https://github.com/hotomski/guidegen | Open source software |
| Pragmatic Ambiguity Detector (S256) | Detection | 2012 | https://github.com/isti-fmt-nlp/Pragmatic-Ambiguity-Detector | Open source software |
| IntelliReq (S602) | Detection | 2014 | http://www.intellireq.org | Website blocked |
| NARCIA (S56) | Detection | 2015 | https://sites.google.com/site/svvnarcia/ | Can be installed on user computers |
| NFR Locator (S758) | Classification | 2013 | https://github.com/RealsearchGroup/NFRLocator | Open source software |
| PUMConf (337) | Classification | 2018 | https://sites.google.com/site/pumconf/ | Can be installed on user computers |
| ReqSimile (S676) | Tracing & Relating | 2005 | http://reqsimile.sourceforge.net | Free open source, Beta version |
| ReqWiki (S711) | Search & Retrieval | 2013 | http://www.semanticsoftware.info/reqwiki | Open source web-based application |

## 5.5 RQ5: NLP Technologies for NLP4RE

From the studies that make explicit use of NLP technologies (including ML and deep learning algorithms), we extracted and synthesized 231 different technologies, and classified them into 140 NLP techniques (Figure 12), 66 NLP tools (Figure 13) and 25 NLP resources (Figure 14). Clearly, NLP resources form the smallest category, a clear indicator of the lack of resources for NLP4RE research. The usage of each category and its relationships with the NLP4RE tasks are discussed as follows.

*Most used and least used NLP techniques.* As shown in Figure 12, the most frequently used NLP technique is POS tagging (used by 187 studies), while the next most used techniques are tokenization (by 81 studies), parsing (by 72 studies), stop-words removal (by 70 studies), term extraction (by 68 studies), and stemming (by 68 studies). Figure 12 reveals that most NLP techniques, including those highly used, are syntactic techniques, a strong indicator of their dominance in NLP4RE research. In addition, a large number of techniques are underused, with 63 of them, accounted for 45.00% of the 140 NLP techniques, being only used once or twice.

*Most used and least used NLP tools*. As shown in Figure 13, the most frequently used NLP tool is Stanford CoreNLP (used by 80 studies), while the next most used tools are GATE (by 35 studies), NLTK (by 23 studies), Apache OpenNLP (by 21 studies), and WEKA (by 13 studies). Among these, apart from WEKA, which is a data-mining tool, the other four tools are general-purpose NLP tools. Figure 13 also reveals a large number of underused tools, of which 42 are only used once and 10 are used twice. This means that 78.79% of the 66 NLP tools have only been used once or twice.

*Most used and least used NLP resources*. As shown in Figure 14, the most frequently used NLP resource is WordNet (by 66 studies), followed by VerbNet (by 9 studies) and British National Corpus (by 7 studies). Most NLP resources listed in Figure 14 are lexical resources. There are a large number of underused resources, including 12 (or 48%) that are only used once or twice.

*Long tail distribution of NLP technologies.* The usage of each category of NLP technologies exhibits a distribution pattern known as the “Long Tail” [82]. It means that only a small number of technologies have been used frequently, as described above, whereas the majority technologies have a very low usage. The most frequently used technologies are called the “hits”, whereas the least used are called “long tails”. Clearly, POS tagging, Stanford CoreNLP and WordNet are the hits, indicating their popularity in NLP4RE research. However, what is more interesting is the huge number of long tail technologies – particularly those that have only been used once or twice. According to the Long Tail theory [82], such technologies are "niches" in the market, but that is not entirely true of the NLP technologies used in NLP4RE research. Some of the long tail technologies are discussed in the following.

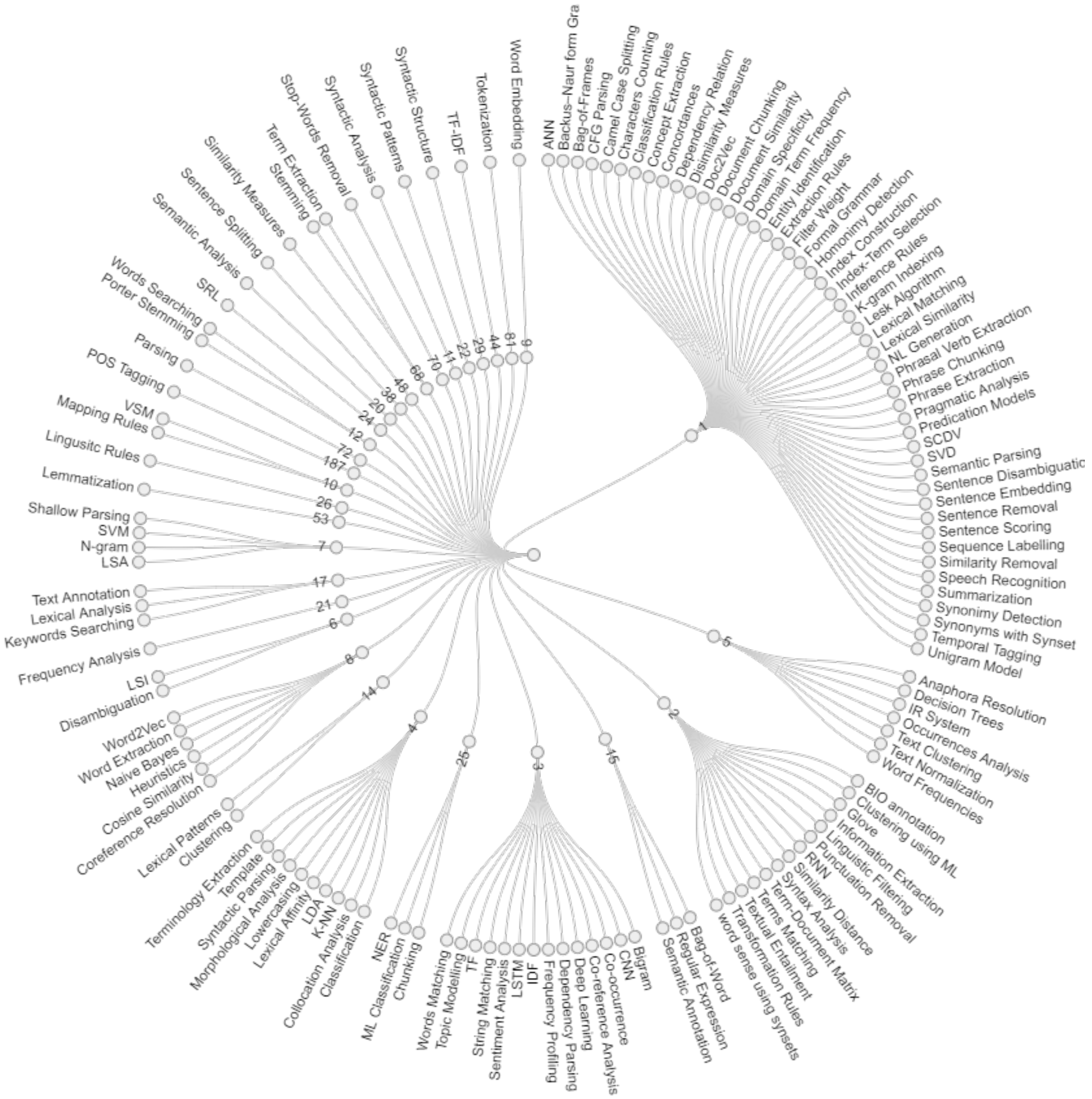

**Figure 12: 140 NLP techniques and their frequency of use.**

*Long tail NLP techniques.* Our initial investigation suggests that most long tail NLP techniques are *nascent*. For example, various deep learning techniques such as Word Embedding, Doc2Vec, LSTM, CNN, and RNN, are novel. It is therefore natural that they have a very low usage. Some long tail techniques are probably out of date or obsolete. For example, Lesk Algorithm, a classical algorithm for word sense disambiguation (WSD), is now being replaced by more advanced WSD techniques [83]. A few long tail technologies are real *niches*, as they are developed to serve specific needs. For example, CFG Parsing is a formal grammar for regulation expression and SCDV is a specialized technique to support faster construction of feature vectors in ML algorithms. For NLP4RE researchers, nascent techniques provide the springboard for innovation and novelty. In the future there will be more and more long tail NLP technologies, but only a small number of them will become future hits, according to the Long Tail theory [84].

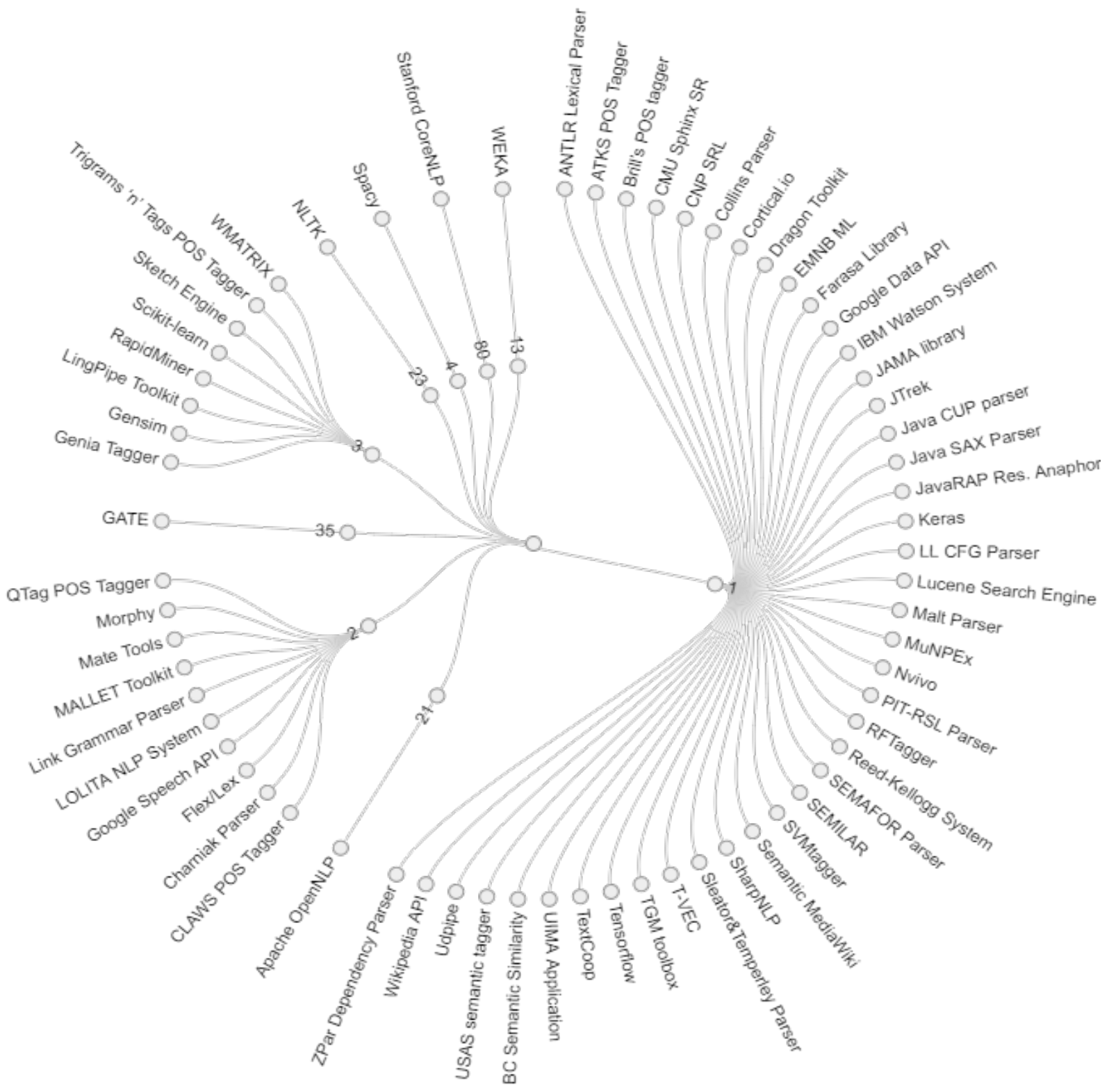

**Figure 13: 66 NLP tools and their frequency of use.**

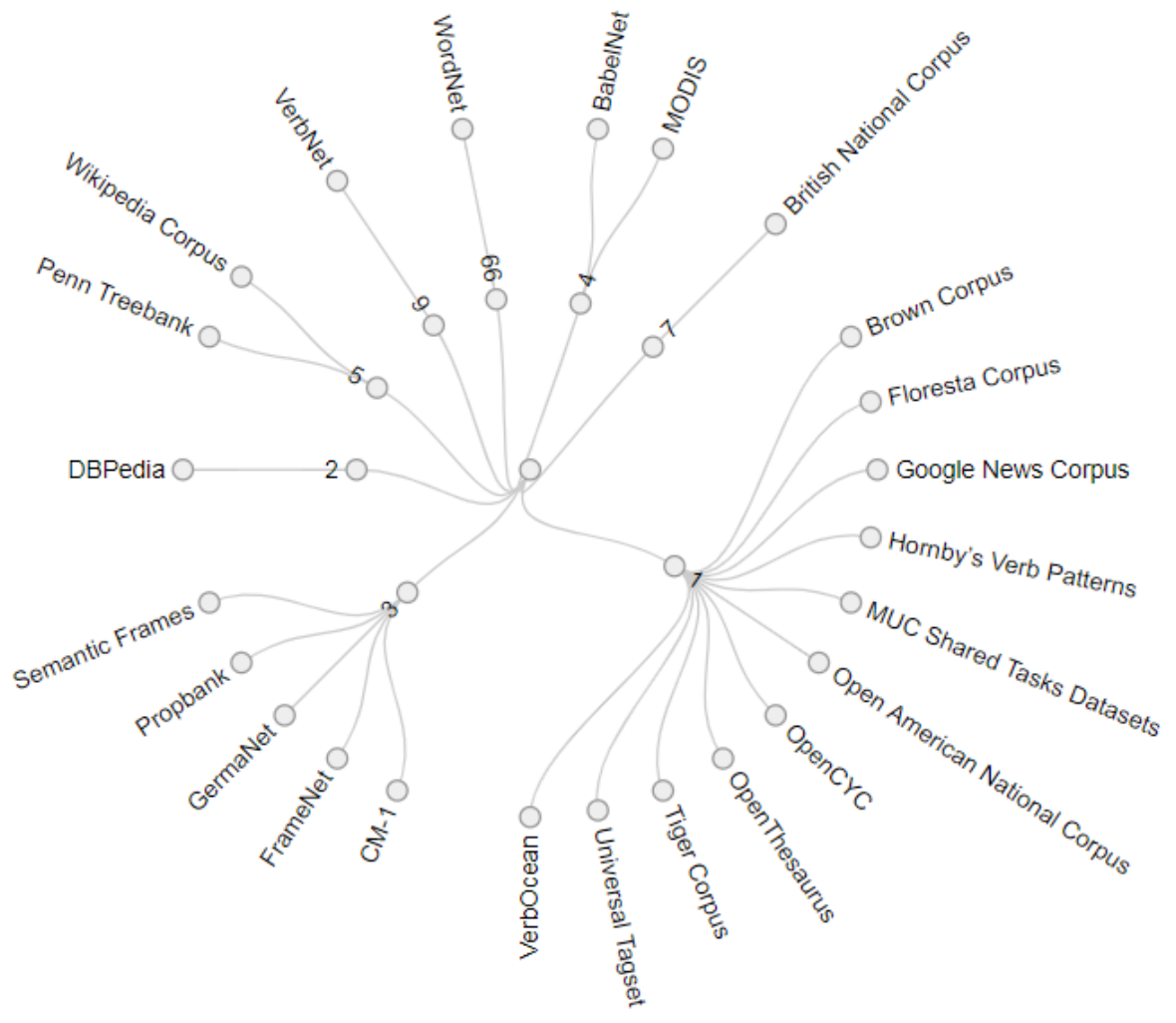

**Figure 14: 25 NLP resources and their frequency of use.**

*Long tail NLP tools*. Most long tail NLP tools are specialized tools. Many of them are taggers (e.g., Genia Tagger, CLAWS POS Tagger and Brill's POS Tagger) and parsers (e.g., ANTLR Lexical Parser, SEMAFOR Parser and Java SAX Parser). There are also other types of tool, such as Gensim for topic modeling and LingPipe for named entity recognition. We therefore deduce that most long tail NLP tools are niches in the sense they are specialized tools.

*Long tail NLP resources.* According to our research, long tail NLP resources tend to be domain specific or language specific. For example, VerbOcean is a lexicon for mining semantic verb relations on the web; MODIS and CM-1 are datasets for software engineering[14]; GermaNet (a lexicon for German language) and Floresta Corpus (a syntactic tree corpus for Portuguese) are Language-specific resources. A few long tail resources are nascent, such as Google News Corpus and DBpedia corpus for Wikipedia. There are also a couple of out of date resources, such as Brown Corpus and Hornby's Verb Patterns.

*Relationship between NLP techniques and NLP4RE tasks.* As it is not possible to show the relationship between all the 140 NLP techniques and their supporting NLP4RE tasks, we have chosen 32 top-ranked NLP techniques for discussion. Figure 15 shows these techniques and their frequency of use – each technique has been used at least 10 times. Figure 16 depicts these techniques and their relationships with the six NLP4RE tasks. It shows that these NLP techniques are evenly distributed across all six NLP4RE tasks, revealing a symmetric pattern. This suggests that all these 32 techniques are generally applicable to the six NLP4RE tasks. However, as these techniques are predominantly word-based, we deduce that most current NLP4RE studies are based on baseline NLP techniques, with little consideration to semantic or discourse analysis.

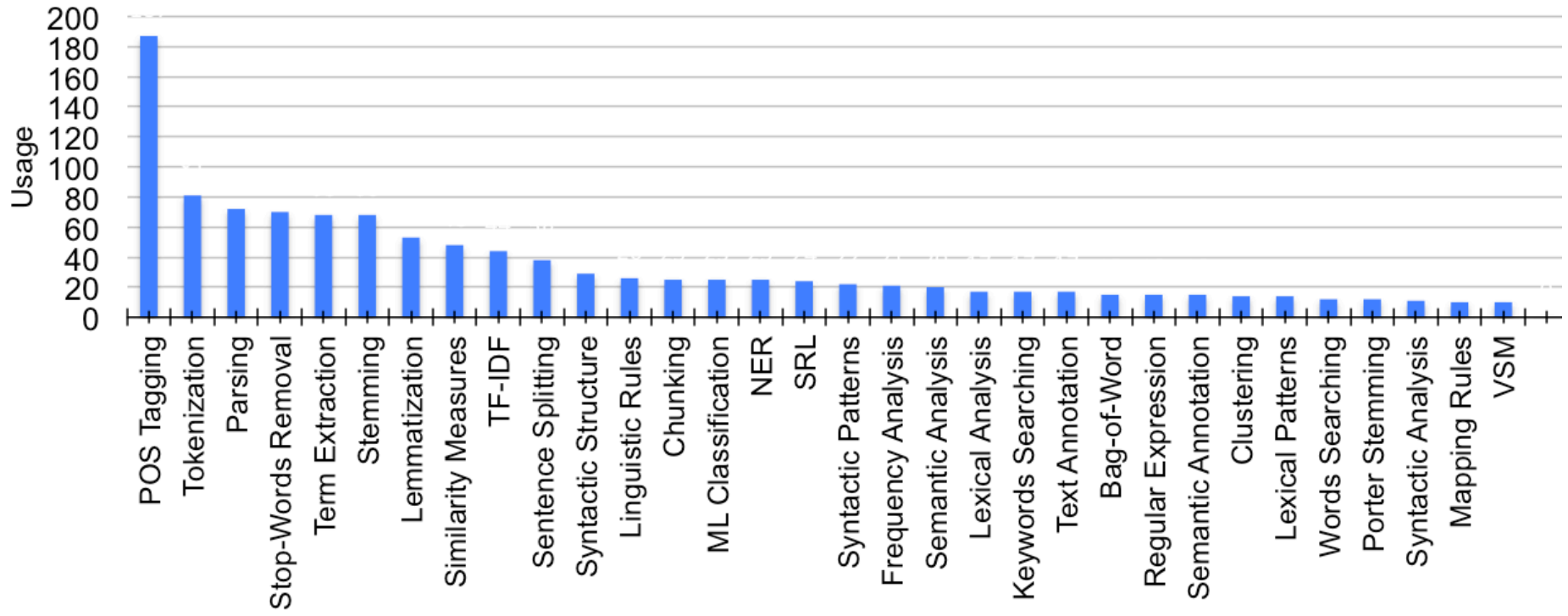

**Figure 15: Frequently used NLP techniques - those that have been used 10 times or more.**

*Relationship between NLP tools and NLP4RE tasks.* As with NLP techniques, we have chosen the top 14 NLP tools for discussion, on the basis that each tool has been used at least three times. Figure 17 shows these tools and their frequency of use. The relationship between these tools and the six NLP4RE tasks is depicted in Figure 18, which shows that most tools are used for detection, classification and extraction, and only a few for modeling, tracing & relating and search & retrieval. The lack of NLP tools for modeling thus supports our early claim that more modeling tools are needed. We noticed that the aforementioned general-purpose NLP tools – that is, Stanford CoreNLP, GATE, NLTK, and Apache OpenNLP – have been used to support all six NLP4RE tasks.

*Relationship between NLP resources and NLP4RE tasks.* We have chosen 14 NLP resources for discussion, on the basis that each resource has been used at least three times. Figure 19 shows these resources and their frequency of use. The relationship between these tools and the six NLP4RE tasks is depicted in Figure 20, which shows that resources are in general scarce for all the tasks, but particularly so for modeling and search & retrieval. Among these NLP resources, only WordNet has been used to support all six NLP4RE tasks; VerbNet used for all tasks but modeling; British National Corpus used in all but modeling and search & retrieval. We observe that there is a general lack of NLP resources suitable for the NLP4RE tasks.

[14] These datasets are located at http://promise.site.uottawa.ca/SERepository/datasets-page.html. Other RE related resources may have been published in short papers and thus have not been included in our study. For example, PURE (See: A. Ferrari, G. O. Spagnolo, and S. Gnesi, "PURE: A Dataset of Public Requirements Docments," RE 2017) and SEM-REQ (see: W. Alhoshan, R. Batista-Navarro, and L. Zhao, "Towards a Corpus of Requirements Documents Enriched with Semantic Frame Annotations," RE2018) have not been included in this mapping study, as they were reported in short papers.

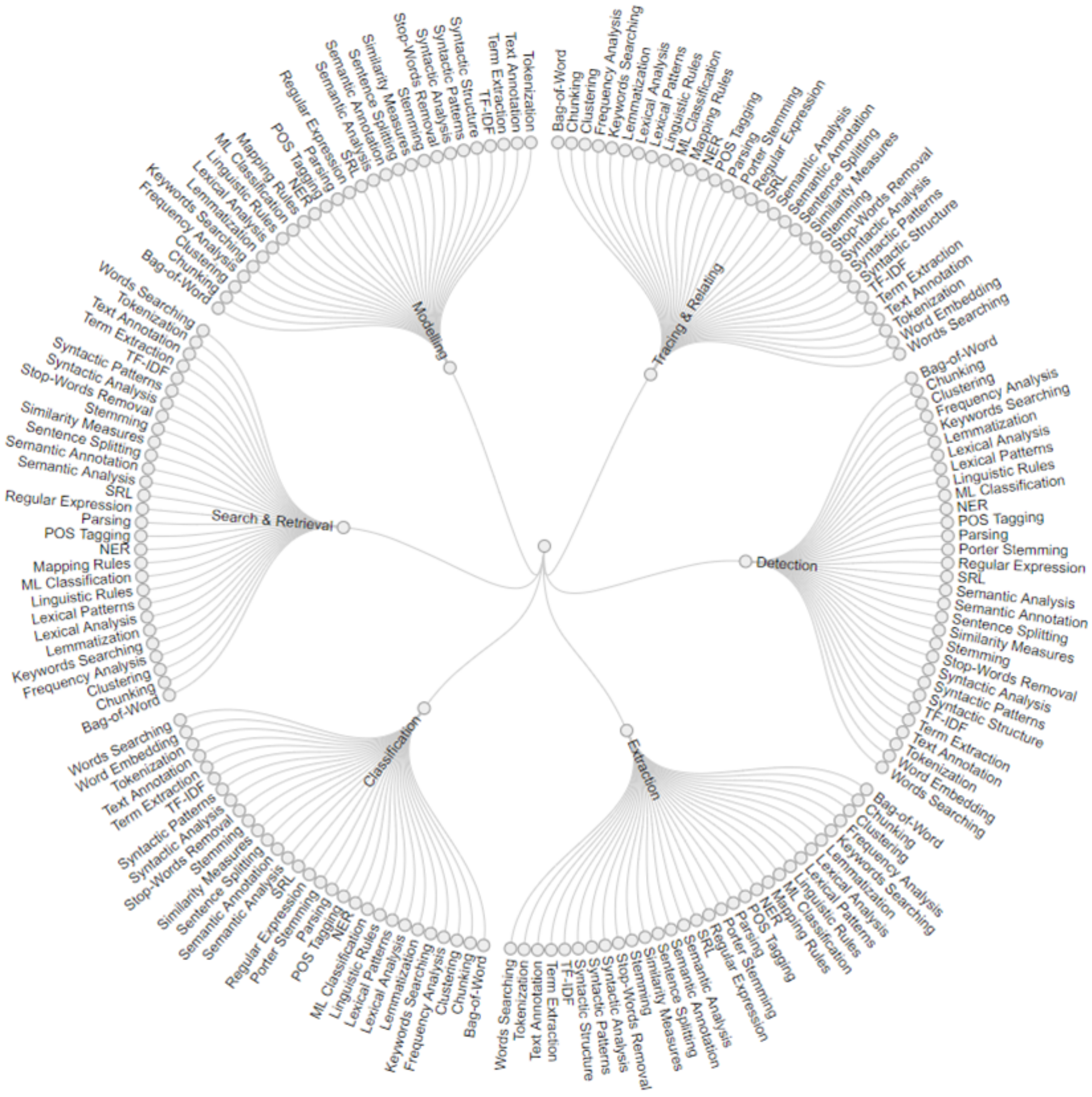

**Figure 16: Relationship between frequently used NLP techniques and the 6 NLP4RE tasks.**

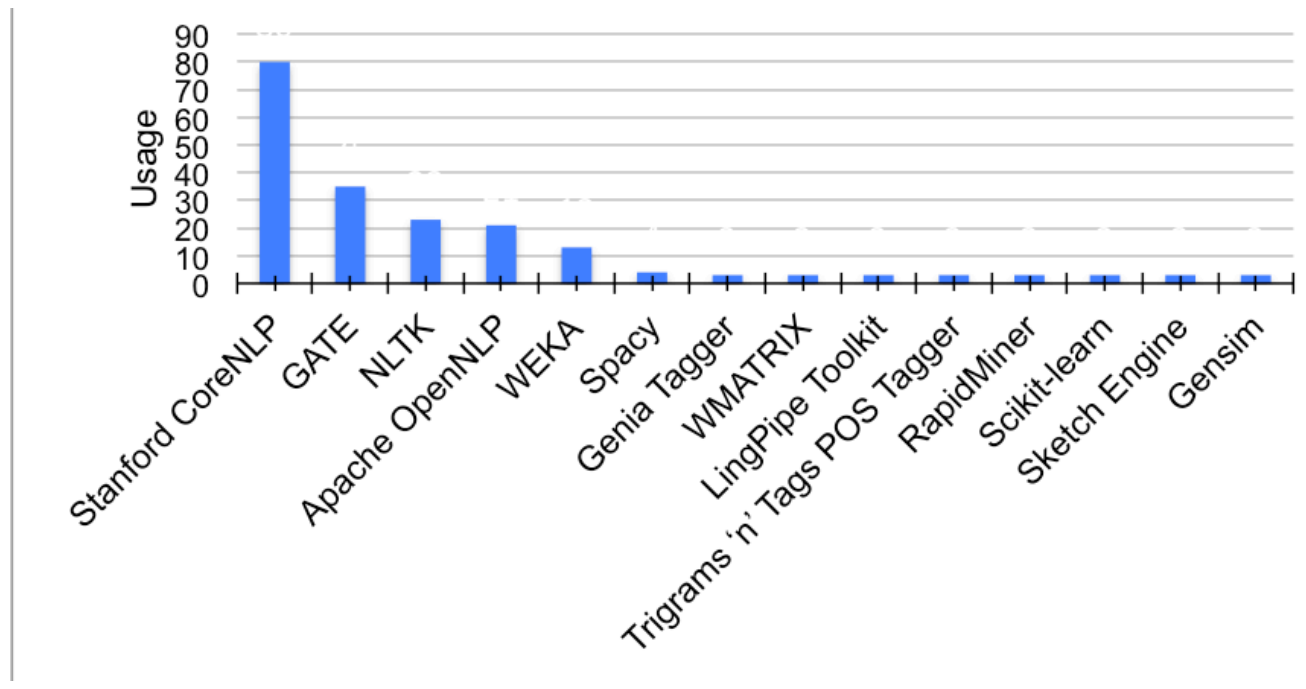

**Figure 17: Frequently used NLP tools - those that have been used three times or more.**

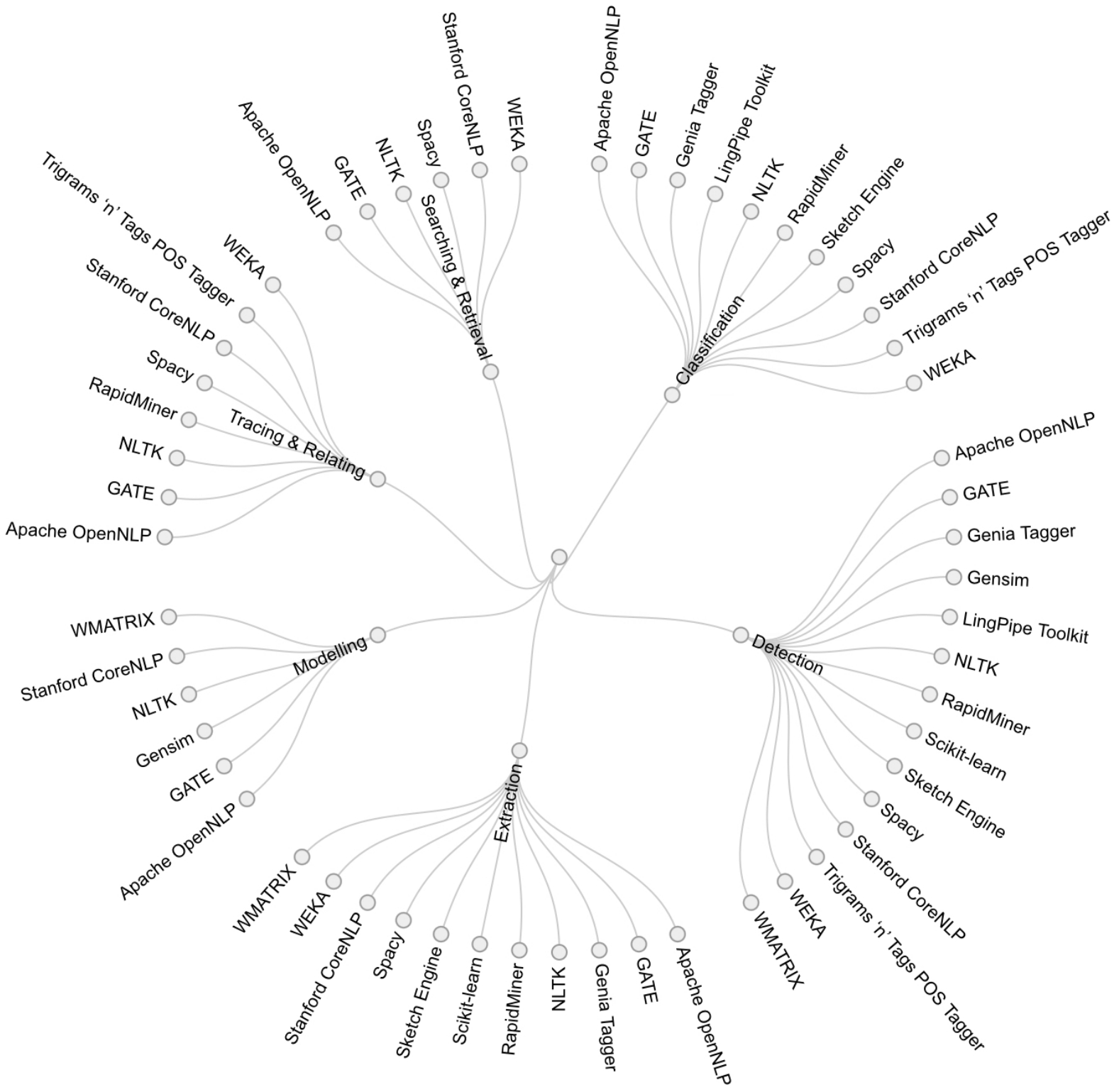

**Figure 18: Relationship between frequently used NLP tools and the corresponding NLP4RE tasks.**

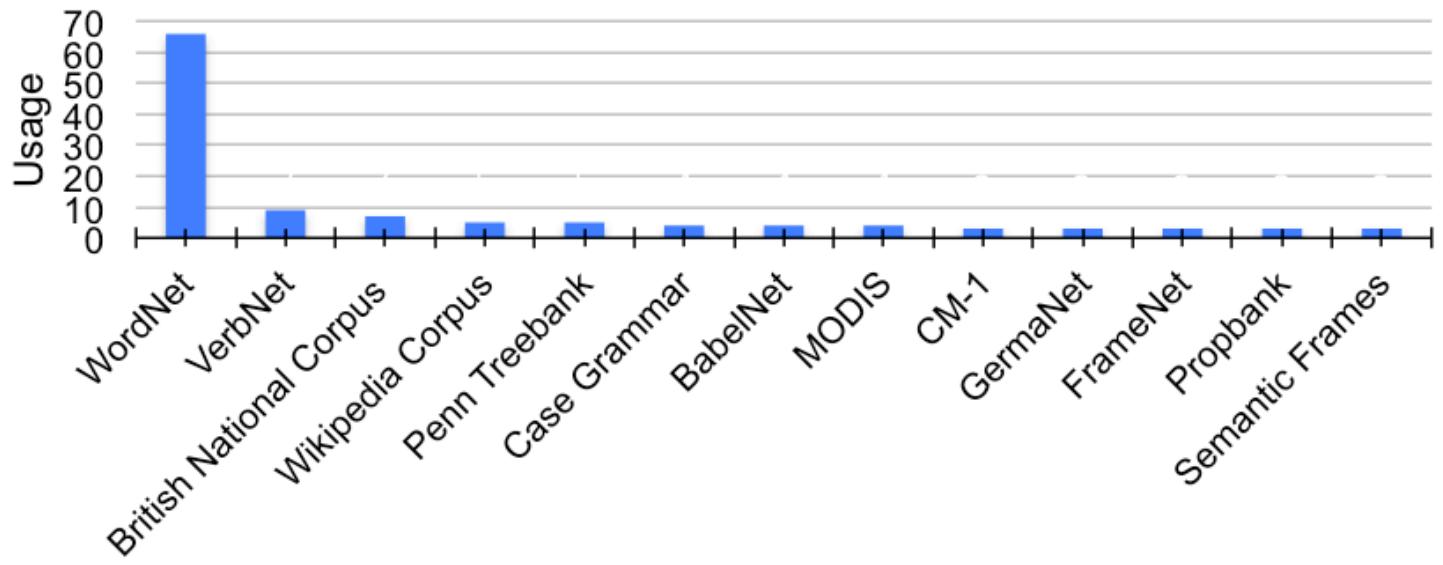

**Figure 19: Frequently used NLP resources - those that have been used three times or more.**

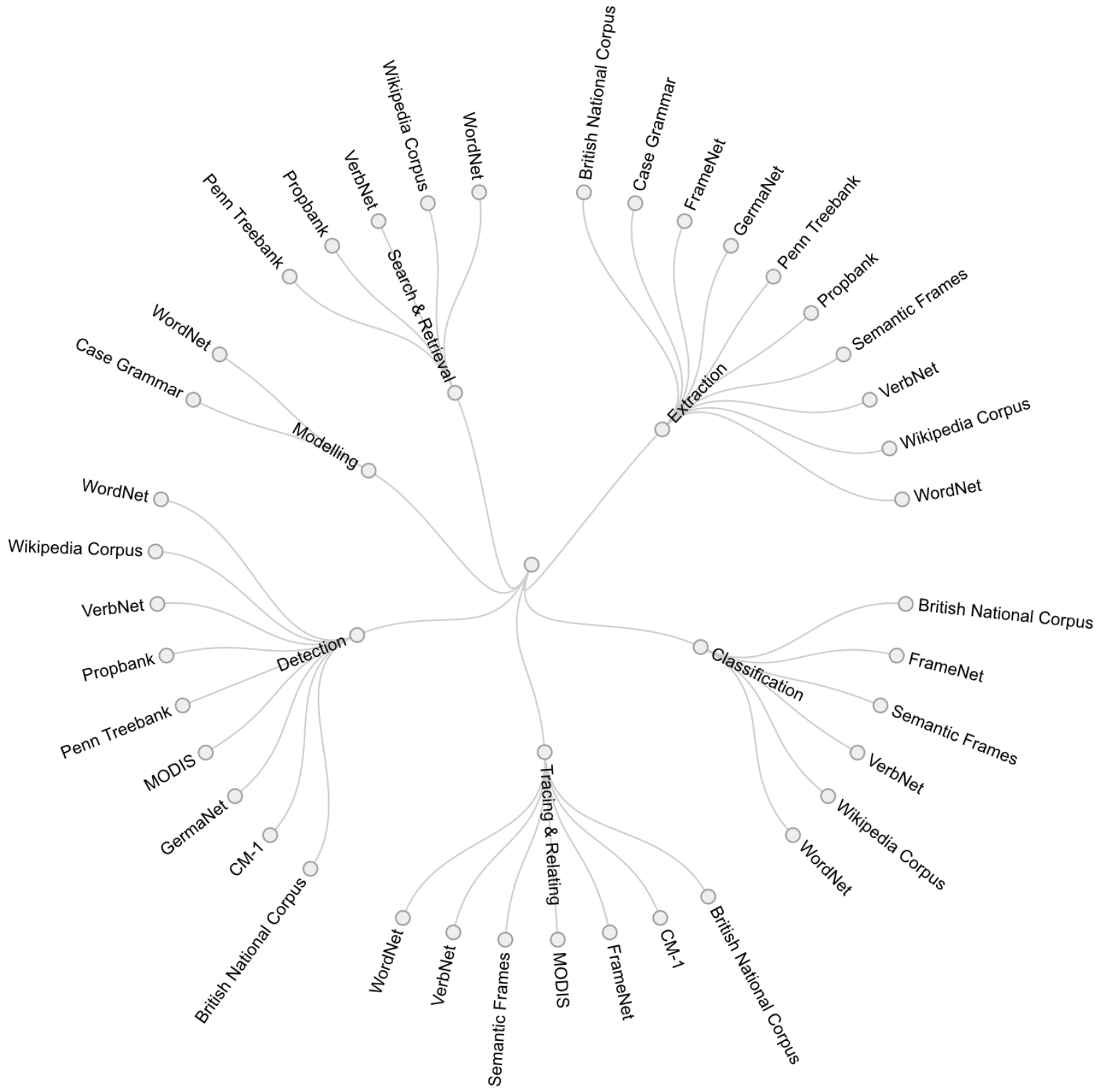

**Figure 20: Relationship between frequently used NLP resources and the corresponding NLP4RE tasks.**

# 6. Reflection on the Key Findings and Implications for Future Work

In this section, we first reflect on our key findings to highlight the trends and gaps in the current NLP4RE research; we then use this reflection to motivate the implications for future research and practice.

## 6.1 Summary of Key Findings and Observations

In this section, we summarize the key findings and observations based on the answers to our five research questions.

### 6.1.1 The State of NLP4RE Literature

This mapping study has identified 404 primary studies relevant to NLP4RE. The number of studies reflects the importance of the field and the attention it has received from researchers. Although publication timelines indicate that NLP4RE research started more than three decades ago, it was only the last 15 years that the field has grown into an active and thriving area, producing 88% of the total identified studies. Technological advancements in the field of NLP during that period have obviously paved the way for the rapid growth and development in NLP4RE.

Of the studies identified, the majority were conference and workshop papers. Such a trend is consistent with the publication pattern in other RE and SE areas, as some RE and SE conferences such as RE and ICSE are highly competitive and comparable to top journals.

The large number of diverse publication venues for these NLP4RE studies shows that NLP4RE has a core base in RE and a strong audience in SE; it has also attracted a general interest from diverse communities. However, while NLP4RE is an application area of NLP, NLP4RE studies were rarely published at NLP venues and this indicates a lack of awareness of NLP4RE work in the NLP research community.

### 6.1.2 The State of Empirical Research in NLP4RE

Although different types of research are found from the reported NLP4RE studies, Solution Proposal is the main research type, adopted by 271 (67.08%) studies, followed by Validation Research (17.33%). The remaining research types are less common: Evaluation Research (7.18%), Philosophical & Opinion (4.70%), and Personal Experience (3.71%). Such a trend has also been observed in other areas of RE and SE, which is likely to repeat in the future.

These different research types were evaluated by different evaluation methods and there is a clear alignment between the research types and the evaluation methods used by them. Around 65% of the Solution Proposal studies (175 out of 271) were evaluated using a laboratory experiment (either a LESS or a LEHS) while around 35% of them using an example application. This suggests that none of the solution proposal studies has been evaluated in a real world environment. Lack of industrial case studies and field experiments is therefore a major challenge for NLP4RE research.

### 6.1.3 The Focus of NLP4RE Research

Of 370 NLP4RE studies relevant to this facet, the majority (42.70%) have targeted at the analysis phase, while only 0.81% have targeted at the design phase. Apart from these two phases, the distribution of the remaining studies puts management and elicitation each at 16.76%, modeling at 16.49%, V&V at 4.05%, and testing at 2.43%. Evidently current NLP4RE research is analysis-centric.

This observation is further supported by our findings that at 26.22%, the detection task, with the main purpose to support the analysis of requirements documents, is the most researched NLP4RE task by the reviewed studies. The remaining studies are more or less evenly distributed among these four tasks: classification (19.73%), extraction (17.03%), modeling (15.95%), and tracing & relating (14.32%). At only 6.76%, search & retrieval is clearly the least studied task.

As we just alluded to, there is a close relationship between the analysis phase and the detection task. This relationship is forged by the main problem in this phase and the NLP solution to this problem. Based on this observation, we identified the following pairs of relationship between the RE phases and the NLP4RE tasks:

- For the analysis phase, the main problem is the detection of language issues in requirements documents. The central NLP4RE task to address this problem is therefore detection. Other tasks, such as classification, extraction, tracing & relating, and search & retrieval, are used to support the detection task.
- For the management phase, the main problem is the identification of traceability relationships between requirements. The central NLP4RE task to address this problem is therefore tracing & relating. Other tasks, including classification, detection, extraction, and search & retrieval, are used to support the tracing & relating task.
- For the elicitation phase, the main problem is the extraction of requirements concepts. The central NLP4RE task to address this problem is therefore extraction. Other tasks, including classification, detection, tracing & relating, and search & retrieval, are used to support this task.
- For the modeling phase, the main problem is the extraction of requirements concepts and the composition of conceptual models. The central NLP4RE task to address this problem is therefore modeling. The extraction task is used to support modeling.

We believe a deep understanding of these intricate relationships between different RE phases and NLP4RE tasks can be valuable for the development of appropriate NLP tools to support RE activities.

After investigating the NLP4RE studies from the perspectives of the RE phases and NLP4RE tasks, we turned our attention to the types of input document processed by these studies. Of the 14 different document types identified, requirements specification is the most commonly processed input document type, used by 239 studies (64.59%). However, we observed that there are some emerging trends towards using other types of document as input, such as user feedback, user-generated content, legal & policy, domain document, and user story. Processing these kinds of document may present new challenges to NLP4RE researchers, not only that such documents would not conform to linguistic or structural standards that are normally expected for requirements specifications [19], but also that such documents would contain unfamiliar vocabulary and concepts to RE. Given years of NLP4RE research, we believe that it is time for researchers to exploit more challenging texts and to explore uncharted territory.

#### 6.1.4 The State of the Practice in NLP Research

From the 370 studies that reported new results, we found 130 tools, of which 26.15% are modeling tools, 23.85% are detection tools, 18.46% are extraction tools, 15.38% are classification tools, 11.54% are tracing & relating tools, and 4.62% are search & retrieval tools. While the development timeline for these tools stretches from 1990 to 2019, the majority of the tools were found between 2004 and 2019. The growth of these tools has followed the same pattern of NLP4RE research over the same period.

However, of the 130 tools, only 15 can be found on the Internet. On closer inspection, one of these tools requires access permission and another is not accessible. This left 13 tools with varying degrees of availability. Furthermore, apart from one tool – ReqSimile – which was developed in 2005, the remaining 14 were developed recently, after 2012. This means that 14 out of 15 available tools have not stood the test of time. As stated in Section 1, some companies have only started to develop NLP tools for RE. This shows that there is a huge discrepancy between the current state of the art and the state of the practice in NLP4RE research.

#### 6.1.5 NLP Technologies for NLP4RE Research

A total of 231 different NLP technologies were identified from the selected studies, with 140 NLP techniques, 66 NLP tools and 25 NLP resources. Such a great number shows that NLP is fundamental to NLP4RE. However, these technologies have not been utilized to full potential, with a large amount being used once or twice, equivalent to 40% of the NLP techniques, 78.79% of the NLP tools and 48% of the NLP resources. The key findings on the usages of these NLP technologies are summarized as follows:

- Across the board, POS tagging, Stanford CoreNLP and WordNet are the most frequently used technologies with respect to their categories (namely, *technique*, *tool* and *resource*) – as might be expected, given the popularity of these technologies.
- Most frequently used NLP techniques were developed in the 1990s. Such techniques typically support the low-level syntactic analysis tasks such as POS tagging, tokenization, parsing, stop-words removal, term extraction, stemming, and lemmatization. This probably explains why the majority of the studies have targeted at the analysis phase and focused on low-level NLP4RE tasks such as detection and extraction.
- By contrast, most of the long tail NLP techniques are recent and novel. For example, various word embedding techniques only emerged around 2010 and, in particular, Google's vector representation of words (Word2Vec) was developed in 2013 [85]. Given the novelty of these techniques, it is therefore natural that only a few studies have used them. On the other hand, the limited attention to these novel techniques may also be attributed to the lack of NLP expertise in NLP4RE.
- General-purpose NLP tools are more popular than specialized tools. Most specialized tools are taggers such as Genia Tagger and CLAWS POS Tagger. This clearly indicates that researchers prefer to use general-purpose tools than specialized tools.
- The number of NLP resources used by NLP4RE studies is small in comparison with the number of NLP techniques and tools; the number of frequently used NLP resources is even smaller, with WordNet as the only predominant resource. There is also a clear lack of RE-specific resources, as apart from MODIS and CM-1, the remaining resources are for general NLP applications.

## 6.2 Implications for Research and Practice

#### 6.2.1 Implications for Research

*Need to collaborate with NLP researchers:* NLP4RE appears to be a lively research field especially in the latest years, and it is recognized in the top RE and SE venues. On the other hand, the penetration of NLP4RE research in the NLP field, which is the source of the technologies for NLP4RE, is scarce (see Sect. 5.1). This calls for more synergies between the RE community and the NLP community, and RE researchers are required to make their problems more appealing for NLP researchers. NLP research is often focused on broader linguistic problems (automatic summarization, machine translation, etc.), oriented to generalization over domains, and based on large datasets. NLP4RE research is typically context-specific, domain dependent and datasets are scarce or limited. Challenging NLP researchers to provide solutions in these contrived contexts may be an opportunity to improve cross-fertilization of the disciplines.

*Need to apply research results to real-world problems:* While NLP4RE counts several solution proposals, evaluated in the lab by means of experiments, the contributions in the form of case studies and even experience reports, which take into account the contexts of organizations, are more limited (Sect. 5.2). This suggests that NLP4RE researchers should take a step forward and apply the large variety of proposed solutions, which have been validated in lab, to real-world industrial problems.

*Need to expand the research scope beyond the typical RE process:* While most of the RE phases are thoroughly investigated, and especially Analysis, there is more limited research on studies that analyze or manipulate RE artifacts in other software engineering phases such as testing or design. This indicates that opportunities exist to better exploit synergies within the SE community to expand the scope of NLP4RE research also outside the typical RE process.

*Need to identify more NLP4RE tasks:* NLP4RE researchers have sought to address NL issues related to a range of NLP4RE tasks, covering Detection, Extraction, Modeling, Classification, Tracing & Relating, and Search & Retrieval (Sect. 5.3). As most efforts have so far been concentrated on the first four tasks, Search & Retrieval provides a possible area for additional investigation. There may also be other tasks that have not been uncovered by this mapping study, which can be investigated in future research.

*Need to analyze a wider range of RE related documents:* Additional areas of research come from the types of documents that can be considered for evaluation: user stories, use cases, domain documents, interview scripts and models are still marginal in the research, while they have a primary role in practice. Furthermore, although recent research is giving relevant attention to artifacts such as user feedback, legal documents and user stories, the field is still open for further investigation.

*Need more publically available NLP4RE tools for research validation:* Researchers should address the lack of public availability of most of the tools developed. While on one hand it is important to share data, it is also extremely relevant to make tools publicly available, especially in a context such as RE in which data are often confidential. This can help other researchers build on top of the work of their peers, and would facilitate also technology transfer with industries, as companies often want to see a working tool to be convinced about the feasibility of a collaboration, for example to adapt the tool to the company context.

*Need to develop RE related language resources:* This mapping study shows that NLP4RE research has mostly utilized lexical resources such as WordNet and VerbNet, whereas the use of corporal resources is still rare. The main reason we believe is lack of RE-specific corpora, as currently there are only a few RE or SE specific datasets available, including MODIS and CM-1. Using general-purpose corporal such as British National Corpus and Wikipedia Corpus to train ML algorithms for processing requirements text would lead to unreliable results. Successful corpus-based NLP (or statistical NLP) for RE will depend on the availability of large, annotated requirements corpora. To help evaluate research results, we also need other types of language resources, including shared datasets, benchmark data and performance metrics.

*Need to introduce NLP literacy into RE education and training:* The top NLP techniques identified in Section 5.5 may be useful to instructors in RE education and training, to focus their teaching on specific techniques, tools and resources that are dominant in NLP4RE. Instead, the technologies in the “long tail” (see Sect. 5.5) can give researchers an indication on the new, most recent, technologies (e.g., Word Embedding, Bag-of-Frames, Doc2Vec) that should be taken into account, as they may not have been fully exploited in NLP4RE research. This mapping study has discovered that advances in NLP technologies have a direct impact on the progress in NLP4RE research. NLP4RE researchers should therefore immerse themselves in the learning of new NLP technologies.

#### 6.2.2 Implications for Practice

*Need to collaborate with industries for research validation:* A mature software technology should be evaluated on real-world applications or industrial projects, to assess its scalability, practicality and usability. Since case study research in industrial settings is still limited as evaluation method, practitioners should open their doors to evaluate the different NLP4RE solutions made available by the researchers, as most of the RE phases are covered by some proposed solution. In particular, practitioners can leverage solutions oriented to the Analysis phase and for the tasks of Detection and Classification. Furthermore, practitioners can leverage solutions mainly for their software requirements specifications, as this is the type of artifact that is mostly considered by current studies (Figure 9).

*Transfer research results to industrial practices:* NLP4RE research has produced a large number of tools, though mostly are in the category of modeling. Practitioners can find the dendogram in Figure 10 particularly useful to explore which tools have been developed for their specific needs, in terms of tasks and RE phase. Unfortunately, practitioners have to contact the tool authors to access most of the tools, as only few of them are made publicly available. This is, however, also an opportunity, as generic, context-independent tools may need to be adapted to the specific company context, and general purpose, freely accessible tools that may not work out of the box, thus leading companies to discard the tool as not appropriate for their needs.

*Transfer technology know-how to industrial practices:* The set of top NLP technologies, including techniques, tools and resources, identified in Section 5.5 may be particularly useful to practitioners who wish to develop *in-house* NLP4RE tools for their needs, by leveraging existing platforms (Stanford CoreNLP, GATE, NLTK, Open NLP). The top techniques, tools and resources identify the basic, well-established, elements that are needed to practice NLP4RE. It is also interesting to notice that knowing about the top 32 NLP techniques identified (POS Tagging, Parsing, etc.) enables to address the whole set of NLP4RE tasks. This provides practitioners with a clear indication of the knowledge needed to develop NLP4RE tools, and can be useful to identify the skills required during recruitment of personnel that may be dedicated to the *in-house* development of NLP4RE tools.

# 7. Study Validity and Limitations

The main threat to the validity of any type of literature review is the question of reliability [86]: if two different studies follow the same research procedures, will they produce the same results [87]? For systematic literature reviews and systematic mapping studies, the threat of reliability can be manifested in the entire review process, from identification of the literature to selection of the relevant papers to the final analysis. To mitigate this threat to the validity of our mapping study, we took some preventive measures in every step of the study process, described as follows.

*Reliability of literature search*: Due to the constraints on resources, time and search engines, it is almost impossible to find the entire population of *all* the relevant papers on NLP4RE [86]. To ensure we found as many relevant papers as possible and as close to the actual population as possible, we followed the recommended guidelines [88], [69] to identify the literature (Sect. 4.2.1), formulate the search terms (Sect. 4.2.2) and perform the search (Sect. 4.2.3). However, it is possible that we may have not found those papers whose authors might have used other terms that have not been included in our search terms, though we have tried to mitigate this problem through initial and targeted searches. As our main search phase relied on the search engines provided by our chosen libraries, the quality of the search engines could have influenced the completeness of the identified primary studies, as reported by many other systematic reviewers [89], [72].

*Reliability of study selection*: To ensure our study selection was as accurate as possible, as free from researcher bias and human errors as possible, we followed a rigorous study selection process, guided by the carefully designed inclusion and exclusion criteria, and enforced by crosschecking and independent checking of the selected and deselected studies (Sect. 4.2.4). We paid attention particularly to the last two stages of study selection to ensure that the data inspectors carefully crosschecked each excluded and included paper. Whenever there was double about the relevance of a paper, we called upon the supervisors for discussion, based on which the final decision was made. We made the decision to exclude short papers in order to obtain a more balanced view of NLP4RE research. This is a limitation of our study. In spite of this, we believe that the study population we identified is close to the actual population and is a good representative sample of the current state of NLP4RE research.

*Reliability of data extraction and classification:* To ensure we extract the required data and organize the selected studies accurately, consistently and uniformly, we followed a faceted classification scheme with a comprehensive set of predefined categories (Sect. 4.3). However, the classification scheme was not foolproof for data extraction, as this process involved subjective interpretations and decisions by the researchers. Lack of sufficient details about the design and execution of the reported studies often hindered data extraction. A particular problem arising from identifying exact NLP technologies from the studies was the lack of precise, explicit and standard description of these technologies in the reported studies. For example, when a study stated that it used a simple syntactic technique to analyze a document, it could mean POS tagging only or both POS tagging and parsing. Worse still, some studies stated that they performed a tokenization task, but did not say which NLP tools were used to perform this task. To mitigate this problem, we compiled our own in-house NLP dictionary with a list of NLP techniques, NLP tools and NLP resources. This dictionary was then used to guide us in extracting NLP technologies from the selected studies. The process of classifying the various aspects of the selected studies (such as research types, evaluation methods, RE phases, NLP4RE tasks) also involved subjective decisions by the researchers. To minimize human errors, we carried out regular checks on each category. Whenever there was double about the classification of a particular study, we would re-assess that study, re-extract the data and re-classify the data if necessary.

*Reliability of data synthesis, analysis and visualization*: To ensure the mapping results were as accurate and error-free as possible, we carefully carried out thematic synthesis, descriptive analysis and frequency counting on the extracted data. Thematic synthesis involved standardizing the names of NLP techniques and establishing the types of input document. To synthesize the extracted NLP techniques, we used our NLP dictionary to normalize the names of NLP techniques or combine similar techniques into one. When we discovered new techniques, we also added them to our dictionary. To synthesize input documents, we relied on our knowledge to identify their common types. The synthesized results were reviewed several times and revisions were made to make them as accurate as possible.

# 8. Conclusion

This article has reported a first-ever systematic mapping study on the landscape of NLP4RE research. From 11,540 search results, 404 primary studies were included in the mapping study and systematically reviewed according to five research questions. These questions interrogate these studies to understand their publication status, state of empirical research, research focus, state of the practice, and finally, usage of NLP technologies. The answers to these research questions show:

– NLP4RE is an active and thriving research area in RE, which has amassed a large number of publications and attracted widespread attention from diverse communities.
– Most NLP4RE studies (67.08%) are solution proposals having only been evaluated using a laboratory experiment or an example application, while only 7.18% of the studies have been evaluated in an industrial setting, which highlights a general lack of industrial validation of NLP4RE research results.
– The biggest proportion (42.70%) of the NLP4RE studies have focused on the analysis phase, with detection as their central linguistic analysis task and requirements specification as their commonly processed document type, indicating the current focus of NLP4RE research.
– A total of 130 new tools have been proposed by the selected studies to support a range of linguistic analysis tasks, but there is limited evidence that these tools have been adopted or accepted by industry, indicating a lack of industrial practice of NLP4RE research results.
– Although a large number of NLP technologies, comprising 140 NLP techniques, 66 NLP tools and 25 NLP resources, have been used by the selected studies, they have not been used to full potential, with a large amount – particularly those novel techniques – being only used once or twice. Frequently used NLP technologies are syntactic analysis techniques, general-purpose tools and generic language lexicons.

These findings have revealed a huge discrepancy between the state of the art and the state of the practice in current NLP4RE research, indicated by insufficient industrial validation of NLP4RE research, little evidence of industrial adoption of the proposed tools, the lack of shared RE-specific language resources, and the lack of NLP expertise in NLP4RE research to advise on the choice of NLP technologies. The findings have also carried many important implications for NLP4RE research and practice. We believe that these implications can be used to drive NLP4RE research agenda. At the top of this agenda should be concrete plans for active collaboration with practitioners to jointly develop and validate NLP4RE tools, for close collaboration with NLP experts to learn and use the cutting-edge NLP technology, and for developing open NLP4RE tools, shared datasets, benchmark data and performance metrics for research evaluation.

In spite of the gaps and limitations, this mapping study also shows that NLP4RE research has made a tremendous progress over the past 15 years, particularly in the areas of publication and tool development. Additionally, recent work in analyzing more challenging documents such as user feedback and legal documents indicates that NLP4RE research has entered a new chapter by taking on more challenging tasks. Furthermore, we noticed that industries have also begun to leverage advanced NLP technologies to develop NLP tools for RE. There is now a real buzz of excitement that NLP4RE research can soon be transformed into a practical technology to support RE practice.

The mapping results can benefit researchers and practitioners in many ways. For researchers (also including students), the selected studies and their categorization can serve as useful references for further study and deeper analysis; the identified publication venues, particularly those 12 leading venues, can be used to narrow the search space for literature review in this area or for publishing relevant work; the trends and gaps identified from this mapping study have provided many new ideas for research opportunities. Practitioners can leverage the proposed NLP4RE tools for future development and research collaboration. Finally, the identified and synthesized NLP technologies, together with their usage and their relationships with the NLP4RE tasks, have provided a roadmap to help both practitioners and researchers gain a better understanding of what NLP technologies are in use in RE and how they are related. To conclude, we believe this mapping study is an important contribution to the field of NLP4RE, as it has provided a comprehensive overview of NLP4RE research and offered insights for future work.

Appendix 1. References for the 404 Selected NLP4RE Studies

| Study ID | Authors Name | Year | Paper Title | Publisher Venue | DOI |
|---|---|---|---|---|---|
| S1 | Han van der Aa, Henrik Leopold, Hajo A. Reijers | 2015 | Detecting Inconsistencies Between Process Models and Textual Descriptions | Lecture Notes in Computer Science | https://doi.org/10.1007/978-3-319-23063-4_6 |
| S3 | Zahra Shakeri Hossein Abad, Oliver Karras, Parisa Ghazi, Martin Glinz, Guenther Ruhe, Kurt Schneider | 2017 | What Works Better? A Study of Classifying Requirements | IEEE 25th International Requirements Engineering Conference | 10.1109/RE.2017.36 |
| S100 | Jaspreet Bhatia, Travis D. Breaux, Florian Schaub | 2016 | Mining Privacy Goals from Privacy Policies Using Hybridized Task Recomposition | ACM Transactions on Software Engineering and Methodology | https://doi.org/10.1145/2907942 |
| S101 | Jaspreet Bhatia, Morgan C. Evans, Sudarshan Wadkar, Travis D. Breaux | 2016 | Automated Extraction of Regulated Information Types Using Hyponymy Relations | IEEE 24th International Requirements Engineering Conference Workshops (REW) | https://doi.org/10.1109/rew.2016.018 |
| S103 | Tanmay Bhowmik, Nan Niu, Juha Savolainen, Anas Mahmoud | 2015 | Leveraging topic modeling and part-of-speech tagging to support combinational creativity in requirements engineering | Requirements Engineering | https://doi.org/10.1007/s00766-015-0226-2 |
| S104 | Daniel Bildhauer, Tassilo Horn, Jurgen Ebert | 2009 | Similarity-driven software reuse | ICSE Workshop on Comparison and Versioning of Software Models | https://doi.org/10.1109/cvsm.2009.5071719 |
| S106 | William J. Black | 1987 | Acquisition of conceptual data models from natural language descriptions | Association for Computational Linguistics | https://doi.org/10.3115/976858.976897 |
| S108 | Ravishankar Boddu, Lan Guo, Supratik Mukhopadhyay, and Bojan Cukic | 2004 | RETNA: from requirements to testing in a natural way | IEEE International Requirements Engineering Conference | https://doi.org/10.1109/icre.2004.1335683 |
| S109 | Mitra Bokaei Hosseini, Travis D. Breaux, Jianwei Niu | 2018 | Inferring Ontology Fragments from Semantic Role Typing of Lexical Variants | Requirements Engineering: Foundation for Software Quality | https://doi.org/10.1007/978-3-319-77243-1_3 |
| S114 | Ekaterina Boutkova, Frank Houdek | 2011 | Semi-automatic identification of features in requirement specifications | International Requirements Engineering Conference | https://doi.org/10.1109/re.2011.6051627 |
| S120 | Antonio Bucchiarone, Stefania Gnesi, Alessandro Fantechi, Gianluca Trentanni | 2010 | An experience in using a tool for evaluating a large set of natural language requirements | ACM Symposium on Applied Computing | https://doi.org/10.1145/1774088.1774148 |
| S121 | Edith Buchholz, Antje Düsterhöft, Bernhard Thalheim | 1997 | Capturing information on behaviour with the RADD-NLI: A linguistic and knowledge-based approach | Data and Knowledge Engineering | https://doi.org/10.1016/S0169-023X(97)00009-8 |
| S128 | Nathan Carlson, Phil Laplante | 2014 | The NASA automated requirements measurement tool: a reconstruction | Innovations in Systems and Software Engineering | https://doi.org/10.1007/s11334-013-0225-8 |
| S129 | Laura V. Galvis Carreno, Kristina Winbladh | 2013 | Analysis of user comments: an approach for software requirements evolution | International Conference on Software Engineering (ICSE) | https://doi.org/10.1109/icse.2013.6606604 |
| S13 | Christine Aguilera, Danie Berry | 1990 | The use of a repeated phrase finder in requirements extraction | Journal of Systems and Software | https://doi.org/10.1016/0164-1212(90)90097-6 |
| S131 | Gustavo Carvalho, Diogo Falcão, Flávia Barros, Augusto Sampaio, Alexandre Mota, Leonardo Motta, Mark Blackburn | 2014 | Test case generation from natural language requirements based on SCR specifications | ACM Symposium on Applied Computing | https://doi.org/10.1145/2480362.2480591 |
| S132 | Erik Casagrande, Selamawit Woldeamlak, Wei Lee Woon, H. H. Zeineldin, Davor Svetinovic | 2014 | NLP-KAOS for Systems Goal Elicitation: Smart Metering System Case Study | IEEE Transactions on Software Engineering | https://doi.org/10.1109/tse.2014.2339811 |
| S133 | Agustin Casamayor, Daniela Godoy, Marcelo Campo | 2010 | Identification of non-functional requirements in textual specifications: A semi-supervised learning approach | Information and Software Technology | https://doi.org/10.1016/j.infsof.2009.10.010 |
| S134 | Agustin Casamayor, Daniela Godoy, Marcelo Campo | 2012 | Functional grouping of natural language requirements for assistance in architectural software design | Knowledge-Based Systems | https://doi.org/10.1016/j.knosys.2011.12.009 |
| S137 | Carlos Castro-Herrera, Chuan Duan, Jane Cleland-Huang, Bamshad Mobasher | 2009 | A recommender system for requirements elicitation in large-scale software projects | ACM symposium on Applied Computing | https://doi.org/10.1145/1529282.1529601 |
| S140 | Carl K. Chang | 2016 | Situation Analytics: A Foundation for a New Software Engineering Paradigm | Computer | https://doi.org/10.1109/mc.2016.21 |
| S141 | Francis Chantree, Bashar Nuseibeh, Anne de Roeck, Alistair Willis | 2006 | Identifying Nocuous Ambiguities in Natural Language Requirements | IEEE International Requirements Engineering Conference | https://doi.org/10.1109/re.2006.31 |
| S145 | Peter Pin-Shan Chen | 1983 | English sentence structure and entity-relationship diagrams | Information Sciences | https://doi.org/10.7717/peerj.6725/table-1 |
| S15 | Fatima Alabdulkareem, Nick Cercone, Sotirios Liaskos | 2015 | Goal and Preference Identification through natural language | International Requirements Engineering Conference (RE) | https://doi.org/10.1109/re.2015.7320408 |
| S153 | Jane Cleland-Huang, Adam Czauderna, Marek Gibiec, John Emenecker | 2010 | A machine learning approach for tracing regulatory codes to product specific requirements | ACM/IEEE International Conference on Software Engineering | https://doi.org/10.1145/1806799.1806825 |
| S154 | Jane Cleland-Huang, Jin Guo | 2014 | Towards more intelligent trace retrieval algorithms | International Workshop on Realizing Artificial Intelligence Synergies in Software Engineering | https://doi.org/10.1145/2593801.2593802 |
| S155 | Jane Cleland-Huang, Raffaella Settimi, Chuan Duan, Xuchang Zou | 2005 | Utilizing supporting evidence to improve dynamic requirements traceability | IEEE International Conference on Requirements Engineering | https://doi.org/10.1109/re.2005.78 |
| S157 | Jane Cleland-Huang, Raffaella Settimi, Xuchang Zou, Peter Solc | 2007 | Automated classification of non-functional requirements | Requirements Engineering | https://doi.org/10.1007/s00766-007-0045-1 |
| S164 | Breno Dantas Cruz, Bargav Jayaraman, Anurag Dwarakanath, Collin McMillan | 2017 | Detecting Vague Words amp;amp; Phrases in Requirements Documents in a Multilingual Environment | International Requirements Engineering Conference (RE) | https://doi.org/10.1109/re.2017.24 |
| S171 | Johan Natt och Dag, Vincenzo Gervasi, Sjaak Brinkkemper, Björn Regnell | 2004 | Speeding up requirements management in a product software company: linking customer wishes to product requirements through linguistic engineering | IEEE International Requirements Engineering Conference | https://doi.org/10.1109/icre.2004.1335685 |
| S172 | Johan Natt och Dag, Thomas Thelin, Björn Regnell | 2006 | An experiment on linguistic tool support for consolidation of requirements from multiple sources in market-driven product development | Empirical Software Engineering | https://doi.org/10.1007/s10664-006-6405-5 |
| S174 | Fabiano Dalpiaz, Micaela Parente | 2019 | RE-SWOT: From User Feedback to Requirements via Competitor Analysis | Requirements Engineering: Foundation for Software Quality | https://doi.org/10.1007/978-3-030-15538-4_4 |
| S175 | Fabiano Dalpiaz, Ivor van der Schalk, Garm Lucassen | 2018 | Pinpointing Ambiguity and Incompleteness in Requirements Engineering via Information Visualization and NLP | Requirements Engineering: Foundation for Software Quality | https://doi.org/10.1007/978-3-319-77243-1_8 |
| S177 | Olawande Daramola, Thomas Moser, Guttorm Sindre, Stefan Biffl | 2012 | Managing Implicit Requirements Using Semantic Case-Based Reasoning Research Preview | Requirements Engineering: Foundation for Software Quality | https://doi.org/10.1007/978-3-642-28714-5_15 |
| S178 | Olawande Daramola, Tor Stalhane, Guttorm Sindre, Inah Omoronyia | 2011 | Enabling hazard identification from requirements and reuse-oriented HAZOP analysis | International Workshop on Managing Requirements Knowledge | https://doi.org/10.1109/mark.2011.6046555 |
| S182 | Jean-Marc Davril, Edouard Delfosse, Negar Hariri, Mathieu Acher, Jane Cleland-Huang, Patrick Heymans | 2013 | Feature model extraction from large collections of informal product descriptions | Joint Meeting on Foundations of Software Engineering | https://doi.org/10.1145/2491411.2491455 |
| S190 | Deva Kumar Deeptimahanti, Ratna Sanyal | 2009 | An Innovative Approach for Generating Static UML Models from Natural Language Requirements | Advances in Software Engineering | https://doi.org/10.1007/978-3-642-10242-4_13 |
| S193 | Alex Dekhtyar, Vivian Fong | 2017 | RE Data Challenge: Requirements Identification with Word2Vec and TensorFlow | International Requirements Engineering Conference (RE) | https://doi.org/10.1109/re.2017.26 |
| S198 | Fangshu Di, Maolin Zhang | 2009 | An Improving Approach for Recovering Requirements-to-Design Traceability Links | International Conference on Computational Intelligence and Software Engineering | https://doi.org/10.1109/cise.2009.5366024 |
| S2 | Zahra Shakeri Hossein Abad, Vincenzo Gervasi, Didar Zowghi, Ken Barker | 2018 | ELICA: An Automated Tool for Dynamic Extraction of Requirements Relevant Information | International Workshop on Artificial Intelligence for Requirements Engineering (AIRE) | https://doi.org/10.1109/aire.2018.00007 |
| S20 | Jean Pierre Alfonso Hoyos, Felipe Restrepo-Calle | 2018 | Fast Prototyping of Web-Based Information Systems Using a Restricted Natural Language Specification | Communications in Computer and Information Science | https://doi.org/10.1007/978-3-319-94135-6_9 |
| S203 | Zuohua Ding, Mingyue Jiang, Jens Palsberg | 2011 | From textual use cases to service component models | International workshop on Principles of engineering service-oriented systems | https://doi.org/10.1145/1985394.1985396 |
| S204 | Julio Cesar Sampaio do Prado Leite, Graciela D. S. Hadad, Jorge Horacio Doorn, Gladys N. Kaplan | 2000 | A scenario construction process | Requirements Engineering | https://doi.org/10.1007/PL00010342 |
| S207 | Markus Dollmann, Michaela Geierhos | 2016 | On- and Off-Topic Classification and Semantic Annotation of User-Generated Software Requirements | Conference on Empirical Methods in Natural Language Processing | https://doi.org/10.18653/v1/d16-1186 |
| S208 | Dov Dori, Nahum Korda, Avi Soffer, Shalom Cohen | 2004 | SMART: System Model Acquisition from Requirements Text | Lecture Notes in Computer Science | https://doi.org/10.1007/978-3-540-25970-1_12 |
| S21 | Waad Alhoshan, Riza Batista-Navarro and Liping Zhao | 2019 | Using Frame Embeddings to Identify Semantically Related Software Requirements | NLP4RE Workshop | http://ceur-ws.org/Vol-2376/NLP4RE19_paper05.pdf |
| S211 | Jaroslav Drazan, Vladimir Mencl | 2007 | Improved Processing of Textual Use Cases: Deriving Behavior Specifications | Lecture Notes in Computer Science | https://doi.org/10.1007/978-3-540-69507-3_74 |
| S212 | Chuan Duan, Jane Cleland-Huang, Bamshad Mobasher | 2008 | A consensus based approach to constrained clustering of software requirements | ACM conference on Information and knowledge mining | https://doi.org/10.1145/1458082.1458225 |
| S213 | Chuan Duan, Paula Laurent, Jane Cleland-Huang, Charles Kwiatkowski | 2009 | Towards automated requirements prioritization and triage | Requirements Engineering | https://doi.org/10.1007/s00766-009-0079-7 |
| S218 | Sebastian Eder, Henning Femmer, Benedikt Hauptmann, Maximilian Junker | 2015 | Configuring latent semantic indexing for requirements tracing | International Workshop on Requirements Engineering and Testing | https://doi.org/10.1109/ret.2015.13 |
| S219 | Mohamed El-Attar | 2012 | Towards developing consistent misuse case models | Journal of Systems and Software | https://doi.org/10.1016/j.jss.2011.08.023 |
| S22 | Busyairah Syd Ali, Zarinah Mohd. Kasirun | 2008 | Developing tool for crosscutting concern identification using NLP | International Symposium on Information Technology | https://doi.org/10.1109/itsim.2008.4632039 |
| S220 | Sarah Saad Eldin, Ammar Mohammed, Hesham Hefny, Ahmed Sharaf Eldin Ahmed | 2019 | An Enhanced Opinion Retrieval Approach on Arabic Text for Customer Requirements Expansion | Journal of King Saud University - Computer and Information Sciences | https://doi.org/10.1016/j.jksuci.2019.01.010 |
| S221 | Roaa Elghondakly, Sherin Moussa, Nagwa Badr | 2015 | Waterfall and agile requirements-based model for automated test cases generation | International Conference on Intelligent Computing and Information Systems (ICICIS) | https://doi.org/10.1109/intelcis.2015.7397285 |
| S223 | Chetan Arora, Mehrdad Sabetzadeh, Lionel Briand, Frank Zimmer, Raul Gnaga | 2013 | Automatic Checking of Conformance to Requirement Boilerplates via Text Chunking: An Industrial Case Study | ACM / IEEE International Symposium on Empirical Software Engineering and Measurement | https://doi.org/10.1109/esem.2013.13 |
| S224 | Muneera Bano, Alessio Ferrari, Didar Zowghi, Vincenzo Gervasi, Stefania Gnesi | 2015 | Automated Service Selection Using Natural Language Processing | Requirements Engineering in the Big Data Era | https://doi.org/10.1007/978-3-662-48634-4_1 |
| S225 | Shuang Liu, Jun Sun, Yang Liu, Yue Zhang, Bimlesh Wadhwa, Jin Song Dong, Xinyu Wang | 2014 | Automatic early defects detection in use case documents | ACM/IEEE international conference on Automated software engineering | https://doi.org/10.1145/2642937.2642969 |
| S226 | Hui Yang, Alistair Willis, Anne De Roeck, Bashar Nuseibeh | 2010 | Automatic detection of nocuous coordination | IEEE/ACM international conference on | https://doi.org/10.1145/1858996.1859007 |
| | | | ambiguities in natural language requirements | Automated software engineering | |
| S227 | Tao Yue, Shaukat Ali, Lionel Briand | 2011 | Automated Transition from Use Cases to UML State Machines to Support State-Based Testing | Modelling Foundations and Applications | https://doi.org/10.1007/978-3-642-21470-7_9 |
| S228 | Onyeka Emebo, Daramola Olawande, Ayo Charles | 2016 | An automated tool support for managing implicit requirements using Analogy-based Reasoning | International Conference on Research Challenges in Information Science (RCIS) | https://doi.org/10.1109/rcis.2016.7549329 |
| S23 | Ishfaq Ali, Muhammad Asif, Muhammad Shahbaz, Adnan Khalid, Mariam Rehman, Aziz Guergachi | 2018 | Text Categorization Approach for Secure Design Pattern Selection Using Software Requirement Specification | IEEE Access | https://doi.org/10.1109/access.2018.2883077 |
| S231 | Morgan C. Evans, Jaspreet Bhatia, Sudarshan Wadkar, Travis D. Breaux | 2017 | An Evaluation of Constituency-Based Hyponymy Extraction from Privacy Policies | International Requirements Engineering Conference (RE) | https://doi.org/10.1109/re.2017.87 |
| S232 | Fabrizio Fabbrini, Mario Fusani, Stefania Gnesi, Giuseppe Lami | 2001 | The linguistic approach to the natural language requirements quality: benefit of the use of an automatic tool | NASA Goddard Software Engineering Workshop | https://doi.org/10.1109/sew.2001.992662 |
| S234 | Davide Falessi, Giovanni Cantone | 2019 | The Effort Savings from Using NLP to Classify Equivalent Requirements | IEEE Software | https://doi.org/10.1109/ms.2018.2874620 |
| S237 | Alessandro Fantechi, Alessio Ferrari, Stefania Gnesi, Laura Semini | 2018 | Hacking an Ambiguity Detection Tool to Extract Variation Points: an Experience Report | International Workshop on Variability Modelling of Software-Intensive Systems | https://doi.org/10.1145/3168365.3168381 |
| S239 | Alessandro Fantechi, Stefania Gnesi, Giuseppe Lami, Alessandro Maccari | 2003 | Applications of linguistic techniques for use case analysis | Requirements Engineering | https://doi.org/10.1007/s00766-003-0174-0 |
| S24 | Nasir Ali, Yann-Gaël Gueheneuc, Giuliano Antoniol | 2011 | Requirements Traceability for Object Oriented Systems by Partitioning Source Code | Working Conference on Reverse Engineering | https://doi.org/10.1109/wcre.2011.16 |
| S241 | Agung Fatwanto | 2013 | Software requirements specification analysis using natural language processing technique | International Conference on Quality in Research (QiR) | https://doi.org/10.1109/qir.2013.6632546 |
| S245 | Henning Femmer, Daniel Méndez Fernández, Stefan Wagner, Sebastian Eder | 2016 | Rapid quality assurance with Requirements Smells | Journal of Systems and Software | https://doi.org/10.1016/j.jss.2016.02.047 |
| S25 | Nasir Ali, Yann-Gaël Gueheneuc, Giuliano Antoniol | 2011 | Trust-Based Requirements Traceability | International Conference on Program Comprehension | https://doi.org/10.1109/icpc.2011.42 |
| S251 | Alessio Ferrari, Felice DellOrletta, Andrea Esuli, Vincenzo Gervasi, Stefania Gnesi | 2017 | Natural Language Requirements Processing: A 4D Vision | IEEE Software | https://doi.org/10.1109/ms.2017.4121207 |
| S252 | Alessio Ferrari, Felice dell'Orletta, Giorgio Oronzo Spagnolo, Stefania Gnesi | 2014 | Measuring and Improving the Completeness of Natural Language Requirements | Requirements Engineering: Foundation for Software Quality | https://doi.org/10.1007/978-3-319-05843-6_3 |
| S254 | Alessio Ferrari, Andrea Esuli, Stefania Gnesi | 2018 | Identification of Cross-Domain Ambiguity with Language Models | International Workshop on Artificial Intelligence for Requirements Engineering (AIRE) | https://doi.org/10.1109/aire.2018.00011 |
| S256 | Alessio Ferrari, Stefania Gnesi | 2012 | Using collective intelligence to detect pragmatic ambiguities | IEEE International Requirements Engineering Conference (RE) | https://doi.org/10.1109/re.2012.6345803 |
| S257 | Alessio Ferrari, Stefania Gnesi, Gabriele Tolomei | 2012 | A clustering-based approach for discovering flaws in requirements specifications | ACM Symposium on Applied Computing | https://doi.org/10.1145/2245276.2231939 |
| S258 | Alessio Ferrari, Stefania Gnesi, Gabriele Tolomei | 2013 | Using Clustering to Improve the Structure of Natural Language Requirements Documents | Requirements Engineering: Foundation for Software Quality | https://doi.org/10.1007/978-3-642-37422-7_3 |
| S259 | Alessio Ferrari, Giuseppe Lipari, Stefania Gnesi, Giorgio O. Spagnolo | 2014 | Pragmatic ambiguity detection in natural language requirements | International Workshop on Artificial Intelligence for Requirements Engineering (AIRE) | https://doi.org/10.1109/aire.2014.6894849 |
| S26 | Nasir Ali, Fehmi Jaafar, Ahmed E. Hassan | 2013 | Leveraging historical co-change information for requirements traceability | Working Conference on Reverse Engineering (WCRE) | https://doi.org/10.1109/wcre.2013.6671311 |
| S261 | Alessio Ferrari, Giorgio O. Spagnolo, Stefania Gnesi, Felice Dell'Orletta | 2015 | CMT and FDE: tools to bridge the gap between natural language documents and feature diagrams | International Conference on Software Product Line | https://doi.org/10.1145/2791060.2791117 |
| S264 | Alessio Ferrari, Paola Spoletini, Stefania Gnesi | 2016 | Ambiguity and tacit knowledge in requirements elicitation interviews | Requirements Engineering | https://doi.org/10.1007/s00766-016-0249-3 |
| S266 | David de Almeida Ferreira, Alberto Rodrigues da Silva | 2012 | RSLingo: An information extraction approach toward formal requirements specifications | International Workshop on Model-Driven Requirements Engineering (MoDRE) | https://doi.org/10.1109/modre.2012.6360073 |
| S271 | Günther Fliedl, Christian Kop, Heinrich C. Mayr | 2005 | From textual scenarios to a conceptual schema | Data and Knowledge Engineering | https://doi.org/10.1016/j.datak.2004.11.007 |
| S272 | Günther Fliedl, Christian Kop, Heinrich C Mayr, Willi Mayerthaler, Christian Winkler | 2000 | Linguistically based requirements engineering: The NIBA-project | Data and Knowledge Engineering | https://doi.org/10.1016/s0169-023x(00)00029-x |
| S273 | Günther Fliedl, Christian Kop, Heinrich C. Mayr, Alexander Salbrechter, Jürgen Vöhringer, Georg Weber, Christian Winkler | 2007 | Deriving static and dynamic concepts from software requirements using sophisticated tagging | Data and Knowledge Engineering | https://doi.org/10.1016/j.datak.2006.06.012 |
| S274 | Jorge J. García Flores | 2004 | Semantic filtering of textual requirements descriptions | Natural Language Processing and Information Systems | https://doi.org/10.1007/978-3-540-27779-8_42 |
| S28 | Yara Alkhader, Amjad Hudaib, Bassam Hammo | 2006 | Experimenting With Extracting Software Requirements Using NLP Approach | International Conference on Information and Automation | https://doi.org/10.1109/icinfa.2006.374136 |
| S283 | F. Friedrich, J. Mendling and F. Puhlmann | 2011 | Process model generation from natural language text | Notes on Numerical Fluid Mechanics and Multidisciplinary Design | https://doi.org/10.1007/978-3-642-21640-4_36 |
| S284 | Ricardo Gacitua, Pete Sawyer, Vincenzo Gervasi | 2010 | On the Effectiveness of Abstraction Identification in Requirements Engineering | International Requirements Engineering Conference | https://doi.org/10.1109/re.2010.12 |
| S291 | Stefan Gartner, Thomas Ruhroth, Jens Burger, Kurt Schneider, Jan Jurjens | 2014 | Maintaining requirements for long-living software systems by incorporating security knowledge | International Requirements Engineering Conference (RE) | https://doi.org/10.1109/re.2014.6912252 |
| S292 | Francesco Garzoli, Danilo Croce, Manuela Nardini, Francesco Ciambra, Roberto Basili | 2013 | Robust Requirements Analysis in Complex Systems through Machine Learning | Trustworthy Eternal Systems via Evolving Software, Data and Knowledge | https://doi.org/10.1007/978-3-642-45260-4_4 |
| S294 | S. Geetha and G. S. A. Mala | 2013 | Extraction of key attributes from natural language requirements specification text | International Conference on Sustainable Energy and Intelligent Systems | https://doi.org/10.1049/ic.2013.0341 |
| S295 | Tim Gemkow, Miro Conzelmann, Kerstin Hartig, Andreas Vogelsang | 2018 | Automatic Glossary Term Extraction from Large-Scale Requirements Specifications | International Requirements Engineering Conference (RE) | https://doi.org/10.1109/re.2018.00052 |
| S297 | Vincenzo Gervasi, Bashar Nuseibeh | 2002 | Lightweight validation of natural language requirements | International Conference on Requirements Engineering | https://doi.org/10.1109/icre.2000.855601 |
| S298 | Vincenzo Gervasi, Didar Zowghi | 2005 | Reasoning about inconsistencies in natural language requirements | ACM Transactions on Software Engineering and Methodology | https://doi.org/10.1145/1072997.1072999 |
| S299 | Smita Ghaisas, Manish Motwani, Preethu Rose Anish | 2013 | Detecting system use cases and validations from documents | IEEE/ACM International Conference on Automated Software Engineering (ASE) | https://doi.org/10.1109/ase.2013.6693114 |
| S30 | Hala Alrumaih, Abdulrahman Mirza, Hessah Alsalamah | 2018 | Toward Automated Software Requirements Classification | Saudi Computer Society National Computer Conference (NCC) | https://doi.org/10.1109/ncg.2018.8593012 |
| S302 | Marek Gibiec, Adam Czauderna, Jane Cleland-Huang | 2010 | Towards mining replacement queries for hard-to-retrieve traces | IEEE/ACM international conference on Automated software engineering | https://doi.org/10.1145/1858996.1859046 |
| S303 | Reynaldo Giganto, Tony Smith | 2008 | Derivation of Classes from Use Cases Automatically Generated by a Three-Level Sentence Processing Algorithm | International Conference on Systems | https://doi.org/10.1109/icons.2008.50 |
| S306 | Gokhan Gokyer, Semih Cetin, Cevat Sener, Meltem T. Yondem | 2008 | Non-functional Requirements to Architectural Concerns: ML and NLP at Crossroads | International Conference on Software Engineering Advances | https://doi.org/10.1109/icsea.2008.28 |
| S307 | Leah Goldin, Danie Berry | 1994 | AbstFinder, a prototype abstraction finder for natural language text for use in requirements elicitation: design, methodology, and evaluation | IEEE International Conference on Requirements Engineering | https://doi.org/10.1109/icre.1994.292399 |
| S309 | Fernando Gomez, Carlos Segami, Carl Delaune | 1999 | A system for the semiautomatic generation of E-R models from natural language specifications | Data and Knowledge Engineering | https://doi.org/10.1016/s0169-023x(98)00032-9 |
| S31 | Lilac A. E. Al-Safadi | 2009 | Natural Language Processing for Conceptual Modeling | International Journal of Digital Content Technology and its Applications | https://doi.org/10.4156/jdcta.vol3.issue3.6 |
| S313 | Nicolas Gorse, Pascale Belanger, Alexandre Chureau, El Mostapha Aboulhamid, Yvon Savaria | 2007 | A high-level requirements engineering methodology for electronic system-level design | System Level Design with .Net Technology | https://doi.org/10.1016/j.compeleceng.2007.02.004 |
| S319 | Eduard C. Groen, Joerg Doerr, and Sebastian Adam | 2015 | Towards Crowd-Based Requirements Engineering: A Research Preview | International Working Conference on Requirements Engineering: Foundation for Software Quality | https://doi.org/10.1007/978-3-319-16101-3_16 |
| S322 | Sarita Gulia, Tanupriya Choudhury | 2016 | An efficient automated design to generate UML diagram from Natural Language Specifications | International Conference - Cloud System and Big Data Engineering (Confluence) | https://doi.org/10.1109/confluence.2016.7508197 |
| S325 | Hui Guo, Ozgur Kafali, Munindar Singh | 2018 | Extraction of Natural Language Requirements from Breach Reports Using Event Inference | International Workshop on Artificial Intelligence for Requirements Engineering (AIRE) | https://doi.org/10.1109/aire.2018.00009 |
| S326 | Qing-lin Guo, Ming Zhang | 2009 | Semantic information integration and question answering based on pervasive agent ontology | Expert Systems with Applications | https://doi.org/10.1016/j.eswa.2009.01.056 |
| S33 | Sousuke Amasaki, Pattara Leelaprute | 2018 | The Effects of Vectorization Methods on Non-Functional Requirements Classification | Euromicro Conference on Software Engineering and Advanced Applications (SEAA) | https://doi.org/10.1109/seaa.2018.00036 |
| S332 | Emitza Guzman, Mohamed Ibrahim, Martin Glinz | 2017 | A Little Bird Told Me: Mining Tweets for Requirements and Software Evolution | International Requirements Engineering Conference (RE) | https://doi.org/10.1109/re.2017.88 |
| S333 | Emitza Guzman, Walid Maalej | 2014 | How Do Users Like This Feature? A Fine Grained Sentiment Analysis of App Reviews | International Requirements Engineering Conference (RE) | https://doi.org/10.1109/re.2014.6912257 |
| S335 | Reiner Hähnle, Kristofer Johannisson, Aarne Ranta | 2002 | An Authoring Tool for Informal and Formal Requirements Specifications | Fundamental Approaches to Software Engineering | https://doi.org/10.1007/3-540-45923-5_16 |
| S337 | Ines Hajri, Arda Goknil, Lionel C. Briand, Thierry Stephany | 2018 | Configuring use case models in product families | Software and Systems Modeling | https://doi.org/10.1007/s10270-016-0539-8 |
| S34 | Vincenzo Ambriola and Vincenzo Gervasi | 1997 | Processing natural language requirements | IEEE International Conference Automated Software Engineering | https://doi.org/10.1109/ase.1997.632822 |
| S341 | Mostafa Hamza, Robert J. Walker | 2015 | Recommending features and feature relationships from requirements documents for software product lines | International Workshop on Realizing Artificial Intelligence Synergies in Software Engineering | https://doi.org/10.1109/raise.2015.12 |
| S343 | H. M. Harmain, Robert Gaizauskas | 2000 | CM-Builder: an automated NL-based CASE tool | IEEE International Conference on Automated Software Engineering | https://doi.org/10.1109/ase.2000.873649 |
| S349 | Jane Huffman Hayes, Giulio Antoniol, Bram Adams, Yann-Gael Gueheneuc | 2015 | Inherent characteristics of traceability artifacts less is more | International Requirements Engineering Conference (RE) | https://doi.org/10.1109/re.2015.7320422 |
| S35 | Ana Paula Ambrosio, Elisabeth Métais, Jean-Noël Meunier | 1997 | The linguistic level: Contribution for conceptual design, view integration, reuse and documentation | Data and Knowledge Engineering | https://doi.org/10.1016/s0169-023x(96)00028-6 |
| S351 | Allenoush Hayrapetian, Rajeev Raje | 2018 | Empirically Analyzing and Evaluating Security Features in Software Requirements | Innovations in Software Engineering Conference | https://doi.org/10.1145/3172871.3172879 |
| S356 | Jose Herrera, Isela Macia, Percy Salas, Rafael Pinho, Ronald Vargas, Alessandro Garcia, Joao Araujo, Karin Breitman | 2012 | Revealing Crosscutting Concerns in Textual Requirements Documents: An Exploratory Study with Industry Systems | Brazilian Symposium on Software Engineering | https://doi.org/10.1109/sbes.2012.10 |
| S36 | Harksoo Kim, Youngjoong Ko, Sooyong Park, Jungyun Seo | 1999 | Informal requirements analysis supporting system for human engineer | IEEE International Conference on Systems, Man, and Cybernetics | https://doi.org/10.1109/icsmc.1999.823367 |
| S362 | E. Ashlee Holbrook, Jane Huffman Hayes, Alex Dekhtyar | 2009 | Toward Automating Requirements Satisfaction Assessment | IEEE International Requirements Engineering Conference | https://doi.org/10.1109/re.2009.10 |
| S363 | Jörg Holtmann, Jan Meyer, Markus von Detten | 2011 | Automatic Validation and Correction of Formalized, Textual Requirements | International Conference on Software Testing, Verification and Validation Workshops | https://doi.org/10.1109/icstw.2011.17 |
| S367 | Bowen Hui, Eric Yu | 2005 | Extracting conceptual relationships from specialized documents | Conceptual Modeling | https://doi.org/10.1007/3-540-45816-6_26 |
| S369 | Ishrar Hussain, Leila Kosseim, Olga Ormandjieva | 2008 | Using linguistic knowledge to classify non-functional requirements in SRS documents | Lecture Notes in Computer Science | https://doi.org/10.1007/978-3-540-69858-6_28 |
| S37 | Barrett R. Bryant and Beum-Seuk Lee | 2002 | Two-level grammar as an object-oriented requirements specification language | Hawaii International Conference on System Sciences | https://doi.org/10.1109/hicss.2002.994486 |
| S370 | Ishrar Hussain, Olga Ormandjieva, Leila Kosseim | 2007 | Automatic Quality Assessment of SRS Text by Means of a Decision-Tree-Based Text Classifier | International Conference on Quality Software | https://doi.org/10.1109/qsic.2007.4385497 |
| S374 | Deniz Iren, Hajo A. Reijers | 2017 | Leveraging business process improvement with natural language processing and organizational semantic knowledge | International Conference on Software and System Process | https://doi.org/10.1145/3084100.3084112 |
| S376 | Prateek Jain, Kunal Verma, Alex Kass, Reymonrod G. Vasquez | 2009 | Automated review of natural language requirements documents: generating useful warnings with user-extensible glossaries driving a simple state machine | Conference on India software engineering conference | https://doi.org/10.1145/1506216.1506224 |
| S377 | Ritika Jain, Smita Ghaisas, Ashish Sureka | 2014 | SANAYOJAN: a framework for traceability link recovery between use-cases in software requirement specification and regulatory documents | International Workshop on Realizing Artificial Intelligence Synergies in Software Engineering | https://doi.org/10.1145/2593801.2593804 |
| S38 | Asbjørn Andersen, Klaus Heje Munch | 1991 | Automatic generation of technical documentation | Expert Systems with Applications | https://doi.org/10.1016/0957-4174(91)90150-d |
| S380 | Simona Jeners, Horst Lichter, Ana Dragomir | 2012 | Towards an Integration of Multiple Process Improvement Reference Models Based on Automated Concept Extraction | Communications in Computer and Information Science | https://doi.org/10.1007/978-3-642-31199-4_18 |
| S381 | Nishant Jha, Anas Mahmoud | 2017 | Mining User Requirements from Application Store Reviews Using Frame Semantics | Requirements Engineering: Foundation for Software Quality | https://doi.org/10.1007/978-3-319-54045-0_20 |
| S385 | Timo Johann, Christoph Stanik, Alireza M. Alizadeh B., Walid Maalej | 2017 | SAFE: A Simple Approach for Feature Extraction from App Descriptions and App Reviews | International Requirements Engineering Conference (RE) | https://doi.org/10.1109/re.2017.71 |
| S388 | Jakub Jurkiewicz, Jerzy R. Nawrocki | 2015 | Automated events identification in use cases | Information and Software Technology | https://doi.org/10.1016/j.infsof.2014.09.011 |
| S395 | Massila Kamalrudin, John Hosking, John Grundy | 2016 | MaramaAIC: tool support for consistency management and validation of requirements | Automated Software Engineering | https://doi.org/10.1007/s10515-016-0192-z |
| S400 | Dhikra Kchaou, Nadia Bouassida, Mariam Mefteh, Hanêne Ben-Abdallah | 2019 | Recovering semantic traceability between requirements and design for change impact analysis | Innovations in Systems and Software Engineering | https://doi.org/10.1007/s11334-019-00330-w |
| S41 | Krasimir Angelov, John J. Camilleri, Gerardo Schneider | 2013 | A framework for conflict analysis of normative texts written in controlled natural language | The Journal of Logic and Algebraic Programming | https://doi.org/10.1016/j.jlap.2013.03.002 |
| S412 | Christoph M. Kirchsteiger, Christoph Trummer, Christian Steger, Reinhold Weiss, Markus Pistauer | 2008 | Specification-based Verification of Embedded Systems by Automated Test Case Generation | Distributed Embedded Systems: Design, Middleware and Resources | https://doi.org/10.1007/978-0-387-09661-2_4 |
| S413 | Hasan Kitapci, Barry Boehm | 2007 | Formalizing informal stakeholder decisions: A hybrid method approach | International Conference on System Sciences | https://doi.org/10.1109/hicss.2007.233 |
| S416 | Nadzeya Kiyavitskaya, Nicola Zeni, Luisa Mich, Danie Berry | 2008 | Requirements for tools for ambiguity identification and measurement in natural language requirements specifications | Requirements Engineering | https://doi.org/10.1007/s00766-008-0063-7 |
| S422 | Youngjoong Ko, Sooyong Park, Jungyun Seo, Soonhwang Choi | 2007 | Using classification techniques for informal requirements in the requirements analysis-supporting system | Information and Software Technology | https://doi.org/10.1016/j.infsof.2006.11.007 |
| S424 | Leonid Kof | 2007 | Scenarios: Identifying Missing Objects and Actions by Means of Computational Linguistics | IEEE International Requirements Engineering Conference | https://doi.org/10.1109/re.2007.38 |
| S427 | Leonid Kof | 2009 | Requirements Analysis: Concept Extraction and Translation of Textual Specifications to Executable Models | Natural Language Processing and Information Systems | https://doi.org/10.1007/978-3-642-12550-8_7 |
| S432 | Sven J. Korner, Torben Brumm | 2009 | RESI - A Natural Language Specification Improver | IEEE International Conference on Semantic Computing | https://doi.org/10.1109/icsc.2009.47 |
| S433 | Sven J. Körner, Mathias Landhäußer | 2010 | Semantic enriching of natural language texts with automatic thematic role annotation | Natural Language Processing and Information Systems | https://doi.org/10.1007/978-3-642-13881-2_9 |
| S438 | Jennifer Krisch, Frank Houdek | 2015 | The myth of bad passive voice and weak words an empirical investigation in the automotive industry | International Requirements Engineering Conference (RE) | https://doi.org/10.1109/re.2015.7320451 |
| S439 | Jaroslaw Kuchta, Priti Padhiyar | 2018 | Extracting Concepts from the Software Requirements Specification Using Natural Language Processing | International Conference on Human System Interaction (HSI) | https://doi.org/10.1109/hsi.2018.8431221 |
| S445 | Zijad Kurtanovic, Walid Maalej | 2017 | Automatically Classifying Functional and Non-functional Requirements Using Supervised Machine Learning | International Requirements Engineering Conference (RE) | https://doi.org/10.1109/re.2017.82 |
| S447 | Mathias Landhäußer, Sven J. Körner, Walter F. Tichy | 2014 | From requirements to UML models and back: how automatic processing of text can support requirements engineering | Software Quality Journal | https://doi.org/10.1007/s11219-013-9210-6 |
| S448 | Mathias Landhausser, Sven J. Korner, Walter F. Tichy, Jan Keim, Jennifer Krisch | 2015 | DeNom: a tool to find problematic nominalizations using NLP | International Workshop on Artificial Intelligence for Requirements Engineering (AIRE) | https://doi.org/10.1109/aire.2015.7337623 |
| S450 | Chetan Arora, Mehrdad Sabetzadeh, Lionel Briand, Frank Zimmer | 2015 | Automated Checking of Conformance to Requirements Templates Using Natural Language Processing | IEEE Transactions on Software Engineering | https://doi.org/10.1109/tse.2015.2428709 |
| S451 | Anurag Dwarakanath, Roshni R. Ramnani, Shubhashis Sengupta | 2013 | Automatic extraction of glossary terms from natural language requirements | IEEE International Requirements Engineering Conference (RE) | https://doi.org/10.1109/re.2013.6636736 |
| S453 | Mirosław Ochodek, Jerzy Nawrocki | 2008 | Automatic Transactions Identification in Use Cases | Balancing Agility and Formalism in Software Engineering | https://doi.org/10.1007/978-3-540-85279-7_5 |
| S455 | Raúl Lapeña, Jaime Font, Carlos Cetina, Óscar Pastor | 2018 | Exploring New Directions in Traceability Link Recovery in Models: The Process Models Case | Advanced Information Systems Engineering | https://doi.org/10.1007/978-3-319-91563-0_22 |
| S457 | Ralf Laue, Wilhelm Koop, Volker Gruhn | 2016 | Indicators for Open Issues in Business Process Models | Requirements Engineering: Foundation for Software Quality | https://doi.org/10.1007/978-3-319-30282-9_7 |
| S458 | Algirdas Laukaitis, Olegas Vasilecas | 2007 | Integrating All Stages of Information Systems Development by Means of Natural Language Processing | Requirements Engineering: Foundation for Software Quality | https://doi.org/10.1007/978-3-540-73031-6_16 |
| S46 | K. M. Annervaz, Vikrant Kaulgud, Shubhashis Sengupta, Milind Savagaonkar | 2013 | Natural language requirements quality analysis based on business domain models | IEEE/ACM International Conference on Automated Software Engineering (ASE) | https://doi.org/10.1109/ase.2013.6693132 |
| S462 | R. Lecoeuche | 2000 | Finding comparatively important concepts between texts | IEEE International Conference on Automated Software Engineering | https://doi.org/10.1109/ase.2000.873650 |
| S463 | Anita Lee, Chun Hung Cheng, Jaydeep Balakrishnan | 1998 | Software development cost estimation: Integrating neural network with cluster analysis | Information and Management | https://doi.org/10.1016/s0378-7206(98)00041-x |
| S464 | Beum-Seuk Lee, Barrett R. Bryant | 2002 | Contextual natural language processing and DAML for understanding software requirements specifications | International conference on Computational linguistics | https://doi.org/10.3115/1072228.1072352 |
| S471 | Henrik Leopold, Rami-Habib Eid-Sabbagh, Jan Mendling, Leonardo Guerreiro Azevedo, Fernanda Araujo Baião | 2013 | Detection of naming convention violations in process models for different languages | Decision Support Systems | https://doi.org/10.1016/j.dss.2013.06.014 |
| S472 | Henrik Leopold, Jan Mendling, Artem Polyvyanyy | 2012 | Generating Natural Language Texts from Business Process Models | Numerical Fluid Mechanics and Multidisciplinary Design | https://doi.org/10.1007/978-3-642-31095-9_5 |
| S473 | Henrik Leopold, Fabian Pittke, Jan Mendling | 2015 | Automatic service derivation from business process model repositories via semantic technology | Journal of Systems and Software | https://doi.org/10.1016/j.jss.2015.06.007 |
| S474 | Henrik Leopold, Sergey Smirnov, Jan Mendling | 2012 | On the refactoring of activity labels in business process models | Information Systems | https://doi.org/10.1016/j.is.2012.01.004 |
| S477 | Tong Li | 2017 | Identifying Security Requirements Based on Linguistic Analysis and Machine Learning | Asia-Pacific Software Engineering Conference (APSEC) | https://doi.org/10.1109/apsec.2017.45 |
| S478 | Tong Li, Fan Zhang, Dan Wang | 2018 | Automatic User Preferences Elicitation: A Data-Driven Approach | Requirements Engineering: Foundation for Software Quality | https://doi.org/10.1007/978-3-319-77243-1_21 |
| S479 | Wenbin Li, Jane Huffman Hayes, Mirosław Truszczyński | 2015 | Towards More Efficient Requirements Formalization: A Study | Requirements Engineering: Foundation for Software Quality | https://doi.org/10.1007/978-3-319-16101-3_12 |
| S480 | Yang Li | 2018 | Feature and variability extraction from natural language software requirements specifications | International Conference on Systems and Software Product Line | https://doi.org/10.1145/3236405.3236427 |
| S481 | Yang Li, Sandro Schulze, Gunter Saake | 2018 | Reverse engineering variability from requirement documents based on probabilistic relevance and word embedding | International Conference on Systems and Software Product Line | https://doi.org/10.1145/3233027.3233033 |
| S482 | Yan Li, Tao Yue, Shaukat Ali, Li Zhang | 2017 | Enabling automated requirements reuse and configuration | Software and Systems Modeling | https://doi.org/10.1007/s10270-017-0641-6 |
| S483 | Zeheng Li, Mingrui Chen, LiGuo Huang, Vincent Ng | 2015 | Recovering Traceability Links in Requirements Documents | Conference on Computational Natural Language Learning | https://doi.org/10.18653/v1/k15-1024 |
| S486 | Sherlock A. Licorish, Bastin Tony Roy Savarimuthu, Swetha Keertipati | 2017 | Attributes that Predict which Features to Fix: Lessons for App Store Mining | International Conference on Evaluation and Assessment in Software Engineering | https://doi.org/10.1145/3084226.3084246 |
| S490 | Lin Liu, Tianying Li, Xiaoxi Kou | 2014 | Eliciting Relations from Natural Language Requirements Documents Based on Linguistic and Statistical Analysis | Computer Software and Applications Conference | https://doi.org/10.1109/compsac.2014.27 |
| S492 | Xueqing Liu, Yue Leng, Wei Yang, Chengxiang Zhai, Tao Xie | 2018 | Mining Android App Descriptions for Permission Requirements Recommendation | International Requirements Engineering Conference (RE) | https://doi.org/10.1109/re.2018.00024 |
| S496 | Claudia López, Víctor Codocedo, Hernán Astudillo, Luiz Marcio Cysneiros | 2012 | Bridging the gap between software architecture rationale formalisms and actual architecture documents: An ontology-driven approach | Science of Computer Programming | https://doi.org/10.1016/j.scico.2010.06.009 |
| S498 | Marco Lormans, Arie van Deursen | 2005 | Reconstructing requirements coverage views from design and test using traceability recovery via LSI | International workshop on Traceability in emerging forms of software engineering | https://doi.org/10.1145/1107656.1107665 |
| S499 | Neil Loughran, Américo Sampaio, Awais Rashid | 2006 | From Requirements Documents to Feature Models for Aspect Oriented Product Line Implementation | Satellite Events at the MoDELS | https://doi.org/10.1007/11663430_27 |
| S500 | Mengmeng Lu, Peng Liang | 2017 | Automatic Classification of Non-Functional Requirements from Augmented App User Reviews | International Conference on Evaluation and Assessment in Software Engineering | https://doi.org/10.1145/3084226.3084241 |
| S501 | Garm Lucassen, Fabiano Dalpiaz, Jan Martijn E. M. van der Werf, Sjaak Brinkkemper | 2016 | Improving agile requirements: the Quality User Story framework and tool | Requirements Engineering | https://doi.org/10.1007/s00766-016-0250-x |
| S503 | Garm Lucassen, Marcel Robeer, Fabiano Dalpiaz, Jan Martijn E. M. van der Werf, Sjaak Brinkkemper | 2017 | Extracting conceptual models from user stories with Visual Narrator | Requirements Engineering | https://doi.org/10.1007/s00766-017-0270-1 |
| S507 | Walid Maalej, Mathias Ellmann, Romain Robbes | 2017 | Using contexts similarity to predict relationships between tasks | Journal of Systems and Software | https://doi.org/10.1016/j.jss.2016.11.033 |
| S508 | Walid Maalej, Zijad Kurtanović, Hadeer Nabil, Christoph Stanik | 2016 | On the automatic classification of app reviews | Requirements Engineering | https://doi.org/10.1007/s00766-016-0251-9 |
| S510 | Yoelle S. Maarek, Danie Berry | 1989 | The use of lexical affinities in requirements extraction | International workshop on Software specification and design | https://doi.org/10.1145/75199.75229 |
| S513 | Kaushik Madala, Danielle Gaither, Rodney Nielsen, Hyunsook Do | 2017 | Automated Identification of Component State Transition Model Elements from Requirements | International Requirements Engineering Conference Workshops (REW) | https://doi.org/10.1109/rew.2017.73 |
| S514 | Kaushik Madala, Shraddha Piparia, Hyunsook Do, Renee Bryce | 2018 | Finding Component State Transition Model Elements Using Neural Networks: An Empirical Study | International Workshop on Artificial Intelligence for Requirements Engineering (AIRE) | https://doi.org/10.1109/aire.2018.00014 |
| S515 | Khalid Mahmood, Hironao Takahashi, Mazen Alobaidi | 2015 | A Semantic Approach for Traceability Link Recovery in Aerospace Requirements Management System | International Symposium on Autonomous Decentralized Systems | https://doi.org/10.1109/isads.2015.33 |
| S517 | Anas Mahmoud, Doris Carver | 2015 | Exploiting online human knowledge in Requirements Engineering | International Requirements Engineering Conference (RE) | https://doi.org/10.1109/re.2015.7320434 |
| S518 | Anas Mahmoud, Nan Niu | 2015 | On the role of semantics in automated requirements tracing | Requirements Engineering | https://doi.org/10.1007/s00766-013-0199-y |
| S52 | Brian Arendse, Garm Lucassen | 2016 | Toward Tool Mashups: Comparing and Combining NLP RE Tools | International Requirements Engineering Conference Workshops (REW) | https://doi.org/10.1109/rew.2016.019 |
| S520 | Neil Maiden, James Lockerbie, Konstantinos Zachos, Antonia Bertolino, Guglielmo De Angelis, Francesca Lonetti | 2014 | A Requirements-Led Approach for Specifying QoS-Aware Service Choreographies: An Experience Report | Requirements Engineering: Foundation for Software Quality | https://doi.org/10.1007/978-3-319-05843-6_18 |
| S523 | Haroon Malik, Elhadi M. Shakshuki | 2016 | Mining Collective Opinions for Comparison of Mobile Apps | Procedia Computer Science | https://doi.org/10.1016/j.procs.2016.08.026 |
| S528 | Ana C. Marcén, Francisca Pérez, Carlos Cetina | 2017 | Ontological Evolutionary Encoding to Bridge Machine Learning and Conceptual Models: Approach and Industrial Evaluation | Conceptual Modeling | https://doi.org/10.1007/978-3-319-69904-2_37 |
| S529 | Matheus Marinho, Danilo Arruda, Fernando Wanderley, Anthony Lins | 2018 | A Systematic Approach of Dataset Definition for a Supervised Machine Learning Using NFR Framework | International Conference on the Quality of Information and Communications Technology (QUATIC) | https://doi.org/10.1109/quatic.2018.00024 |
| S532 | Beniamino Di Martino, Jessica Pascarella, Stefania Nacchia, Salvatore Augusto Maisto, Pietro Iannucci, Fabio Cerri | 2018 | Cloud Services Categories Identification from Requirements Specifications | International Conference on Advanced Information Networking and Applications Workshops (WAINA) | https://doi.org/10.1109/waina.2018.00125 |
| S534 | Aaron K. Massey, Jacob Eisenstein, Annie I. Anton, Peter P. Swire | 2013 | Automated text mining for requirements analysis of policy documents | IEEE International Requirements Engineering Conference (RE) | https://doi.org/10.1109/re.2013.6636700 |
| S535 | Satoshi Masuda, Tohru Matsuodani, Kazuhiko Tsuda | 2016 | Automatic Generation of UTP Models from Requirements in Natural Language | International Conference on Software Testing, Verification and Validation Workshops (ICSTW) | https://doi.org/10.1109/icstw.2016.27 |
| S538 | Jin Matsuoka, Yves Lepage | 2011 | Ambiguity spotting using wordnet semantic similarity in support to recommended practice for Software Requirements Specifications | International Conference on Natural Language Processing and Knowledge Engineering | https://doi.org/10.1109/nlpke.2011.6138247 |
| S546 | Thorsten Merten, Matus Falis, Paul Hubner, Thomas Quirchmayr, Simone Bursner, Barbara Paech | 2016 | Software Feature Request Detection in Issue Tracking Systems | International Requirements Engineering Conference (RE) | https://doi.org/10.1109/re.2016.8 |
| S547 | Thorsten Merten, Bastian Mager, Simone Bürsner, Barbara Paech | 2016 | Do Information Retrieval Algorithms for Automated Traceability Perform Effectively on Issue Tracking System Data? | Working Conference on Mining Software Repositories | https://doi.org/10.1145/2597073.2597112 |
| S548 | Elisabeth Métais | 2002 | Enhancing information systems management with natural language processing techniques | Data and Knowledge Engineering | https://doi.org/10.1016/s0169-023x(02)00043-5 |
| S55 | Chetan Arora, Mehrdad Sabetzadeh, Lionel Briand, Frank Zimmer | 2017 | Automated Extraction and Clustering of Requirements Glossary Terms | IEEE Transactions on Software Engineering | https://doi.org/10.1109/tse.2016.2635134 |
| S550 | Manel Mezghani, Juyeon Kang, Florence Sedes | 2018 | Industrial Requirements Classification for Redundancy and Inconsistency Detection in SEMIOS | International Requirements Engineering Conference (RE) | https://doi.org/10.1109/re.2018.00037 |
| S551 | Farid Meziane, Nikos Athanasakis, Sophia Ananiadou | 2008 | Generating Natural Language specifications from UML class diagrams | Requirements Engineering | https://doi.org/10.1007/s00766-007-0054-0 |
| S552 | Farid Meziane, Yacine Rezgui | 2004 | A document management methodology based on similarity contents | Information Sciences | https://doi.org/10.1016/j.ins.2003.08.009 |
| S553 | Luisa Mich | 1996 | NL-OOPS: from natural language to object oriented requirements using the natural language processing system LOLITA | Natural Language Engineering | https://doi.org/10.1017/S1351324996001337 |
| S555 | Nasir Mehmood Minhas, Shahla Majeed, Ziaul Qayyum, Muhammad Aasem | 2011 | Controlled vocabulary based software requirements classification | Malaysian Conference in Software Engineering | https://doi.org/10.1109/mysec.2011.6140639 |
| S56 | Chetan Arora, Mehrdad Sabetzadeh, Arda Goknil, Lionel C. Briand, Frank Zimmer | 2015 | Change impact analysis for Natural Language requirements: An NLP approach | International Requirements Engineering Conference (RE) | https://doi.org/10.1109/re.2015.7320403 |
| S560 | Janardan Misra | 2016 | Terminological inconsistency analysis of natural language requirements | Information and Software Technology | https://doi.org/10.1016/j.infsof.2015.11.006 |
| S570 | Itzel Morales-Ramirez, Fitsum Meshesha Kifetew, Anna Perini | 2017 | Analysis of Online Discussions in Support of Requirements Discovery | Advanced Information Systems Engineering | https://doi.org/10.1007/978-3-319-59536-8_11 |
| S575 | Laurens Müter, Tejaswini Deoskar, Max Mathijssen, Sjaak Brinkkemper, Fabiano Dalpiaz | 2019 | Refinement of User Stories into Backlog Items: Linguistic Structure and Action Verbs | Requirements Engineering: Foundation for Software Quality | https://doi.org/10.1007/978-3-030-15538-4_7 |
| S582 | Masoud Narouei, Hassan Takabi | 2015 | Automatic Top-Down Role Engineering Framework Using Natural Language Processing Techniques | ACM Symposium on Access Control Models and Technologies | https://doi.org/10.1145/2752952.2752958 |
| S583 | Mirza Muhammad Naseer, Khalid Mahmood | 2014 | Subject dispersion of LIS research in Pakistan | Library and Information Science Research | https://doi.org/10.1016/j.lisr.2013.10.005 |
| S584 | Sana Ben Nasr, Guillaume Bécan, Mathieu Acher, João Bosco Ferreira Filho, Nicolas Sannier, Benoit Baudry, Jean-Marc Davril | 2017 | Automated extraction of product comparison matrices from informal product descriptions | Journal of Systems and Software | https://doi.org/10.1016/j.jss.2016.11.018 |
| S585 | Johan Natt och Dag, Björn Regnell, Pär Carlshamre, Michael Andersson, Joachim Karlsson | 2002 | A Feasibility Study of Automated Natural Language Requirements Analysis in Market-Driven Development | Requirements Engineering | https://doi.org/10.1007/s007660200002 |
| S586 | Falak Nawaz, Omar Hussain, Farookh Khadeer Hussain, Naeem Khalid Janjua, Morteza Saberi, Elizabeth Chang | 2019 | Proactive management of SLA violations by capturing relevant external events in a Cloud of Things environment | Future Generation Computer Systems | https://doi.org/10.1016/j.future.2018.12.034 |
| S598 | Tuong Huan Nguyen, John Grundy, Mohamed Almorsy | 2015 | Rule-based extraction of goal-use case models from text | Joint Meeting on Foundations of Software Engineering | https://doi.org/10.1145/2786805.2786876 |
| S600 | Remco A. Niemeijer, Bauke de Vries, Jakob Beetz | 2014 | Freedom through constraints: User-oriented architectural design | Advanced Engineering Informatics | https://doi.org/10.1016/j.aei.2013.11.003 |
| S601 | Allen P. Nikora, Galen Balcom | 2009 | Automated Identification of LTL Patterns in Natural Language Requirements | International Symposium on Software Reliability Engineering | https://doi.org/10.1109/issre.2009.15 |
| S602 | Gerald Ninaus, Florian Reinfrank, Martin Stettinger, Alexander Felfernig | 2014 | Content-based recommendation techniques for requirements engineering | International Workshop on Artificial Intelligence for Requirements Engineering (AIRE) | https://doi.org/10.1109/aire.2014.6894853 |
| S604 | Nan Niu, Steve Easterbrook | 2008 | Extracting and Modeling Product Line Functional Requirements | IEEE International Requirements Engineering Conference | https://doi.org/10.1109/re.2008.49 |
| S609 | Florian Noyrit, Sébastien Gérard, François Terrier | 2013 | Computer Assisted Integration of Domain-Specific Modeling Languages Using Text Analysis Techniques | Lecture Notes in Computer Science | https://doi.org/10.1007/978-3-642-41533-3_31 |
| S61 | Muesluem Atas, Ralph Samer, Alexander Felfernig | 2018 | Automated Identification of Type-Specific Dependencies between Requirements | IEEE/WIC/ACM International Conference on Web Intelligence (WI) | https://doi.org/10.1109/wi.2018.00-10 |
| S612 | Khenaidoo Nursimulu, Robert L. Probert | 1995 | Cause-effect graphing analysis and validation of requirements | Conference of the Centre for Advanced Studies on Collaborative research | https://dl.acm.org/citation.cfm?id=781961 |
| S616 | Olga Ormandjieva, Ishrar Hussain, Leila Kosseim | 2007 | Toward a text classification system for the quality assessment of software requirements written in natural language | International workshop on Software quality assurance in conjunction with the 6th ESEC/FSE joint meeting - SOQUA | https://doi.org/10.1145/1295074.1295082 |
| S617 | Mike Osborne, Cara K. MacNish | 1996 | Processing natural language software requirement specifications | International Conference on Requirements Engineering | https://doi.org/10.1109/icre.1996.491451 |
| S618 | Mohd Hafeez Osman, Mohd Firdaus Zaharin | 2018 | Ambiguous software requirement specification detection: an automated approach | International Workshop on Requirements Engineering and Testing | https://doi.org/10.1145/3195538.3195545 |
| S623 | Scott P. Overmyer, Benoit Lavoie, Owen Rambow | 2001 | Conceptual modeling through linguistic analysis using LIDA | International Conference on Software Engineering | https://doi.org/10.1109/icse.2001.919113 |
| S626 | Vincenzo Pallotta, Afzal Ballim | 2001 | Agent-Oriented Language Engineering for Robust NLP | Engineering Societies in the Agents World II | https://doi.org/10.1007/3-540-45584-1_7 |
| S629 | Fabio Palomba, Mario Linares-Vásquez, Gabriele Bavota, Rocco Oliveto, Massimiliano Di Penta, Denys Poshyvanyk, Andrea De Lucia | 2018 | Crowdsourcing user reviews to support the evolution of mobile apps | Journal of Systems and Software | https://doi.org/10.1016/j.jss.2017.11.043 |
| S635 | Sooyong Park, Harksoo Kim, Youngjoong Ko, Jungyun Seo | 2000 | Implementation of an efficient requirements-analysis supporting system using similarity measure techniques | Information and Software Technology | https://doi.org/10.1016/s0950-5849(99)00102-0 |
| S636 | Eugenio Parra, Christos Dimou, Juan Llorens, Valentín Moreno, Anabel Fraga | 2015 | A methodology for the classification of quality of requirements using machine learning techniques | Information and Software Technology | https://doi.org/10.1016/j.infsof.2015.07.006 |
| S637 | Anna Perini, Angelo Susi, Paolo Avesani | 2013 | A Machine Learning Approach to Software Requirements Prioritization | IEEE Transactions on Software Engineering | https://doi.org/10.1109/tse.2012.52 |
| S641 | Barbara Plank, Thomas Sauer, Ina Schaefer | 2013 | Supporting Agile Software Development by Natural Language Processing | Trustworthy Eternal Systems via Evolving Software, Data and Knowledge | https://doi.org/10.1007/978-3-642-45260-4_7 |
| S642 | Jantima Polpinij | 2009 | An ontology-based text processing approach for simplifying ambiguity of requirement specifications | IEEE Asia-Pacific Services Computing Conference (APSCC) | https://doi.org/10.1109/apscc.2009.5394119 |
| S644 | Daniel Popescu, Spencer Rugaber, Nenad Medvidovic, Danie Berry | 2007 | Reducing Ambiguities in Requirements Specifications Via Automatically Created Object-Oriented Models | Lecture Notes in Computer Science | https://doi.org/10.1007/978-3-540-89778-1_10 |
| S645 | Daniel Port, Allen Nikora, Jane Huffman Hayes and LiGuo Huang | 2011 | Text Mining Support for Software Requirements: Traceability Assurance | International Conference on System Sciences | https://doi.org/10.1109/hicss.2011.399 |
| S646 | Daniel Port, Allen Nikora, Jairus Hihn, LiGuo Huang | 2011 | Experiences with text mining large collections of unstructured systems development artifacts at jpl | International conference on Software engineering | https://doi.org/10.1145/1985793.1985891 |
| S647 | Roxana L. Q. Portugal, Tong Li, Lyrene Silva, Eduardo Almentero, Julio Cesar S. do Prado Leite | 2018 | NFRfinder: a knowledge based strategy for mining non-functional requirements | Brazilian Symposium on Software Engineering | https://doi.org/10.1145/3266237.3266269 |
| S651 | Piotr Pruski, Sugandha Lohar, Rundale Aquanette, Greg Ott, Sorawit Amornborvornwong, Alexander Rasin, Jane Cleland-Huang | 2014 | TiQi: Towards natural language trace queries | International Requirements Engineering Conference (RE) | https://doi.org/10.1109/re.2014.6912254 |
| S655 | Thomas Quirchmayr, Barbara Paech, Roland Kohl, Hannes Karey, Gunar Kasdepke | 2017 | Semi-automatic Software Feature-Relevant Information Extraction from Natural Language User Manuals | Empirical Software Engineering | https://doi.org/10.1007/s10664-018-9597-6 |
| S661 | Alejandro Rago, Claudia Marcos, J. Andrés Diaz-Pace | 2013 | Uncovering quality-attribute concerns in use case specifications via early aspect mining | Requirements Engineering | https://doi.org/10.1007/s00766-011-0142-z |
| S662 | Alejandro Rago, Claudia Marcos, J. Andres Diaz-Pace | 2014 | Assisting requirements analysts to find latent concerns with REAssistant | Automated Software Engineering | https://doi.org/10.1007/s10515-014-0156-0 |
| S663 | Alejandro Rago, Claudia Marcos, J. Andres Diaz-Pace | 2016 | Identifying duplicate functionality in textual use cases by aligning semantic actions | Software & Systems Modeling | https://doi.org/10.1109/models.2015.7338276 |
| S665 | Mona Rahimi, Mehdi Mirakhorli, Jane Cleland-Huang | 2014 | Automated extraction and visualization of quality concerns from requirements specifications | International Requirements Engineering Conference (RE) | https://doi.org/10.1109/re.2014.6912267 |
| S667 | Michael Rath, David Lo, Patrick Mäder | 2018 | Analyzing requirements and traceability information to improve bug localization | International Conference on Mining Software Repositories | https://doi.org/10.1145/3196398.3196415 |
| S672 | Iris Reinhartz-Berger, Mark Kemelman | 2019 | Extracting core requirements for software product lines | Requirements Engineering | https://doi.org/10.1007/s00766-018-0307-0 |
| S674 | Maria Riaz, Jason King, John Slankas, Laurie Williams | 2014 | Hidden in plain sight: Automatically identifying security requirements from natural language artifacts | International Requirements Engineering Conference (RE) | https://doi.org/10.1109/re.2014.6912260 |
| S675 | David Ribes, Geoffrey C. Bowker | 2009 | Between meaning and machine: Learning to represent the knowledge of communities | Information and Organization | https://doi.org/10.1016/j.infoandorg.2009.04.001 |
| S676 | Johan Natt och Dag , Björn Regnell, Vincenzo Gervasi, Sjaak Brinkkemper | 2005 | A linguistic-engineering approach to large-scale requirements management | IEEE Software | https://doi.org/10.1109/ms.2005.1 |
| S677 | Benedikt Gleich, Oliver Creighton, Leonid Kof | 2010 | Ambiguity Detection: Towards a Tool Explaining Ambiguity Sources | Requirements Engineering: Foundation for Software Quality | https://doi.org/10.1007/978-3-642-14192-8_20 |
| S678 | Colette Rolland and C. Proix | 1992 | A natural language approach for Requirements Engineering | Seminal Contributions to Information Systems Engineering | https://doi.org/10.1007/978-3-642-36926-1_3 |
| S679 | Hui Yang, Anne de Roeck, Vincenzo Gervasi, Alistair Willis, Bashar Nuseibeh | 2011 | Analysing anaphoric ambiguity in natural language requirements | Requirements Engineering | https://doi.org/10.1007/s00766-011-0119-y |
| S683 | Marcel Robeer, Garm Lucassen, Jan Martijn E. M. van der Werf, Fabiano Dalpiaz, Sjaak Brinkkemper | 2016 | Automated Extraction of Conceptual Models from User Stories via NLP | International Requirements Engineering Conference (RE) | https://doi.org/10.1109/re.2016.40 |
| S684 | Christopher L. Robinson-Mallett, Robert M. Hierons | 2017 | Integrating Graphical and Natural Language Specifications to Support Analysis and Testing | International Requirements Engineering Conference Workshops (REW) | https://doi.org/10.1109/rew.2017.50 |
| S686 | Danissa V. Rodriguez, Doris L. Carver, Anas Mahmoud | 2018 | An efficient wikipedia-based approach for better understanding of natural language text related to user requirements | IEEE Aerospace Conference | https://doi.org/10.1109/aero.2018.8396645 |
| S688 | Colette Rolland, Camille Ben Achour | 1998 | Guiding the construction of textual use case specifications | Data and Knowledge Engineering | https://doi.org/10.1016/s0169-023x(97)86223-4 |
| S689 | Colette Rolland, V. Plihon | 1996 | Using generic method chunks to generate process models fragments | International Conference on Requirements Engineering | https://doi.org/10.1109/icre.1996.491442 |
| S690 | Lorijn van Rooijen, Frederik Simon Baumer, Marie Christin Platenius, Michaela Geierhos, Heiko Hamann, Gregor Engels | 2017 | From User Demand to Software Service: Using Machine Learning to Automate the Requirements Specification Process | International Requirements Engineering Conference Workshops (REW) | https://doi.org/10.1109/rew.2017.26 |
| S691 | Vivien M. Rooney, Simon N. Foley | 2018 | What Users Want: Adapting Qualitative Research Methods to Security Policy Elicitation | Computer Security | https://doi.org/10.1007/978-3-319-72817-9_15 |
| S692 | Benedetta Rosadini, Alessio Ferrari, Gloria Gori, Alessandro Fantechi, Stefania Gnesi, Iacopo Trotta, Stefano Bacherini | 2017 | Using NLP to Detect Requirements Defects: An Industrial Experience in the Railway Domain | Requirements Engineering: Foundation for Software Quality | https://doi.org/10.1007/978-3-319-54045-0_24 |
| S693 | Michael Roth, Themistoklis Diamantopoulos, Ewan Klein, Andreas Symeonidis | 2015 | Parsing Software Requirements with an Ontology-based Semantic Role Labeler | Workshop on Semantic Parsing | https://doi.org/10.3115/v1/w14-2410 |
| S695 | Daniel Russo, Vincenzo Lomonaco, Paolo Ciancarini | 2018 | A Machine Learning Approach for Continuous Development | Advances in Intelligent Systems and Computing | https://doi.org/10.1007/978-3-319-70578-1_11 |
| S699 | Motoshi Saeki, Hisayuki Horai, Hajime Enomoto | 1989 | Software development process from natural language specification | International conference on Software engineering | https://doi.org/10.1145/74587.74594 |
| S70 | Imran Sarwar Bajwa, M. Abbas Choudhary | 2012 | From Natural Language Software Specifications to UML Class Models | Enterprise Information Systems | https://doi.org/10.1007/978-3-642-29958-2_15 |
| S700 | Vidhu Bhala R. Vidya Sagar, S. Abirami | 2014 | Conceptual modeling of natural language functional requirements | Journal of Systems and Software | https://doi.org/10.1016/j.jss.2013.08.036 |
| S701 | Nicolas Sannier, Morayo Adedjouma, Mehrdad Sabetzadeh, Lionel Briand | 2017 | An automated framework for detection and resolution of cross references in legal texts | Requirements Engineering | https://doi.org/10.1007/s00766-015-0241-3 |
| S702 | Nicolas Sannier, Morayo Adedjouma, Mehrdad Sabetzadeh, Lionel Briand, John Dann, Marc Hisette, Pascal Thill | 2017 | Legal Markup Generation in the Large: An Experience Report | International Requirements Engineering Conference (RE) | https://doi.org/10.1109/re.2017.10 |
| S704 | João Santos, Ana Moreira, João Ara újo, Vasco Amaral, Mauricio Alf érez, Uirá Kulesza | 2008 | Generating Requirements Analysis Models from Textual Requirements | International Workshop on Managing Requirements Knowledge | https://doi.org/10.1109/mark.2008.4 |
| S707 | Edgar Sarmiento, Julio Cesar Sampaio do Prado Leite, Eduardo Almentero | 2014 | CandL: Generating model based test cases from natural language requirements descriptions | International Workshop on Requirements Engineering and Testing (RET) | https://doi.org/10.1109/ret.2014.6908677 |
| S709 | Federica Sarro, Afnan A. Al-Subaihin, Mark Harman, Yue Jia, William Martin, Yuanyuan Zhang | 2015 | Feature lifecycles as they spread, migrate, remain, and die in App Stores | International Requirements Engineering Conference (RE) | https://doi.org/10.1109/re.2015.7320410 |
| S71 | Noor Hasrina Bakar, Zarinah M. Kasirun, Norsaremah Salleh, Hamid A. Jalab | 2016 | Extracting features from online software reviews to aid requirements reuse | Applied Soft Computing | https://doi.org/10.1016/j.asoc.2016.07.048 |
| S710 | Federica Sarro, Mark Harman, Yue Jia, Yuanyuan Zhang | 2018 | Customer Rating Reactions Can Be Predicted Purely using App Features | International Requirements Engineering Conference (RE) | https://doi.org/10.1109/re.2018.00018 |
| S711 | Bahar Sateli, Elian Angius, Rene Witte | 2013 | The ReqWiki Approach for Collaborative Software Requirements Engineering with Integrated Text Analysis Support | Computer Software and Applications Conference | https://doi.org/10.1109/compsac.2013.68 |
| S714 | Kiran Prakash Sawant, Suman Roy, Deepti Parachuri, François Plesse, Pushpak Bhattacharya | 2014 | Enforcing structure on textual use cases via annotation models | India Software Engineering Conference | https://doi.org/10.1145/2590748.2590766 |
| S716 | Pete Sawyer, Paul Rayson, Ken Cosh | 2005 | Shallow knowledge as an aid to deep understanding in early phase requirements engineering | IEEE Transactions on Software Engineering | https://doi.org/10.1109/tse.2005.129 |
| S717 | Pete Sawyer, Paul Rayson, Roger Garside | 2002 | REVERE: Support for Requirements Synthesis from Documents | Information Systems Frontiers | https://doi.org/10.1023/A:1019918908208 |
| S718 | Kurt Schneider, Eric Knauss, Siv Houmb, Shareeful Islam, Jan Jürjens | 2012 | Enhancing security requirements engineering by organizational learning | Requirements Engineering | https://doi.org/10.1007/s00766-011-0141-0 |
| S719 | Peter Schnupp | 1985 | Specification languages | Computer Physics Communications | https://doi.org/10.1016/0010-4655(85)90083-9 |
| S721 | Mathias Schraps, Maximilian Peters | 2014 | Semantic annotation of a formal grammar by Semantic Patterns | International Workshop on Requirements Patterns (RePa) | https://doi.org/10.1109/repa.2014.6894838 |
| S723 | Gunnar Schulze, Joanna Chimiak-Opoka, Jim Arlow | 2012 | An Approach for Synchronizing UML Models and Narrative Text in Literate Modeling | Model Driven Engineering Languages and Systems | https://doi.org/10.1007/978-3-642-33666-9_38 |
| S726 | Matt Selway, Georg Grossmann, Wolfgang Mayer, Markus Stumptner | 2015 | Formalising natural language specifications using a cognitive linguistic/configuration based approach | Information Systems | https://doi.org/10.1016/j.is.2015.04.003 |
| S727 | Jia-Lang Seng, S.Bing Yao, Alan R. Hevner | 2005 | Requirements-driven database systems benchmark method | Decision Support Systems | https://doi.org/10.1016/j.dss.2003.06.002 |
| S728 | Shubhashis Sengupta, Roshni R. Ramnani, Subhabrata Das, Anitha Chandran | 2015 | Verb-based Semantic Modelling and Analysis of Textual Requirements | India Software Engineering Conference | https://doi.org/10.1145/2723742.2723745 |
| S732 | Faiz Ali Shah, Kairit Sirts, Dietmar Pfahl | 2019 | Is the SAFE Approach Too Simple for App Feature Extraction? A Replication Study | Requirements Engineering: Foundation for Software Quality | https://doi.org/10.1007/978-3-030-15538-4_2 |
| S733 | Valerie L. Shalin, Edward J. Wisniewski, Keith R. Levi, Paul D. Scott | 1988 | A formal analysis of machine learning systems for knowledge acquisition | International Journal of Man-Machine Studies | https://doi.org/10.1016/s0020-7373(88)80004-x |
| S736 | Richa Sharma, Jaspreet Bhatia, Kanad K. Biswas | 2014 | Machine learning for constituency test of coordinating conjunctions in requirements specifications | International Workshop on Realizing Artificial Intelligence Synergies in Software Engineering | https://doi.org/10.1145/2593801.2593806 |
| S737 | Richa Sharma, Sarita Gulia, Kanad K. Biswas | 2014 | Automated generation of activity and sequence diagrams from natural language requirements | International Conference on Evaluation of Novel Approaches to Software Engineering | https://doi.org/10.5220/0004893600690077 |
| S738 | Richa Sharma, Nidhi Sharma, Kanad K. Biswas | 2016 | Machine Learning for Detecting Pronominal Anaphora Ambiguity in NL Requirements | Intl Conf on Applied Computing and Information Technology/3rd Intl Conf on Computational Science/Intelligence and Applied Informatics/1st Intl Conf on Big Data, Cloud Computing, Data Science and Engineering (ACIT-CSII-BCD) | https://doi.org/10.1109/acit-csii-bcd.2016.043 |
| S740 | Vibhu Saujanya Sharma, Roshni R. Ramnani, Shubhashis Sengupta | 2014 | A framework for identifying and analyzing non-functional requirements from text | International Workshop on Twin Peaks of Requirements and Architecture | https://doi.org/10.1145/2593861.2593862 |
| S741 | Yonghee Shin, Jane Cleland-Huang | 2012 | A comparative evaluation of two user feedback techniques for requirements trace retrieval | ACM Symposium on Applied Computing | https://doi.org/10.1145/2245276.2231943 |
| S743 | Alberto Rodrigues da Silva | 2014 | SpecQua: Towards a Framework for Requirements Specifications with Increased Quality | Enterprise Information Systems | https://doi.org/10.1007/978-3-319-22348-3_15 |
| S744 | Bruno Cesar F. Silva, Gustavo Carvalho, Augusto Sampaio | 2016 | Test Case Generation from Natural Language Requirements Using CPN Simulation | Lecture Notes in Computer Science | https://doi.org/10.1007/978-3-319-29473-5_11 |
| S747 | Maninder Singh | 2018 | Automated Validation of Requirement Reviews: A Machine Learning Approach | International Requirements Engineering Conference (RE) | https://doi.org/10.1109/re.2018.00062 |
| S748 | Maninder Singh, Vaibhav Anu, Gursimran S. Walia, Anurag Goswami | 2018 | Validating Requirements Reviews by Introducing Fault-Type Level Granularity: A Machine Learning Approach | Innovations in Software Engineering Conference | https://doi.org/10.1145/3172871.3172880 |
| S75 | Flore Barcellini, Camille Albert, Corinne Grosse, Patrick Saint-Dizier | 2012 | Risk Analysis and Prevention: LELIE, a Tool dedicated to Procedure and Requirement Authoring | International Conference on Language Resources and Evaluation | http://www.lrec-conf.org/proceedings/lrec2012/pdf/139_Paper.pdf |
| S750 | Sandeep K. Singh, Reetesh Gupta, Sangeeta Sabharwal, J.P. Gupta | 2009 | Automatic extraction of events from Textual Requirements specification | World Congress on Nature and Biologically Inspired Computing (NaBIC) | https://doi.org/10.1109/nabic.2009.5393565 |
| S751 | Avik Sinha, Amit Paradkar, Hironori Takeuchi, Taiga Nakamura | 2010 | Extending Automated Analysis of Natural Language Use Cases to Other Languages | IEEE International Requirements Engineering Conference | https://doi.org/10.1109/re.2010.52 |
| S757 | John Slankas, Laurie Williams | 2013 | Access Control Policy Extraction from Unconstrained Natural Language Text | International Conference on Social Computing | https://doi.org/10.1109/socialcom.2013.68 |
| S758 | John Slankas, Laurie Williams | 2013 | Automated extraction of non-functional requirements in available documentation | International Workshop on Natural Language Analysis in Software Engineering (NaturaLiSE) | https://doi.org/10.1109/naturalise.2013.6611715 |
| S76 | Fahmi Bargui, Hanene Ben-Abdallah, Jamel Feki | 2009 | Multidimensional concept extraction and validation from OLAP requirements in NL | International Conference on Natural Language Processing and Knowledge Engineering | https://doi.org/10.1109/nlpke.2009.5313769 |
| S760 | Amin Sleimi, Nicolas Sannier, Mehrdad Sabetzadeh, Lionel Briand, John Dann | 2018 | Automated Extraction of Semantic Legal Metadata using Natural Language Processing | International Requirements Engineering Conference (RE) | https://doi.org/10.1109/re.2018.00022 |
| S763 | Harry M. Sneed | 2018 | Requirement-Based Testing - Extracting Logical Test Cases from Requirement Documents | Lecture Notes in Business Information Processing | https://doi.org/10.1007/978-3-319-71440-0_4 |
| S764 | Harry M. Sneed, Chris Verhoef | 2013 | Natural language requirement specification for web service testing | IEEE International Symposium on Web Systems Evolution (WSE) | https://doi.org/10.1109/wse.2013.6642410 |
| S765 | Fábio Soares, João Araújo, Fernando Wanderley | 2015 | VoiceToModel: an approach to generate requirements models from speech recognition mechanisms | ACM Symposium on Applied Computing | https://doi.org/10.1145/2695664.2695724 |
| S767 | Mathias Soeken, Robert Wille, Rolf Drechsler | 2012 | Assisted Behavior Driven Development Using Natural Language Processing | Objects, Models, Components, Patterns | https://doi.org/10.1007/978-3-642-30561-0_19 |
| S768 | Anastasis A. Sofokleous, Andreas S. Andreou | 2008 | Automatic, evolutionary test data generation for dynamic software testing | Journal of Systems and Software | https://doi.org/10.1016/j.jss.2007.12.809 |
| S771 | George Spanoudakis, Andrea Zisman, Elena Pérez-Miñana, Paul Krause | 2004 | Rule-based generation of requirements traceability relations | Journal of Systems and Software | https://doi.org/10.1016/s0164-1212(03)00242-5 |
| S772 | Anjali Sree-Kumar, Elena Planas, Robert Clarisó | 2018 | Extracting software product line feature models from natural language specifications | International Conference on Systems and Software Product Line | https://doi.org/10.1145/3233027.3233029 |
| S777 | Cara Stein, Letha Etzkorn, Dawn Utley | 2004 | Computing software metrics from design documents | ACM Southeast Regional Conference (ACM-SE) | https://doi.org/10.1145/986537.986571 |
| S778 | Michal Steinberger, Iris Reinhartz-Berger, Amir Tomer | 2016 | A Tool for Analyzing Variability Based on Functional Requirements and Testing Artifacts | Lecture Notes in Computer Science | https://doi.org/10.1007/978-3-319-47717-6_21 |
| S784 | Xiaomeng Su, Jon Atle Gulla | 2006 | An information retrieval approach to ontology mapping | Data and Knowledge Engineering | https://doi.org/10.1016/j.datak.2005.05.012 |
| S786 | Hakim Sultanov, Jane Huffman Hayes | 2013 | Application of reinforcement learning to requirements engineering: requirements tracing | IEEE International Requirements Engineering Conference (RE) | https://doi.org/10.1109/re.2013.6636705 |
| S787 | Hakim Sultanov, Jane Huffman Hayes, Wei-Keat Kong | 2011 | Application of swarm techniques to requirements tracing | Requirements Engineering | https://doi.org/10.1007/s00766-011-0121-4 |
| S788 | Dong Sun, Rong Peng | 2015 | A Scenario Model Aggregation Approach for Mobile App Requirements Evolution Based on User Comments | Requirements Engineering in the Big Data Era | https://doi.org/10.1007/978-3-662-48634-4_6 |
| S79 | Frederik S. Bäumer, Michaela Geierhos | 2016 | Running Out of Words: How Similar User Stories Can Help to Elaborate Individual Natural Language Requirement Descriptions | Communications in Computer and Information Science | https://doi.org/10.1007/978-3-319-46254-7_44 |
| S792 | Sahar Tahvili, Marcus Ahlberg, Eric Fornander, Wasif Afzal, Mehrdad Saadatmand, Markus Bohlin, Mahdi Sarabi | 2018 | Functional Dependency Detection for Integration Test Cases | IEEE International Conference on Software Quality, Reliability and Security Companion (QRS-C) | https://doi.org/10.1109/qrs-c.2018.00047 |
| S799 | Jitendra Singh Thakur, Atul Gupta | 2016 | Identifying domain elements from textual specifications | IEEE/ACM International Conference on Automated Software Engineering | https://doi.org/10.1145/2970276.2970323 |
| S800 | U. Thayasivam, K. Verma, A. Kass and R. G. Vasquez | 2011 | Automatically Mapping Natural Language Requirements to Domain-Specific Process Models | Innovative Applications of Artificial Intelligence; Twenty-Third IAAI Conference | https://www.aaai.org/ocs/index.php/IAAI/IAAI-11/paper/view/3440 |
| S802 | Keerthi Thomas, Arosha K. Bandara, Blaine A. Price, Bashar Nuseibeh | 2014 | Distilling privacy requirements for mobile applications | International Conference on Software Engineering | https://doi.org/10.1145/2568225.2568240 |
| S803 | Stephen W. Thomas, Bram Adams, Ahmed E. Hassan, Dorothea Blostein | 2014 | Studying software evolution using topic models | Science of Computer Programming | https://doi.org/10.1016/j.scico.2012.08.003 |
| S805 | Andre Di Thommazo, Thiago Ribeiro, Guilherme Olivatto, Vera Werneck, Sandra Fabbri | 2013 | An Automatic Approach to Detect Traceability Links Using Fuzzy Logic | Brazilian Symposium on Software Engineering | https://doi.org/10.1109/sbes.2013.11 |
| S809 | Walter F. Tichy, Sven J. Koerner | 2010 | Text to software: developing tools to close the gaps in software engineering | FSE/SDP workshop on Future of software engineering research | https://doi.org/10.1145/1882362.1882439 |
| S810 | Saurabh Tiwari, Deepti Ameta, Asim Banerjee | 2019 | An Approach to Identify Use Case Scenarios from Textual Requirements Specification | Innovations on Software Engineering Conference | https://doi.org/10.1145/3299771.3299774 |
| S811 | A Min Tjoa and Linda Berger | 1993 | Transformation of requirement specifications expressed in natural language into an EER model | International Conference on Conceptual Modeling (ER) | https://doi.org/10.1007/bfb0024368 |
| S812 | Sri Fatimah Tjong, Danie Berry | 2013 | The Design of SREE - A Prototype Potential Ambiguity Finder for Requirements Specifications and Lessons Learned | Requirements Engineering: Foundation for Software Quality | https://doi.org/10.1007/978-3-642-37422-7_6 |
| S813 | Maurizio De Tommasi, Angelo Corallo | 2006 | SBEAVER: A Tool for Modeling Business Vocabularies and Business Rules | Lecture Notes in Computer Science | https://doi.org/10.1007/11893011_137 |
| S816 | Reut Tsarfaty, Ilia Pogrebezky, Guy Weiss, Yaarit Natan, Smadar Szekely, David Harel | 2014 | Semantic Parsing Using Content and Context: A Case Study from Requirements Elicitation | Empirical Methods in Natural Language Processing (EMNLP) | https://doi.org/10.3115/v1/d14-1136 |
| S82 | Martin Beckmann, Andreas Vogelsang, Christian Reuter | 2017 | A Case Study on a Specification Approach Using Activity Diagrams in Requirements Documents | International Requirements Engineering Conference (RE) | https://doi.org/10.1109/re.2017.28 |
| S820 | Yoshihisa Udagawa | 2011 | An Augmented Vector Space Information Retrieval for Recovering Requirements Traceability | International Conference on Data Mining Workshops | https://doi.org/10.1109/icdmw.2011.27 |
| S821 | Ashfa Umber, Imran Sarwar Bajwa | 2011 | Minimizing ambiguity in natural language software requirements specification | International Conference on Digital Information Management | https://doi.org/10.1109/icdim.2011.6093363 |
| S823 | Michael Unterkalmsteiner, Tony Gorschek | 2017 | Requirements Quality Assurance in Industry: Why, What and How? | Requirements Engineering: Foundation for Software Quality | https://doi.org/10.1007/978-3-319-54045-0_6 |
| S828 | Varsha Veerappa, Rachel Harrison | 2013 | Assessing the maturity of requirements through argumentation: a good enough approach | IEEE/ACM International Conference on Automated Software Engineering (ASE) | https://doi.org/10.1109/ase.2013.6693131 |
| S83 | B. Belkhouche and J. Kozma | 1993 | Semantic case analysis of informal requirements | The 4th Workshop on the Next Generation of CASE Tools | https://faculty.uaeu.ac.ae/b_belkhouche/Belkhouche/bb_dir/Papiers_publies/Conferences/ootse.pdf |
| S830 | Perla Velasco-Elizondo, Rosario Marín-Piña, Sodel Vazquez-Reyes, Arturo Mora-Soto, Jezreel Mejia | 2016 | Knowledge representation and information extraction for analysing architectural patterns | Science of Computer Programming | https://doi.org/10.1016/j.scico.2015.12.007 |
| S831 | Carlos Videira, David Ferreira, Alberto Rodrigues Da Silva | 2006 | A linguistic patterns approach for requirements specification | EUROMICRO Conference on Software Engineering and Advanced Applications | https://doi.org/10.1109/euromicro.2006.8 |
| S834 | Radu Vlas, William N. Robinson | 2011 | A rule-based natural language technique for requirements | International Conference on System Sciences | https://doi.org/10.1109/hicss.2011.28 |
| | | | discovery and classification in open-source software development projects | | |
| S838 | Jürgen Vöhringer, Günther Fliedl | 2011 | Adapting the Lesk Algorithm for Calculating Term Similarity in the Context of Requirements Engineering | Information Systems Development | https://doi.org/10.1007/978-1-4419-9790-6_63 |
| S841 | Waad Alhoshan, Riza Batista-Navarro and Liping Zhao | 2019 | Semantic Frame Embeddings for Detecting Relations between Software Requirements | 13th International Conference on Computational Semantics (IWCS) | https://www.aclweb.org/anthology/W19-0606 |
| S845 | Chunhui Wang, Fabrizio Pastore, Lionel Briand | 2018 | Automated Generation of Constraints from Use Case Specifications to Support System Testing | International Conference on Software Testing, Verification and Validation (ICST) | https://doi.org/10.1109/icst.2018.00013 |
| S846 | Jian Wang, Neng Zhang, Cheng Zeng, Zheng Li, Keqing He | 2013 | Towards Services Discovery Based on Service Goal Extraction and Recommendation | IEEE International Conference on Services Computing | https://doi.org/10.1109/scc.2013.16 |
| S849 | Wentao Wang, Nan Niu, Hui Liu, Zhendong Niu | 2018 | Enhancing Automated Requirements Traceability by Resolving Polysemy | International Requirements Engineering Conference (RE) | https://doi.org/10.1109/re.2018.00-53 |
| S850 | Yinglin Wang | 2015 | Semantic information extraction for software requirements using semantic role labeling | IEEE International Conference on Progress in Informatics and Computing (PIC) | https://doi.org/10.1109/pic.2015.7489864 |
| S852 | Yinglin Wang, Jianzhang Zhang | 2016 | Experiment on automatic functional requirements analysis with the EFRF's semantic cases | International Conference on Progress in Informatics and Computing (PIC) | https://doi.org/10.1109/pic.2016.7949577 |
| S853 | Kimberly S. Wasson, K. N. Schmid, Robyn R. Lutz, John C. Knight | 2005 | Using occurrence properties of defect report data to improve requirements | IEEE International Conference on Requirements Engineering | https://doi.org/10.1109/re.2005.77 |
| S854 | Jens H. Weber-Jahnke, Adeniyi Onabajo | 2009 | Finding Defects in Natural Language Confidentiality Requirements | IEEE International Requirements Engineering Conference | https://doi.org/10.1109/re.2009.41 |
| S856 | Alexander Weissman, Martin Petrov, Satyandra K. Gupta | 2011 | A computational framework for authoring and searching product design specifications | Advanced Engineering Informatics | https://doi.org/10.1016/j.aei.2011.02.001 |
| S858 | Marcel Wever, Lorijn van Rooijen, Heiko Hamann | 2017 | Active coevolutionary learning of requirements specifications from examples | Genetic and Evolutionary Computation Conference | https://doi.org/10.1145/3071178.3071258 |
| S86 | Allan Berrocal Rojas and Gabriela Barrantes Sliesarieva | 2010 | Automated Detection of Language Issues Affecting Accuracy, Ambiguity and Verifiability in Software Requirements Written in Natural Language | the NAACL HLT 2010 Young Investigators Workshop on Computational Approaches to Languages of the Americas. | http://dl.acm.org/citation.cfm?id=1868701.1868715 |
| S861 | William M. Wilson, Linda H. Rosenberg, Lawrence E. Hyatt | 1997 | Automated analysis of requirement specifications | International conference on Software engineering | https://doi.org/10.1145/253228.253258 |
| S862 | Jonas Winkler, Andreas Vogelsang | 2016 | Automatic Classification of Requirements Based on Convolutional Neural Networks | International Requirements Engineering Conference Workshops (REW) | https://doi.org/10.1109/rew.2016.021 |
| S864 | Stefan Winkler | 2009 | Trace retrieval for evolving artifacts | ICSE Workshop on Traceability in Emerging Forms of Software Engineering | https://doi.org/10.1109/tefse.2009.5069583 |
| S865 | Karolin Winter, Stefanie Rinderle-Ma, Wilfried Grossmann, Ingo Feinerer, Zhendong Ma | 2017 | Characterizing Regulatory Documents and Guidelines Based on Text Mining | On the Move to Meaningful Internet Systems | https://doi.org/10.1007/978-3-319-69462-7_1 |
| S87 | Daniel Berry, Ricardo Gacitua, Pete Sawyer, Sri Fatimah Tjong | 2012 | The Case for Dumb Requirements Engineering Tools | Requirements Engineering: Foundation for Software Quality | https://doi.org/10.1007/978-3-642-28714-5_18 |
| S871 | Ming Xiao, Gang Yin, Tao Wang, Cheng Yang, Mengwen Chen | 2015 | Requirement Acquisition from Social Q&A Sites | Requirements Engineering in the Big Data Era | https://doi.org/10.1007/978-3-662-48634-4_5 |
| S872 | Xusheng Xiao, Amit Paradkar, Suresh Thummalapenta, Tao Xie | 2012 | Automated extraction of security policies from natural-language software documents | International Symposium on the Foundations of Software Engineering | https://doi.org/10.1145/2393596.2393608 |
| S877 | Hui Yang, Anne de Roeck, Vincenzo Gervasi, Alistair Willis, Bashar Nuseibeh | 2010 | Extending Nocuous Ambiguity Analysis for Anaphora in Natural Language Requirements | IEEE International Requirements Engineering Conference | https://doi.org/10.1109/re.2010.14 |
| S878 | Hui Yang, Anne De Roeck, Vincenzo Gervasi, Alistair Willis, Bashar Nuseibeh | 2012 | Speculative requirements: Automatic detection of uncertainty in natural language requirements | IEEE International Requirements Engineering Conference (RE) | https://doi.org/10.1109/re.2012.6345795 |
| S88 | Daniel Berry | 2008 | Ambiguity in Natural Language Requirements Documents | Lecture Notes in Computer Science | https://doi.org/10.1007/978-3-540-89778-1_1 |
| S880 | Li Yi, Wei Zhang, Haiyan Zhao, Zhi Jin, Hong Mei | 2012 | Mining binary constraints in the construction of feature models | IEEE International Requirements Engineering Conference (RE) | https://doi.org/10.1109/re.2012.6345798 |
| S881 | Huishi Yin, Dietmar Pfahl | 2017 | A Method to Transform Automatically Extracted Product Features into Inputs for Kano-Like Models | Product-Focused Software Process Improvement | https://doi.org/10.1007/978-3-319-69926-4_17 |
| S884 | Fatima Zait, Nacereddine Zarour | 2018 | Addressing Lexical and Semantic Ambiguity in Natural Language Requirements | International Symposium on Innovation in Information and Communication Technology (ISIICT) | https://doi.org/10.1109/isiict.2018.8613726 |
| S887 | Yong Zeng | 2008 | Recursive object model (ROM) - Modelling of linguistic information in engineering design | Computers in Industry | https://doi.org/10.1016/j.compind.2008.03.002 |
| S888 | Nicola Zeni, Nadzeya Kiyavitskaya, Luisa Mich, James R. Cordy, John Mylopoulos | 2013 | GaiusT: supporting the extraction of rights and obligations for regulatory compliance | Requirements Engineering | https://doi.org/10.1007/s00766-013-0181-8 |
| S89 | Danie Berry, Nancy Yavne, Moshe Yavne | 1987 | Application of program design language tools to abbott's method of program design by informal natural language descriptions | Journal of Systems and Software | https://doi.org/10.1016/0164-1212(87)90044-6 |
| S891 | Jiansong Zhang, Nora M. El-Gohary | 2017 | Integrating semantic NLP and logic reasoning into a unified system for fully-automated code checking | Automation in Construction | https://doi.org/10.1016/j.autcon.2016.08.027 |
| S892 | Li Zhang, Xin-Yue Huang, Jing Jiang, Ya-Kun Hu | 2017 | CSLabel: An Approach for Labelling Mobile App Reviews | Journal of Computer Science | https://doi.org/10.1007/s11390-017-1784-1 |
| S897 | Hao Zhong, Lu Zhang, Tao Xie, Hong Mei | 2009 | Inferring Resource Specifications from Natural Language API Documentation | IEEE/ACM International Conference on Automated Software Engineering | https://doi.org/10.1109/ase.2009.94 |
| S898 | Jiale Zhou, Yue Lu, Kristina Lundqvist | 2013 | A Context-based Information Retrieval Technique for Recovering Use-Case-to-Source-Code Trace Links in Embedded Software Systems | Euromicro Conference on Software Engineering and Advanced Applications | https://doi.org/10.1109/seaa.2013.30 |
| S901 | Jie Zou, Ling Xu, Mengning Yang, Xiaohong Zhang, Dan Yang | 2017 | Towards comprehending the non-functional requirements through DevelopersÍ eyes: An exploration of Stack Overflow using topic analysis | Information and Software Technology | https://doi.org/10.1016/j.infsof.2016.12.003 |
| S902 | Chuan Duan, Jane Cleland-Huang | 2007 | A Clustering Technique for Early Detection of Dominant and Recessive Cross-Cutting Concerns | IEarly Aspects at ICSE: Workshops in Aspect-Oriented Requirements Engineering and Architecture Design | https://doi.org/10.1109/earlyaspects.2007.1 |
| S903 | Gonzalo Génova, José M. Fuentes, Juan Llorens, Omar Hurtado, Valentín Moreno | 2013 | A framework to measure and improve the quality of textual requirements | Requirements Engineering | https://doi.org/10.1007/s00766-011-0134-z |
| S905 | Eric Knauss, Daniel Ott | 2014 | (Semi-) automatic Categorization of Natural Language Requirements | Requirements Engineering: Foundation for Software Quality | https://doi.org/10.1007/978-3-319-05843-6_4 |
| S906 | Inah Omoronyia, Guttorm Sindre, Tor Stålhane, Stefan Biffl, Thomas Moser, Wikan Sunindyo | 2010 | A Domain Ontology Building Process for Guiding Requirements Elicitation | Requirements Engineering: Foundation for Software Quality | https://doi.org/10.1007/978-3-642-14192-8_18 |
| S907 | Sofija Hotomski, Martin Glinz | 2019 | GuideGen: An approach for keeping requirements and acceptance tests aligned via automatically generated guidance | Information and Software Technology | https://doi.org/10.1016/j.infsof.2019.01.011 |
| S908 | Walid Maalej, Hadeer Nabil; | 2015 | Bug report, feature request, or simply praise? on automatically classifying app reviews | International Requirements Engineering Conference (RE) | https://doi.org/10.1109/re.2015.7320414 |
| S909 | Tao Yue, Lionel C. Briand, Yvan Labiche | 2015 | aToucan: An Automated Framework to Derive UML Analysis Models from Use Case Models | ACM Transactions on Software Engineering and Methodology | https://doi.org/10.1145/2699697 |
| S91 | Valdis Berzins, Craig Martell, Luqi, Paige Adams | 2008 | Innovations in Natural Language Document Processing for Requirements Engineering | Lecture Notes in Computer Science | https://doi.org/10.1007/978-3-540-89778-1_11 |
| S910 | Andrea Di Sorbo, Sebastiano Panichella, Carol V. Alexandru, Junji Shimagaki, Corrado A. Visaggio, Gerardo Canfora, Harald C. Gall | 2016 | What would users change in my app? summarizing app reviews for recommending software changes | International Symposium on the Foundations of Software Engineering | https://doi.org/10.1145/2950290.2950299 |
| S911 | Jin Guo, Jinghui Cheng, Jane Cleland-Huang | 2017 | Semantically enhanced software traceability using deep learning techniques | International Conference on Software Engineering | https://doi.org/10.1109/icse.2017.9 |
| S912 | Alessio Ferrari, Gloria Gori, Benedetta Rosadini, Iacopo Trotta, Stefano Bacherini, Alessandro Fantechi, Stefania Gnesi | 2018 | Detecting requirements defects with NLP patterns: an industrial experience in the railway domain | Empirical Software Engineering | https://doi.org/10.1007/s10664-018-9596-7 |
| S913 | Hui Yang, Anne de Roeck, Vincenzo Gervasi, Alistair Willis, Bashar Nuseibeh | 2011 | Analysing anaphoric ambiguity in natural language requirements | Requirements Engineering | https://doi.org/10.1007/s00766-011-0119-y |
| S914 | Davide Falessi, Giovanni Cantone, Gerardo Canfora | 2013 | Empirical Principles and an Industrial Case Study in Retrieving quivalent Requirements via Natural Language | IEEE Transactions on Software Engineering | https://doi.org/10.1109/tse.2011.122 |
| S915 | Shalini Ghosh, Daniel Elenius, Wenchao Li, Patrick Lincoln, Natarajan Shankar, Wilfried Steiner | 2016 | ARSENAL: Automatic Requirements Specification Extraction from Natural Language | NASA Formal Methods Symposium | https://doi.org/10.1007/978-3-319-40648-0_4 |
| S916 | Alessio Ferrari, Felice dell'Orletta, Giorgio Oronzo Spagnolo, Stefania Gnesi | 2014 | Measuring and Improving the Completeness of Natural Language Requirements | Requirements Engineering: Foundation for Software Quality | https://doi.org/10.1007/978-3-319-05843-6_3 |
| S917 | Vincenzo Gervasi, Didar Zowghi | 2014 | Supporting traceability through affinity mining | International Requirements Engineering Conference | https://doi.org/10.1109/re.2014.6912256 |
| S919 | Horatiu Dumitru, Marek Gibiec, Negar Hariri, Jane Cleland-Huang, Bamshad Mobasher, Carlos Castro-Herrera, Mehdi Mirakhorli | 2011 | On-demand feature recommendations derived from mining public product descriptions | International Conference on Software Engineering | https://doi.org/10.1145/1985793.1985819 |
| S920 | Anas Mahmoud, Nan Niu | 2010 | An experimental investigation of reusable requirements retrieval | IEEE International Conference on Information Reuse and Integration | https://doi.org/10.1109/iri.2010.5558914 |
| S921 | Leonid Kof, Ricardo Gacitua, Mark Rouncefield, Pete Sawyer | 2010 | Concept mapping as a means of requirements tracing | Third International Workshop on Managing Requirements Knowledge | doi: 10.1109/MARK.2010.5623813 |
| S922 | Leah Goldin, Danie Berry | 1994 | AbstFinder, a prototype abstraction finder for natural language text for use in requirements elicitation: design, methodology, and evaluation | Automated Software Engineering | https://doi.org/10.1023/a:1008617922496 |
| S923 | Jane Cleland-Huang , Carl K Chang, Gaurav Sethi, Kumar Javvaji, Haijian Hu, Jinchun Xia | 2002 | Automating speculative queries through event-based requirements traceability | International Conference on Requirements Engineering | https://doi.org/10.1109/ICRE.2002.1048540 |
| S924 | Eduard C. Groen, Jacqueline Schowalter, Sylwia Kopczynska | 2018 | Is there Really a Need for NLP in RE? A Benchmarking Study to Assess Scalability of Manual Analysis | NLP4RE Workshop | http://ceur-ws.org/Vol-2075/NLP4RE_paper11.pdf |
| S925 | Aaron Schlutter, Andreas Vogelsang | 2018 | Knowledge Representation of Requirements Documents Using Natural Language Processing | NLP4RE Workshop | http://ceur-ws.org/Vol-2075/NLP4RE_paper9.pdf |
| S926 | Henning Femmer | 2018 | Requirements Quality Defect Detection with the Qualicen Requirements Scout | NLP4RE Workshop | http://ceur-ws.org/Vol-2075/NLP4RE_paper2.pdf |
| S927 | Daniel Toews, Leif Van Holland | 2019 | Determining Domain Specific Differences of Polysemic Words Using Context Information | NLP4RE Workshop | http://ceur-ws.org/Vol-2376/NLP4RE19_paper02.pdf |
| S928 | Rubens Santos, Eduard C. Groen, Karina Villela | 2019 | A Taxonomy for User Feedback Classifications | NLP4RE Workshop | http://ceur-ws.org/Vol-2376/NLP4RE19_paper10.pdf |
| S929 | Alessandro Fantechi, Stefania Gnesi, Laura Semini | 2019 | From generic requirements to variability | NLP4RE Workshop | http://ksuweb.kennesaw.edu/~pspoleti/REFSQJP19/NLP4RE19_paper16.pdf |
| S930 | Bert de Brock | 2019 | An NL-based Foundation for Increased Traceability, Transparency, and Speed in Continuous Development of Information Systems | NLP4RE Workshop | http://ceur-ws.org/Vol-2376/NLP4RE19_paper18.pdf |
| S931 | Russell J. Abbott | 1983 | Program design by informal English descriptions | Communications of the ACM | https://doi.org/10.1145/182.358441 |
| S933 | Jane Huffman Hayes, Alex Dekhtyar, James Osborne | 2003 | Improving requirements tracing via information retrieval | Journal of Lightwave Technology | https://doi.org/10.1109/icre.2003.1232745 |
| S934 | Jane Huffman Hayes, Alex Dekhtyar, Senthil K. Sundaram | 2006 | Advancing candidate link generation for requirements tracing: the study of methods | IEEE Transactions on Software Engineering | https://doi.org/10.1109/tse.2006.3 |
| S935 | Kevin Ryan | 1993 | The role of natural language in requirements engineering. | IEEE International Symposium on Requirements Engineering | https://doi.org/10.1109/ISRE.1993.324852 |
| S936 | Walid Maalej , Maleknaz Nayebi, Timo Johann , and Guenther Ruhe | 2016 | Toward data-driven requirements engineering | IEEE Software | https://doi.org/10.1109/MS.2015.153 |
| S96 | Jaspreet Bhatia, Travis D. Breaux | 2015 | Towards an information type lexicon for privacy policies | IEEE Eighth International Workshop on Requirements Engineering and Law (RELAW) | https://doi.org/10.1109/relaw.2015.7330207 |
| S851 | Ye Wang, Bo Jiang, Ting Wang | 2016 | Using Workflow Patterns to Model and Validate Service Requirements | 6th IEEE International Workshop on Requirements Patterns (RePa'16) | https://doi.org/10.1109/rew.2016.053 |

Appendix 2. 170 Publication Venues for NLP4RE Studies

| **Publication Venue** |
|---|
| Expert Systems with Applications (ESA) |
| ACM / IEEE International Symposium on Empirical Software Engineering and Measurement (ESEM) |
| ACM Conference on Information and knowledge mining (CIKM) |
| ACM Symposium on Applied Computing (SAC) |
| ACM Transactions on Software Engineering and Methodology (TOSEM) |
| ACM/IEEE International Conference on Software Engineering (ICSE) |
| ACM/IEEE International Conference on Software Engineering Advances |
| Advanced Engineering Informatics |
| Advanced Information Systems Engineering |
| Advances in Intelligent Systems and Computing |
| Advances in Software Engineering |
| Applied Soft Computing |
| Asia-Pacific Software Engineering Conference (APSEC) |
| Association for Computational Linguistics |
| Automated Software Engineering |
| Automation in Construction |
| Balancing Agility and Formalism in Software Engineering |
| Brazilian Symposium on Formal Methods |
| Brazilian Symposium on Software Engineering |
| Communications in Computer and Information Science |
| Communications of the ACM (CACM) |
| Computer |
| Computer Physics Communications |
| Computer Security |
| Computer Software and Applications Conference |
| Computers in Industry |
| Conceptual Modeling |
| Conference of the Centre for Advanced Studies on Collaborative research |
| Conference on Computational Natural Language Learning (CoNLL) |
| Conference on Empirical Methods in Natural Language Processing (EMNLP) |
| Data & Knowledge Engineering (DKE) |
| Decision Support Systems |
| Distributed Embedded Systems: Design, Middleware and Resources |
| Early Aspects at ICSE: Workshops in Aspect-Oriented Requirements Engineering and Architecture Design |
| Empirical Methods in Natural Language Processing (EMNLP) |
| Empirical Software Engineering |
| Enterprise Information Systems |
| International Conference on Conceptual Modeling (ER) |
| Euromicro Conference on Software Engineering and Advanced Applications (SEAA) |
| Expert Systems with Applications (ESA) |
| FSE/SDP Workshop on Future of Software Engineering Research |
| Fundamental Approaches to Software Engineering (FASE) |
| Future Generation Computer Systems (FGCS) |
| Genetic and Evolutionary Computation Conference (GECCO) |
| ICSE Workshop on Comparison and Versioning of Software Models |
| ICSE Workshop on Traceability in Emerging Forms of Software Engineering (TEFSE) |
| IEEE Access |
| IEEE Aerospace Conference |
| IEEE Asia-Pacific Services Computing Conference (APSCC) |
| IEEE International Conference on Information Reuse and Integration (IRI) |
| IEEE International Conference on Progress in Informatics and Computing (PIC) |
| IEEE International Conference on Semantic Computing (ICSC) |
| IEEE International Conference on Services Computing (SCC) |
| IEEE International Conference on Software Quality, Reliability and Security Companion (QRS-C) |
| IEEE International Conference on Systems, Man, and Cybernetics (SMC) |
| IEEE International Requirements Engineering Conference (RE) |
| IEEE International Symposium on Web Systems Evolution (WSE) |
| IEEE International Workshop on Requirements Engineering and Law (RELAW) |
| IEEE International Workshop on Requirements Patterns (RePa) |
| IEEE Software |
| IEEE Transactions on Software Engineering (TSE) |
| IEEE/ACM International Conference on Automated Software Engineering (ASE) |
| IEEE/WIC/ACM International Conference on Web Intelligence (ICWI) |
| India Software Engineering Conference |
| Information and Organization |
| Information and Software Technology (IST) |
| Information Sciences |
| Information Systems |
| Information Systems Development |
| Information Systems Frontiers |
| Innovations in Systems and Software Engineering |
| Innovations on Software Engineering Conference |
| Innovative Applications of Artificial Intelligence |
| International Conference - Cloud System and Big Data Engineering (Confluence) |
| International Conference on Advanced Information Networking and Applications Workshops (WAINA) |
| International Conference on Advanced Information Systems Engineering |
| International Conference on Application of Natural Language to Information Systems |
| International Conference on Business Process Management |
| International Conference on Computational Intelligence and Software Engineering |
| International conference on Computational linguistics |
| International Conference on Computational Semantics (IWCS) |
| International Conference on Conceptual Modeling |
| International Conference on Current Trends in Theory and Practice of Computer Science |
| International Conference on Data Mining Workshops |
| International Conference on Digital Information Management |
| International Conference on Evaluation and Assessment in Software Engineering (EASE) |
| International Conference on Evaluation of Novel Approaches to Software Engineering |
| International Conference on Human System Interaction (HSI) |
| International Conference on Information and Automation |
| International Conference on Information Security Theory and Practice |
| International Conference on Intelligent Computing and Information Systems (ICICIS) |
| International Conference on Knowledge-Based and Intelligent Information and Engineering Systems |
| International Conference on Language Resources and Evaluation |
| International Conference on Mining Software Repositories |
| International Conference on Model Driven Engineering Languages and Systems (MODELS) |
| International Conference on Modelling Techniques and Tools for Computer Performance Evaluation |
| International Conference on Natural Language Processing and Knowledge Engineering |
| International Conference on Product-Focused Software Process Improvement |
| International Conference on Program Comprehension |
| International Conference on Progress in Informatics and Computing (PIC) |
| International Conference on Quality in Research (QiR) |
| International Conference on Quality Software |
| International Conference on Research Challenges in Information Science (RCIS) |
| International Conference on Social Computing |
| International Conference on Software and System Process (ICSSP) |
| International Conference on Software Product Line (SPLC) |
| International Conference on Software Quality |
| International Conference on Software Testing, Verification and Validation (ICST) |
| International Conference on Software Testing, Verification and Validation Workshops |
| International Conference on Sustainable Energy and Intelligent Systems (SEISCON) |
| International Conference on System Sciences |
| International Conference on Systems |
| International Conference on Systems and Software Product Line |
| International Conference on the Quality of Information and Communications Technology (QUATIC) |
| International Journal of Digital Content Technology and its Applications (JDCTA) |
| International Journal of Man-Machine Studies |
| International Symposium on Autonomous Decentralized Systems (ISADS) |
| International Symposium on Information Technology (ICIT) |
| International Symposium on Innovation in Information and Communication Technology (ISIICT) |
| International Symposium on Software Reliability Engineering (ISSRE) |
| International Symposium on the Foundations of Software Engineering |
| International Working Conference on Requirements Engineering: Foundation for Software Quality (REFSQ) |
| International Workshop on Artificial Intelligence for Requirements Engineering (AIRE) |
| International Workshop on Engineering Societies in the Agents World (ESAW) |
| International Workshop on Managing Requirements Knowledge |
| International Workshop on Model-Driven Requirements Engineering (MoDRE) |
| International Workshop on Natural Language Analysis in Software Engineering (NaturaLiSE) |
| International Workshop on Principles of engineering service-oriented systems (PESOS) |
| International Workshop on Realizing Artificial Intelligence Synergies in Software Engineering (RAISE) |
| International Workshop on Requirements Engineering and Testing (RET) |
| International Workshop on Requirements Patterns (RePa) |
| International workshop on Software quality assurance in conjunction with the 6th ESEC/FSE joint meeting - SOQUA |
| International Workshop on Software specification and design (IWSSD) |
| International Workshop on Twin Peaks of Requirements and Architecture |
| International Workshop on Variability Modelling of Software-Intensive Systems (VaMoS) |
| Intl Conf on Applied Computing and Information Technology/3rd Intl Conf on Computational Science/Intelligence and Applied Informatics/1st Intl Conf on Big Data, Cloud Computing, Data Science and Engineering (ACIT-CSII-BCD) |
| Joint Meeting on Foundations of Software Engineering |
| Journal of Computer Science |
| Journal of King Saud University - Computer and Information Sciences |
| Journal of Lightwave Technology (JLT) |
| Journal of Systems and Software (JSS) |
| Knowledge-Based Systems |
| Library and Information Science Research |
| Malaysian Conference in Software Engineering (MySEC) |
| Model Driven Engineering Languages and Systems (MODELS) |
| Modelling Foundations and Applications |
| NASA Formal Methods Symposium (NFM) |
| NASA Goddard Software Engineering Workshop |
| Natural Language Engineering |
| NLP4RE Workshop |
| On the Move to Meaningful Internet Systems (OTM) |
| Procedia Computer Science |
| Requirements Engineering in the Big Data Era |
| Requirements Engineering Journal (REJ) |
| Satellite Events at the MoDELS |
| Saudi Computer Society National Computer Conference (NCC) |
| Science of Computer Programming |
| Seminal Contributions to Information Systems Engineering |
| Software and Systems Modeling (SSM) |
| Software Quality Journal |
| ACM Southeast Regional Conference (ACM-SE) |
| System Level Design with .Net Technology |
| The 4th Workshop on the Next Generation of CASE Tools |
| The Journal of Logic and Algebraic Programming |
| The NAACL HLT 2010 Young Investigators Workshop on Computational Approaches to Languages of the Americas. |
| Trustworthy Eternal Systems via Evolving Software, Data and Knowledge (EternalS) |
| Working Conference on Mining Software Repositories (MSR) |
| Working Conference on Reverse Engineering (WCRE) |
| Workshop on Semantic Parsing |
| World Congress on Nature and Biologically Inspired Computing (NaBIC) |